\documentclass[superscriptaddress,11pt]{revtex4}

\usepackage{color}

\usepackage{amssymb,amsmath,array,verbatim,amsthm}
\usepackage[dvipdfmx]{graphicx}
\usepackage{epsfig}

\newcommand{\eq} {equation}
\newcommand{\eqa} {eqnarray}
\newcommand{\NN} {\nonumber}

\newcommand{\bpsi} {\bar{\psi}}

\newcommand{\blam} {\bar{\lambda}}

\newcommand{\be}{\begin{eqnarray}}
\newcommand{\ee}{\end{eqnarray}}

\newcommand{\nl}{\nonumber \\}
\newcommand{\pd}{\partial}
\newcommand{\Tr}{{\rm Tr}}

 \makeatletter
 
 \@addtoreset{equation}{section}
 \makeatother
    
\begin{document}
\begin{flushright}
{\footnotesize WIS/09/27-May-DPPA}
\end{flushright}

\title{\Large
{Resurgence and Lefschetz thimble in 
3d $\mathcal{N}=2$ supersymmetric Chern-Simons matter theories}
}

\author{Toshiaki Fujimori}
\email{toshiaki.fujimori018(at)gmail.com}
\affiliation{
Department of Physics, 
and Research and 
Education Center for Natural Sciences, 
Keio University, 4-1-1 Hiyoshi, Yokohama, Kanagawa 223-8521, Japan
}

\author{Masazumi Honda}
\email{masazumi.honda(at)weizmann.ac.il}
\affiliation{
Department of Particle Physics, Weizmann Institute of Science, Rehovot 7610001, Israel}

\author{Syo Kamata}
\email{skamata(at)ncsu.edu}
\affiliation{
Department of Physics, North Carolina State University, Raleigh, NC 27695, USA}

\author{Tatsuhiro Misumi}
\email{misumi(at)phys.akita-u.ac.jp}
\affiliation{
Department of Mathematical Science, Akita 
University,  Akita 010-8502, Japan
}
\affiliation{
Department of Physics, 
and Research and 
Education Center for Natural Sciences, 
Keio University, 4-1-1 Hiyoshi, Yokohama, Kanagawa 223-8521, Japan
}
\affiliation{
iTHEMS, RIKEN,
2-1 Hirasawa, Wako, Saitama 351-0198, Japan
}

\author{Norisuke Sakai}
\email{norisuke.sakai(at)gmail.com}
\affiliation{
Department of Physics, 
and Research and 
Education Center for Natural Sciences, 
Keio University, 4-1-1 Hiyoshi, Yokohama, Kanagawa 223-8521, Japan
}

\begin{abstract}
We study a certain class of supersymmetric (SUSY) observables 
in 3d $\mathcal{N}=2$ SUSY Chern-Simons (CS) matter theories
and investigate how their exact results are related to
the perturbative series with respect to coupling constants given by inverse CS levels.
We show that the observables have nontrivial resurgent structures 
by expressing the exact results 
as a full transseries consisting of perturbative and non-perturbative parts. 
As real mass parameters are varied, 
we encounter Stokes phenomena at an infinite number of points, 
where the perturbative series becomes non-Borel-summable due to singularities 
on the positive real axis of the Borel plane.
We also investigate the Stokes phenomena when the phase of the coupling constant is varied.
For these cases, 
we find that the Borel ambiguities in the perturbative sector
are canceled by those in nonperturbative sectors
and end up with an unambiguous result 
which agrees with the exact result
even on the Stokes lines.
We also decompose the Coulomb branch localization formula,
which is an integral representation for the exact results, 
into Lefschetz thimble contributions and study how they are related to the resurgent transseries.
We interpret the non-perturbative effects appearing 
in the transseries as contributions of complexified SUSY solutions which formally satisfy the SUSY conditions 
but are not on the original path integral contour.
\end{abstract}

\maketitle
\newpage

\tableofcontents

\newpage

\section{Introduction}

Perturbative series in quantum field theory (QFT)
is usually divergent \cite{Dyson:1952tj}.
One of standard procedures to take resummation of divergent series is Borel resummation. 
Given a perturbative series
\be
F_{pert}(g)=\sum_{\ell =0}^\infty c_\ell \, g^{a+\ell},
\ee
Borel resummation of $F_{pert}(g)$ 
along the direction $\varphi$ is defined by 
\begin{\eq}
\mathcal{S}_\varphi F_{pert} (g)
=\int_0^{\infty e^{i\varphi}} dt\ 
e^{-\frac{t}{g}} \mathcal{B} \hspace{-1pt} F_{pert}(t) ,
\label{eq:Borel} 
\end{\eq}
where $\mathcal{B} \hspace{-1pt} F_{pert}(t)$ is 
the analytic continuation of the formal Borel transformation  
$\sum_{\ell =0}^\infty \frac{c_\ell}{\Gamma (a+\ell )} t^{a+\ell -1}$
and $\varphi$ is usually taken as $\varphi ={\rm arg}(g)$.
It is known (or expected) that
$\mathcal{B} \hspace{-1pt} F_{pert}(t)$ in typical QFT
has singularities along the positive real axis $\mathbb{R}_+$ 
in complex $t$-plane called the Borel plane. 
Some of the famous examples are quantum mechanics 
with degenerate classical vacua and 
asymptotically free field theories \cite{tHooft:1977xjm}.
In this situation,
the integral \eqref{eq:Borel} with $\varphi =0$ is ill-defined
and hence we have to deform the integration contour 
or equivalently complexify the parameter $g$ to avoid the singularities. 
Consequently, the integral becomes ambiguous
depending on the way of avoiding the singularities.
In resurgence theory \cite{Ec1}, 
which is often useful in such situations, 
one considers the following ansatz called a ``transseries" 
for the exact result of the physical quantity 
\begin{\eq}
F(g)\ =\ C_0 \sum_{\ell} c_\ell^{(0)} g^\ell
+\sum_{I\in {\rm saddles}} C_I  e^{-\frac{S_I}{g}} 
\sum_{\ell} c_\ell^{(I)} g^\ell ,
\label{eq:transseries}
\end{\eq}
where $I$ labels nonperturbative saddle points and $S_I$ are 
the actions at the saddle points. 
$C_I$ denotes a transseries parameter which can jump at certain 
values of parameters called ``Stokes lines".
It is expected that the ambiguities of perturbative Borel 
resummation are canceled by those of the nonperturbative saddles 
and one can obtain an unambiguous answer 
which is equivalent to the exact result.
Typically a divergent perturbative series and 
non-perturbative contributions are related with each other 
via the cancellation of the ambiguities. 
Such a significant relation, 
called a ``resurgent relation", 
enables us to reconstruct non-perturbative terms 
from divergent perturbative series and vice versa
\cite{Brezin:1977ab,Bogomolny:1980ur,Ec1,Voros1,Pham1, BH1,Howls1,DH1,K1}.

Resurgence theory has a long history 
in quantum mechanics and differential equations.
There have been various applications 
in a variety of physical systems 
including quantum mechanics (QM) 
\cite{Alvarez1,ZinnJustin:2004ib,Dunne:2013ada, Basar:2013eka,
Dunne:2014bca,Escobar-Ruiz:2015nsa,Misumi:2015dua,Behtash:2015zha,Gahramanov:2015yxk,Dunne:2016qix,Fujimori:2016ljw,Dunne:2016jsr,Serone:2016qog,Basar:2017hpr,Alvarez:2017sza,Behtash:2018voa}, 
hydrodynamics \cite{Basar:2015ava}, 
non-critical \cite{noncritical} and topological string theory \cite{Marino:2006hs,Grassi:2014cla} as well as QFT 
\footnote{See also reviews on math side  \cite{Costin1} 
and physics side \cite{Marino:2012zq}.
}.
There are various types of applications to QFT
such as in weak coupling expansions, strong coupling expansions \cite{Aniceto:2015rua}, $1/N$-expansions \cite{Grassi:2014cla},
large-$N_f$ expansions \cite{Gukov:2016tnp} and expansions by geometric parameters of space \cite{Gukov:2017kmk}. 
In this paper we make further progress 
in understanding the applications of resurgence theory 
to the weak coupling expansions in QFT with Lagrangians. 
The weak coupling expansion of QFT in the context of resurgence 
theory has been studied in 2D quantum field theories 
\cite{Dunne:2012ae,Cherman:2013yfa,Cherman:2014ofa,Misumi:2014jua,Nitta:2014vpa,Behtash:2015kna,Dunne:2015ywa,Buividovich:2015oju,Demulder:2016mja,Sulejmanpasic:2016llc}, 
3D pure Chern-Simons theory \cite{Gukov:2016njj,Gang:2017hbs},
4D non-SUSY QFT 
\cite{Argyres:2012vv,Dunne:2015eoa,Yamazaki:2017ulc}
and supersymmetric (SUSY) gauge theories in various dimensions 
\cite{Russo:2012kj,Aniceto:2014hoa,Honda:2016mvg,Honda:2016vmv,
Honda:2017qdb,Dorigoni:2017smz,Honda:2017cnz}. 
In all the known examples with sufficient data, 
observables have resurgent structures 
with respect to the coupling parameter
and unambiguous transseries expressions, 
which agree with exact results.
However it is currently unclear 
which observables/theories have resurgent structures.
In other words,
we do not know
when one obtains an unambiguous answer by the resurgence procedure
and when the answer obtained in this manner agrees with the exact result. 
If we can identify such a class,
then we can obtain ``semi-classical decoding" \cite{decoding} 
of exact results or conversely, may use the resurgent structure
to define QFT for this class.

In general, it is much harder to study 
the resummation problem in QFT than in quantum mechanics
since Schr\"odinger equations are not available 
and we have to confront 
the saddle-point analysis of path integrals ``seriously". 
According to the recent progress in understanding 
the resurgent structure of QM 
from the path integral viewpoint \cite{Basar:2013eka},
what we have to do is as follows: 
\begin{itemize}
\item Find all critical points including complex saddles.
\item See which critical points contribute 
in terms of Lefschetz thimble decompositions.
\item Study perturbative expansions around contributing critical points.
\end{itemize}
We know that the first step is already technically hard in typical QFT
and the second step is harder than the first step.
Indeed there are only few known examples of 
physical quantities satisfying the following ideal conditions: 
\begin{enumerate}
\item physical quantities in $d$-dimensional QFT ($d\geq 2$),
\item quantities for which mathematically well-defined descriptions for their exact results are known 
\footnote{
This does not necessarily mean that closed expressions for the exact results are explicitly known. 
For example, localization method typically provides finite dimensional integral representations for the exact results
but we likely do not know how to perform the integrals analytically for gauge theories with multiple finite ranks. 
In this situation, we know the mathematically well-defined descriptions for the exact results
but do not know their final closed expressions. 
},
\item quantities with the non-trivial resurgent structure.
\end{enumerate}
To the best of our knowledge, 
the only examples satisfying all these three conditions
are 2d pure YM theory \cite{Ahmed:2017lhl} and 
pure CS theory \cite{Gukov:2016njj}
\footnote{
If we count so-called Cheshire cat resurgence \cite{Dunne:2016jsr},
then the $S^2$ partition function of 
the 2d $\mathcal{N}=(2,2)$ $\mathbb{CP}^N$ model
 (and correlators generated by it)
also provides the example \cite{Dorigoni:2017smz},
where an expansion parameter is inverse of FI-parameter.
}.
Main reasons for the difficulties to find such examples are
that the condition 2 is not satisfied in most cases at present 
and it is too complicated to check 
whether or not they satisfy the condition 3. 
Although exactly solvable quantities
trivially satisfy the condition 2, 
they often do not satisfy the condition 3.
Namely, they usually have truncated, 
convergent or Borel summable weak-coupling perturbative series, 
which has the trivial resurgent structure and gives an unambiguous result.
A certain class of models becomes solvable in the large-$N$ limit 
but perturbative series with respect to the 't Hooft coupling 
in large-$N$ QFTs is typically convergent \cite{tHooft:1982uvh}
\footnote{
When we have IR renormalons, this would not be true.
}.  
In some supersymmetric gauge theories, 
we have non-renormalizable theorems 
which imply that some observables are tree-level or 1-loop exact.  
The prepotentials of 4d $\mathcal{N}=2$ theories 
receive an infinite number of instanton corrections
but its perturbative series in each sector is truncated \cite{Seiberg:1994rs}.
One of more non-trivial examples is a class of SUSY observables in 4d $\mathcal{N}=2$ theories
which also receive instanton corrections and have an asymptotic perturbative series in every sector,
but all the perturbative series are Borel summable 
and hence unambiguous \cite{Russo:2012kj,Aniceto:2014hoa,Honda:2016mvg,Honda:2017cnz}.

In this paper we propose an infinite number of examples satisfying 
all the above conditions 1, 2 and 3.
The examples are a certain class of supersymmetric observables
in 3d $\mathcal{N}=2$ SUSY Chern-Simons (CS) theories coupled to matters,
which appear in a broad context of theoretical physics
such as AdS/CFT, M-theory, duality, higher spin gauge theory,
condensed matter physics and so on.
A typical quantity of this class is the partition function on $S^3$.
Although the partition function is originally defined by the infinite-dimensional path integral,
it is known that the partition function of 3d $\mathcal{N}=2$ theory on $S^3$ 
has a finite-dimensional integral representation obtained 
by the SUSY localization method \cite{Pestun:2007rz} whose dimension is a rank of gauge group
\footnote{
Partition function on $S^3$ (more generally odd-dimensional sphere)
is physical in the following sense:
First, there is no IR divergence since sphere is compact.
Second, $\log{|Z|}$ has power-law UV divergence
but does not have $\log$-divergence in odd dimensions.
Therefore $\mathcal{O}(1)$ part of $\log{|Z|}$ cannot be changed 
by counter terms and is physical,
though there are counter terms to shift phase of $Z$
to some extent \cite{Closset:2012vp}.
$Z_{S^3}$ in the main text of the present paper means this $\mathcal{O}(1)$ part.  
}
\begin{\eq}
Z_{S^3}(g,m) 
=\int_{\mathbb{R}^N} d^N \sigma\ e^{-S[\sigma ]} ,
\label{eq:Clocalization}
\end{\eq}
where $g$ is a coupling constant proportional to the inverse of CS level $k$,
$N$ is rank of gauge group and $\sigma$ is a Coulomb branch parameter.
The integrand is uniquely determined by specifying the gauge group, the representation of matters, $U(1)_R$ charges, CS levels, FI parameters and real masses 
\footnote{
A real mass is given by a constant background of the flavor vector multiplet.
}.
Since this is just a finite-dimensional integral,
it obviously satisfies the condition 2.
Furthermore, we will discuss that it is a resurgent function of $g$
and has non-trivial resurgent structures.

Another motivation of this paper comes from mysterious results in the same setup
previously found by one of the present authors \cite{Honda:2017qdb}.
First, the work \cite{Honda:2017qdb} found 
an explicit finite-dimensional integral representation
of perturbative Borel transformation for the $S^3$ partition functions
in 3d $\mathcal{N}=2$ SUSY Chern-Simons matter theory 
\footnote{
More precisely 
a class of theory considered in \cite{Honda:2017qdb}
is theory with well-defined sphere partition functions
though ill-defined cases is also interesting \cite{Morita:2011cs}.
}.
Second, Borel summability along $\mathbb{R}_+$ on the Borel plane
depends on matter contents and values of real masses.
Third, the exact result is always the same as
the Borel resummation along half imaginary axis:
\begin{\eq}
Z_{S^3} =\int_{0}^{-i\infty} dt\ e^{-\frac{t}{g}} \mathcal{B}Z(t)\, \quad\quad\quad (k>0).
\end{\eq}
Technically these results were obtained by rewriting the exact result 
and we did not have appropriate interpretations for them.
To obtain more precise understanding of these results,
we decompose the integration path of the Coulomb branch localization formula \eqref{eq:Clocalization}
into a sum of Lefschetz thimbles (steepest descent contours)
\footnote{
Decomposition of localization formula by Lefschetz thimble
has been considered in \cite{Behtash:2017rqj}
in the context of Witten index of SUSY QM.
}
which has been recently applied in a variety of contexts
such as analytic continuation of path integral \cite{Witten:2010cx,Witten:2010zr},
real time path integral \cite{Tanizaki:2014xba,Alexandru:2016gsd},
black hole information problem \cite{Fitzpatrick:2016ive},
cosmology \cite{Feldbrugge:2017mbc}, 
the sign problems in Monte Carlo simulation
\cite{Cristoforetti:2013wha,Fujii:2015bua,Alexandru:2016gsd}, 
and of course resurgence theory \cite{Basar:2013eka}.
The advantage to use Lefschetz thimbles in our problem is that
one can systematically express the exact result as a sum over contributions from critical points.
In particular, we can determine which critical points contribute to the integral
by looking at the intersections of dual thimbles (steepest ascent contours) and
the original integral contour even if the critical points are not on the original integration contour. 
As we will see, in our setups, the intersection numbers depend on the values of the real masses
and precisely describe the step-function behavior of the transseries parameter.
We will discuss how the Lefschetz thimble decomposition
is related to the resurgent transseries.

We explicitly demonstrate the above arguments
based on partition functions of a certain class of rank-1 3d $\mathcal{N}=2$ CS matter theories on $S^3$.
Let us briefly summarize our results in the simplest nontrivial theory:
the $\mathcal{N}=3$ CS SQED
which is $\mathcal{N}=3$ $U(1)$ CS theory coupled to a charge-$1$ hyper multiplet with a real mass $m$.
This model can be regarded as a special case of the 3d $\mathcal{N}=2$ theories.
The exact result for the sphere partition function of this theory is simply given by
\begin{\eq}
Z 
=\int_{-\infty}^\infty d\sigma \  \frac{e^{\frac{ik}{4\pi}\sigma^2}}{2\cosh{\frac{\sigma -m}{2}} }\, .
\end{\eq}
It has been shown \cite{Honda:2016vmv} that
this expression is regarded as the Borel resummation along the direction $\varphi =-\pi /2$:
\begin{\eq}
Z =\int_{0}^{-i\infty} dt\ e^{-\frac{t}{g}} \mathcal{B}Z(t)\, ,
\end{\eq}
where the Borel transformation $\mathcal{B}Z(t)$ will be
explicitly given in \eqref{eq:U1BorelT} later.
By changing the integration contour,
we can also write this as
the Borel resummation along $\mathbb{R}_+$ 
plus residues in the 4th quadrant of Borel plane:
\begin{\eq}
Z 
= \int_{0}^\infty dt\ e^{-\frac{t}{g}} \mathcal{B}Z(t)
+\sum_{{\rm poles}\in {\rm 4th\ quadrant}} {\rm Res}_{t=t_{\rm pole}} 
\Bigl[ e^{-\frac{t}{g}} \mathcal{B}Z(t) \Bigr] ,
\end{\eq}
where the second term generates non-perturbative corrections.
The most important point here is that
distribution of the poles depends on the mass $m$.
We will discuss that
the number of the poles in the 4th quadrant is $|n|$ when $(2n-1)\pi<m<(2n+1)\pi$
and find that $Z$ has the following transseries expression
\begin{\eq}
Z
=\sum_{q} c_q^{(0)} (m) g^q
+\sum_{n =1}^\infty \theta ( m -(2n -1)\pi )  
e^{\frac{i}{g}[m+(2n-1)\pi i]^2} 
\sum_{q} c_q^{(n )}(m) g^q ,
\end{\eq}
where $\theta (x)$ is step function
and the perturbative coefficients $c_q^{(n)}(m)$
will be given in \eqref{eq:coeffP}.
The second term consists of exponentially suppressed corrections
which are identified as the non-perturbative contributions.
We will show that the transseries has a nontrivial resurgent structure
and hence gives the unambiguous answer in agreement with the exact result. 
We will also decompose the Coulomb branch localization formula \eqref{eq:Clocalization}
in terms of Lefschetz thimbles 
and discuss relations between the transseries expression
and the thimble decomposition.
We will first find critical points around the origin and singularities of the integrand,
which are interpreted as perturbative and non-perturbative critical points respectively.
It will be shown that the value of real mass $m$ determines
which thimbles associated with the nonperturbative critical points are contributing
while the perturbative thimble always contributes to the partition function.
However, it will be also shown that the correspondence between each thimble integral and each of the building blocks of the transseries is complicated for finite $g$.
We will argue that one building block of the transseries can be given by the multiple thimble integrals.
For example, a sum of the perturbative thimble integral and one of the nonperturbative thimble integrals coincides with the perturbative Borel resummation along $\mathbb{R}_+$ in a certain region of the real mass $m$.

We also discuss path integral interpretation 
of the non-perturbative contributions appearing in the resurgent transseries.
Recently, one of the present authors has found complexified supersymmetric solutions 
in general 3d $\mathcal{N}=2$ SUSY field theory on $S^3$
which formally satisfy SUSY conditions but are not on the original path integral contour,
and then proposed that these solutions correspond
to the singularities of the Borel transformation of the perturbative series (Borel singularities)
in 3d $\mathcal{N}=2$ SUSY Chern-Simons matter theory \cite{Honda:2017qdb}.
We discuss possible interpretation of the nonperturbative effects in terms of the complexified SUSY solutions.

This paper is organized as follows:
In Sec.~\ref{SQED}, we first obtain 
the full transseries expression of the partition function 
in the $\mathcal{N}=3$ CS SQED.
Next we discuss the thimble decomposition of the partition function expressed as the integral with respect to the Coulomb branch parameter, with emphasis on the Stokes phenomena at the special values of real mass.
In Sec.~\ref{SQCD}, we obtain the full transseries of partition function 
in $SU(2)$ vector multiplet with the Chern-Simons term coupled with hyper multiplets (CS SQCD), 
where we discuss the thimble decomposition and the Stokes phenomena.
In Sec.~\ref{sec:generalization}, we discuss generalization to more generic theories and other observables.
In Sec.~\ref{Int}, we propose an interpretation on the relation between the complex saddles of the Coulomb branch parameter and the complex SUSY solutions of the CS SQED and CS SQCD.
Sec.~\ref{SD} is devoted to summary and discussion.

\section{$\mathcal{N}=3$ Chern-Simons SQED}
\label{SQED}
In this section, we study the $S^3$ partition function of 
3D $\mathcal{N}=3$ $U(1)_k$ CS theory with $N_f$ charge $+1$ hyper multiplets,
which we call $\mathcal{N}=3$ CS SQED
\footnote{
By ``$G_k$ CS theory", we mean CS theory with gauge group $G$ and CS level $k$.
}.
In the 3D $\mathcal{N}=2$ language,
this theory consists of an ${\mathcal N}=2$ vector multiplet,
an adjoint chiral multiplet with $U(1)_R$ charge 1
and $N_f$ pairs of charge $+1$ and $-1$ chiral multiplets 
with $U(1)_R$ charge $1/2$
\footnote{
The adjoint chiral multiplet with $U(1)_R$ charge 1 is technically irrelevant
because this contributes to the integrand by 1.
}.
We also turn on real masses $m_a$ $(a=1,2,...,N_{f})$
associated with the $U(N_f )$ flavor transformation of the hyper multiplets
\footnote{
In 3d $\mathcal{N}=2$ language,
we have $U(N_f ) \times U(N_f )$ flavor symmetry
whose first one rotates the charge $+1$ chirals with $N_f$ representation
while second one rotates the charge $-1$ chirals with $\bar{N_f}$ representation.
If we denote real masses associated with these two $U(N_f )$ symmetries
by $m^f_a$ and $\bar{m}^f_a$ respectively,
then we are taking $m_a =m^f_a =\bar{m}^f_a$ 
which corresponds to so-called vector real mass.
Since we are considering $U(1)$ gauge theory,
the diagonal part of the vector real mass $m$ 
is absorbed by shifting $\sigma \rightarrow \sigma +m$
but this absorption gives FI-term with the coefficient $km /2\pi$
and $U(1)$-flavor CS term with level $k$ because of the $U(1)$ gauge CS term.
Thus one can also say that
this setup is $U(1)_k$ CS theory with FI-parameter
and vector real masses associated with $SU(N_f )$ flavor symmetry. 
}.
 
Applying the SUSY localization \cite{Pestun:2007rz} to the present theory,
the partition function 
is expressed as \cite{Kapustin:2009kz,Jafferis:2010un}
\begin{\eq}
Z 
=\int_{-\infty}^\infty d\sigma \  \frac{e^{\frac{ik}{4\pi}\sigma^2}}{\prod_{a=1}^{N_f} 2\cosh{\frac{\sigma -m_a}{2}} }\,,
\label{eq:Zexact}
\end{\eq}
where $\sigma$ is the Coulomb branch parameter
given by constant configuration of the adjoint scalar in 3d $\mathcal{N}=2$ vector multiplet
\footnote{We are taking radius of $S^3$ to be 1.
The dependence on the radius can be recovered by $\sigma \rightarrow R_{S^3} \sigma$ and $m_a \rightarrow R_{S^3}m_a$. 
It is also known that $S^3$ partition function of 3d $\mathcal{N}=2$ theory is independent of Yang-Mills coupling
because of $Q$-exactness.
Therefore even if we add super Yang-Mills action, we still have the same partition function \eqref{eq:Zexact}.}. 
 
In Sec.~\ref{sec:SQED_trans}, we show that the exact partition 
function obtained by the localization technique with respect to 
the Coulomb branch parameter can be written as a full 
transseries with non-perturbative exponential contributions.
In Sec.~\ref{sec:SQED_thimble}, we argue the thimble decomposition 
of the integral with respect to the Coulomb branch 
parameter. 
In both cases, we discuss the Stokes phenomena 
at the special real masses.

\subsection{Exact results as resurgent transseries}
\label{sec:SQED_trans}
\begin{figure}[t]
\begin{center}
\includegraphics[clip, width=70mm]{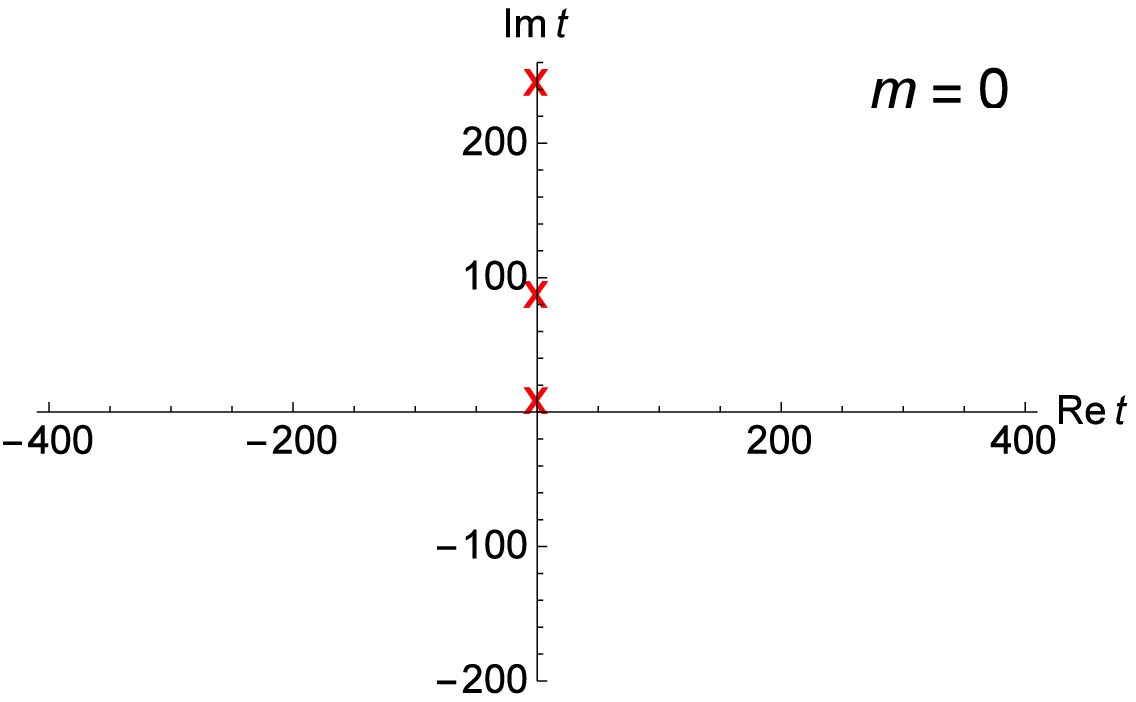}
\includegraphics[clip, width=70mm]{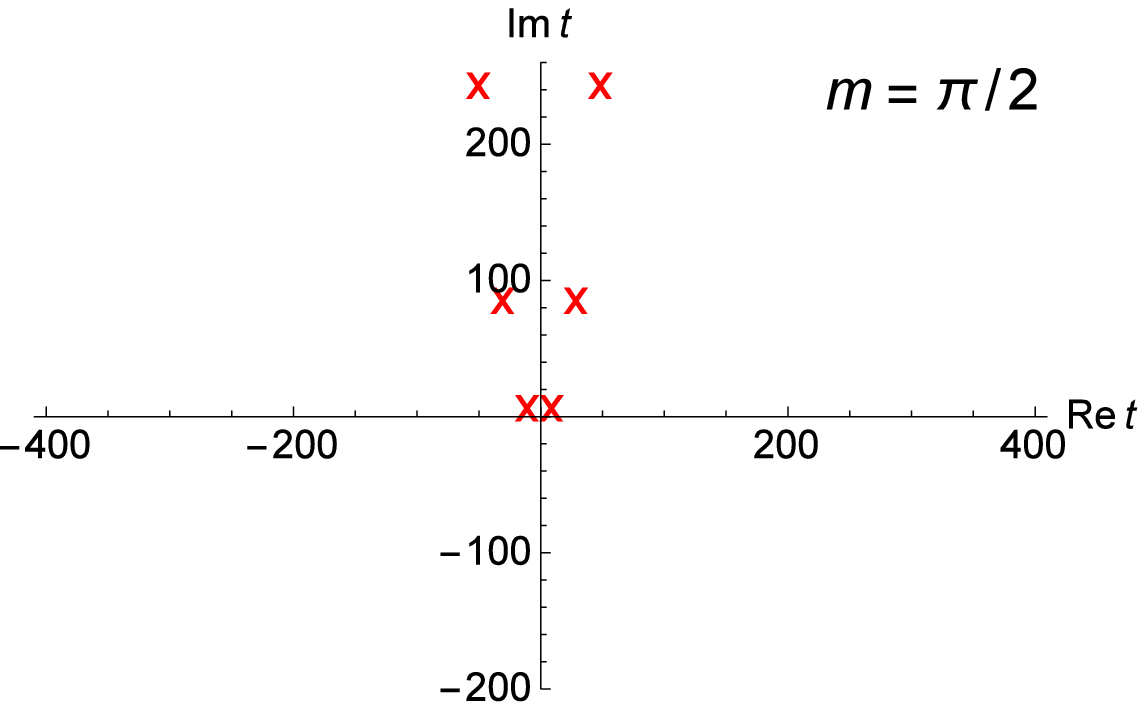}\\
\includegraphics[clip, width=70mm]{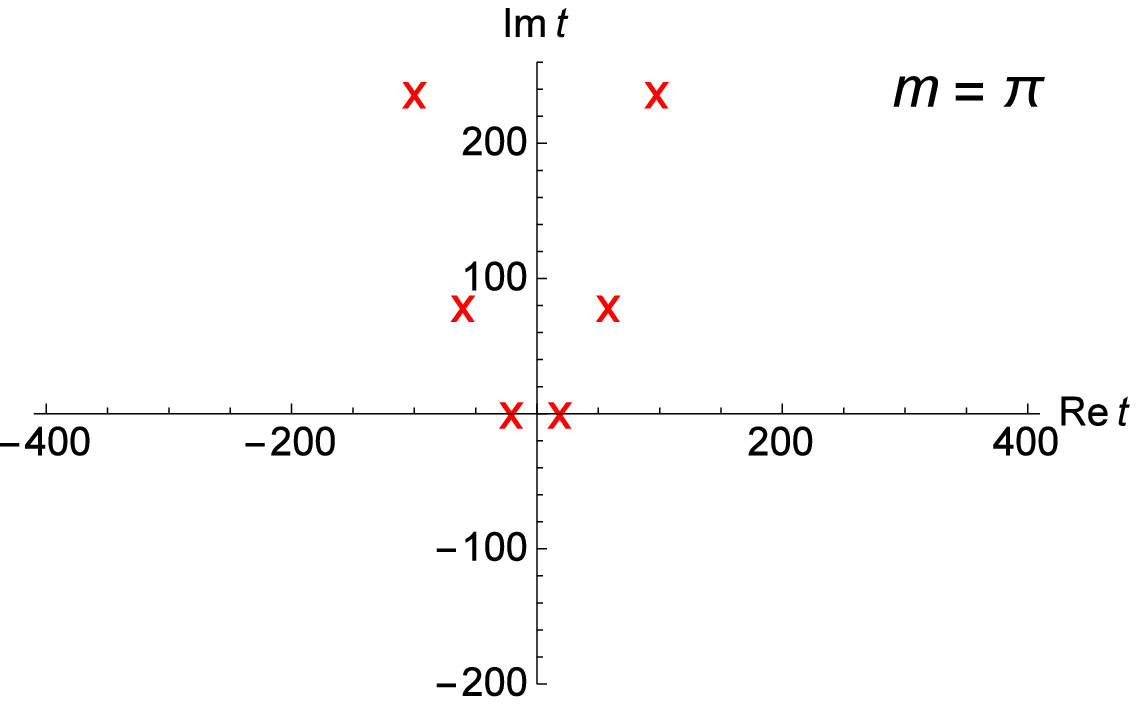}
\includegraphics[clip, width=70mm]{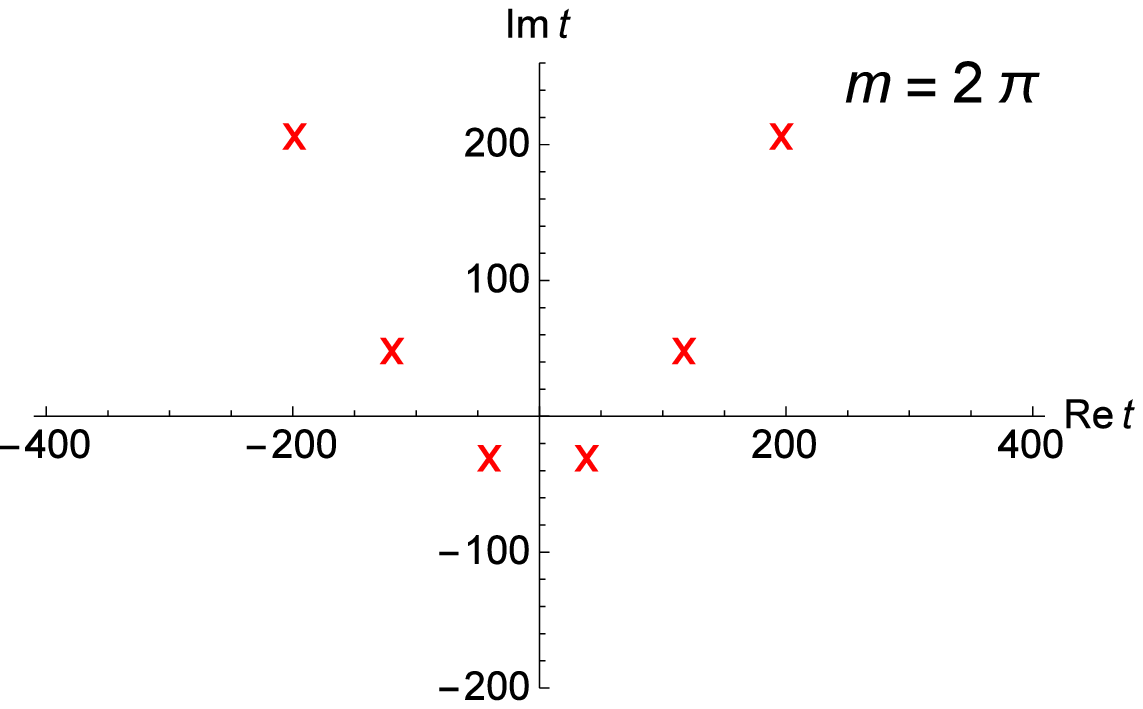}\\
\includegraphics[clip, width=70mm]{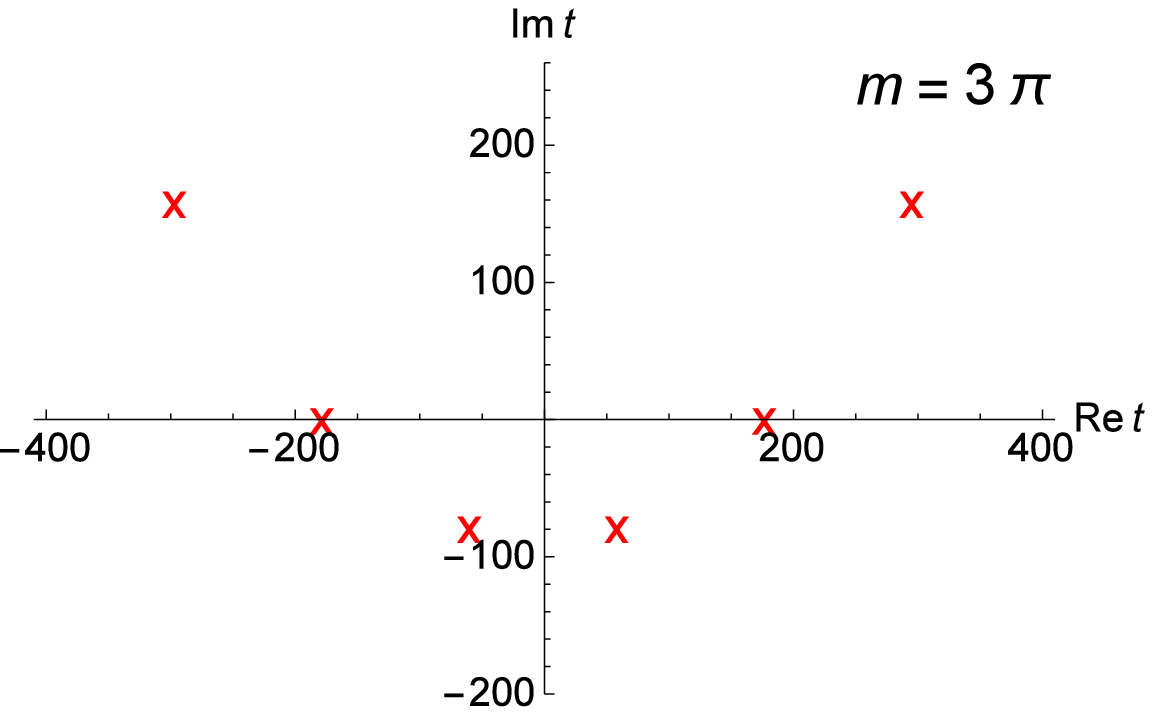}
\includegraphics[clip, width=70mm]{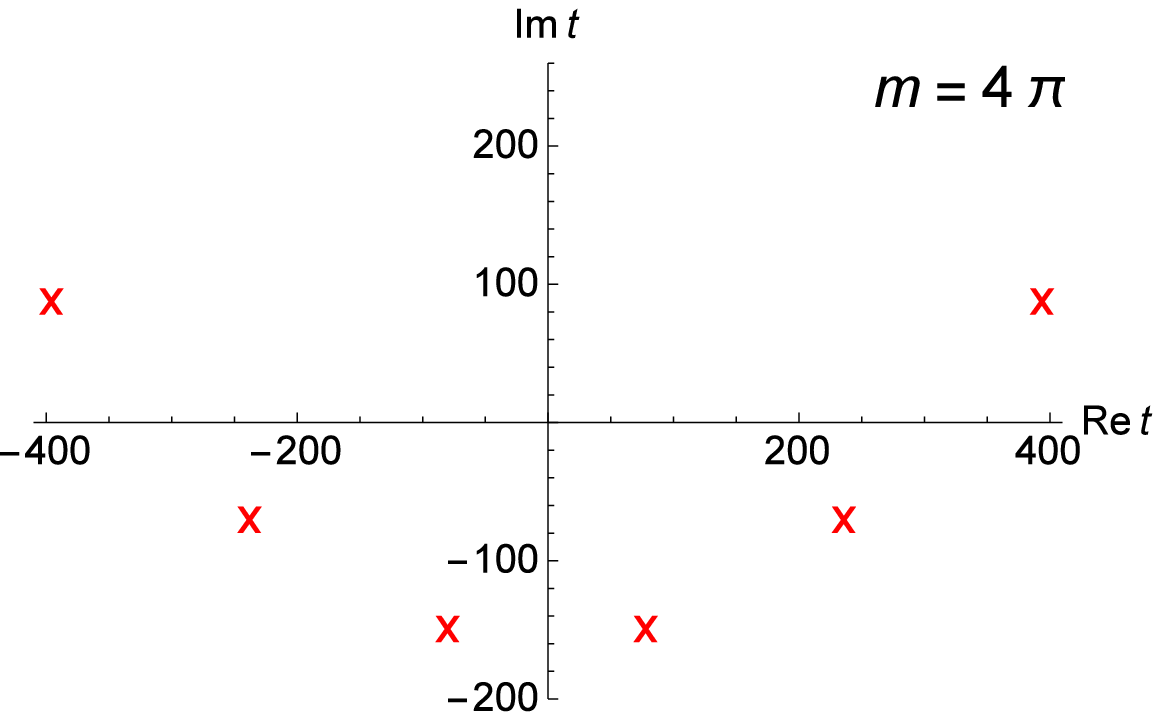}
\caption{Borel singularities (red crosses) are depicted for 
$N_{f}=1$ ${\mathcal N}=2$ CS SQED with the real masses 
$m=0,\pi/2, \pi, 2\pi ,3\pi, 4\pi$. 
Among these choices, $m=\pi,3\pi$ are Stokes lines.}
    \label{fig:borel}
  \end{center}
\end{figure}

Let us take $k>0$ and $m_{a}\geq0$ for simplicity
\footnote{Generalization to $k<0$ 
and $m_{a}<0$ 
is straightforward.
}.
By changing the variables as $g=\frac{4\pi}{k}$ and $\sigma =\sqrt{i t}$,
we rewrite the partition function as
\begin{\eq}
Z =\int_{0}^{-i\infty} dt\ e^{-\frac{t}{g}} \mathcal{B}Z(t)\,,
\end{\eq}
where 
\begin{align}
\mathcal{B}Z(t)
=\frac{i}{4\sqrt{it}}
\left[
\frac{1}
{\prod_{a=1}^{N_f} \cosh{\frac{\sqrt{it}-m_a}{2}} }
+
\frac{1}
{\prod_{a=1}^{N_f} \cosh{\frac{\sqrt{it}+m_a}{2}} }
\right]\,.
\label{eq:U1BorelT}
\end{align}
Note that 
this expression is similar to Borel resummation \eqref{eq:Borel} along $\varphi =-\pi /2$.
Indeed it has been proved in \cite{Honda:2016vmv} that 
the function $\mathcal{B}Z(t)$ is rigorously
the same as the Borel transformation of the perturbative series of $Z$.
This shows that the exact result is equivalent to the Borel resummation along $-i\mathbb{R}_+$.
The Borel transformation has simple poles at
\begin{\eq}
t_{n_{a}}^{*}= -i \left[m_{a} \pm (2n_{a}-1)\pi i \right]^2 \,,
\label{eq:bsing}
\end{\eq}
with $n_{a}\in {\mathbb N}$ for each of flavors.
We can easily see that
${\rm arg}(t)$ of the poles depends on the values of the real masses 
as depicted in Fig.~\ref{fig:borel} for $N_{f}=1$.
In particular, with $m_a =(2n_{a}-1)\pi$, we have Borel 
singularities on the real axis 
\be
\left. t_{n_{a}}^{*} \right|_{m_a =(2n_{a}-1)\pi}
= \pm 2(2n_{a}-1)^2 \pi^2, 
\ee
which leads to non-Borel-summability of the perturbative series along $\mathbb{R}_+$.
This means that $m_a =(2n_{a}-1)\pi$ is the Stokes line, where the Stokes phenomena occur.
We depict the Borel singularities for $N_{f}=1$ with $m=m_{a}=0$ and $n=n_{a}$ 
in Fig.~\ref{fig:borel};
As we turn on the real mass,
the degenerate singularities (double poles) on the positive imaginary axis for $m_{a}=0$ 
get lifted and move to positive and negative real directions.
When the real mass goes beyond $m=(2n-1)\pi$, a singularity crosses the positive real axis
and come into the fourth quadrant from the first quadrant.

Since the exact result is given by the integral along $-i\mathbb{R}_+$
in the Borel resummation,
by use of Cauchy integration theorem,
the exact result turns out to be composed of the Borel resummation along $\mathbb{R}_+$ ({\it perturbative part}) and
the residue of all the singularities in the fourth quadrant of the Borel plane ({\it non-perturbative part}):
\begin{\eq}
Z 
= \int_{0}^\infty dt\ e^{-\frac{t}{g}} \mathcal{B}Z(t)
+\sum_{{\rm poles}\in {\rm 4th\ quadrant}} {\rm Res}_{t=t_{\rm pole}} 
\Bigl[ e^{-\frac{t}{g}} \mathcal{B}Z(t) \Bigr] ,
\label{eq:decompose}
\end{\eq}
where ${\rm Res}_{z=z_0} \left[ f(z) \right]$ denotes residue of $f(z)$ at $z=z_0$
in the normalization
\begin{\eq}
{\rm Res}_{z=z_0} \left[ \frac{1}{z-z_0} \right] =2\pi i .
\end{\eq}
The number of the singularities in this region 
is $|n_{a}|$ for the real mass 
$(2n_{a}-1)\pi<m_{a}<(2n_{a}+1)\pi$ ($n_{a}\in {\mathbb N}^{0}$) 
for each of the flavors.
This is also a correct statement even for negative $m_{a}$ with $n_{a}\in {\mathbb Z}$.
When the real mass $m_{a}$ crosses $(2n_{a}+1)\pi$, 
we start to receive a contribution from another Borel singularity 
which leads to ambiguity of the perturbative Borel resummation at $m_{a}=(2n_{a}+1)\pi$ 
as we have discussed above.
This is how the Stokes phenomena emerge in the present example.
For the degenerate mass $m=m_{a}$ for all flavors, 
the singularities are also degenerate, where the order of their poles is $N_{f}$.

To show these results explicitly, first let us focus on $N_{f}=1$.
For $(2n-1)\pi<m<(2n+1)\pi$, 
the second term in \eqref{eq:decompose} is given by
\begin{\eq}
\sum_{\ell=1}^{n}(-1)^{\ell-1}\,2\pi \,e^{\frac{i}{g}[m+ (2\ell-1)\pi i]^{2}}
\quad {\rm for}\,\,\, (2n-1)\pi<m<(2n+1)\pi .
\end{\eq}
Note that it vanishes for $n=0$, where we just have the perturbative part. 
By use of the step function $\theta (x)$,
the partition function $Z$ is also written as
\begin{align}
& Z =Z_{\rm pt}+\sum_{n =1}^\infty Z_{\rm np}^{(n )}\,, 
 \label{eq:Z_Z0_Zk} \\
& Z_{\rm pt}
=  \int_{0}^\infty dt\ e^{-\frac{t}{g}} \mathcal{B}Z(t) ,\quad\quad
 Z_{\rm np}^{(n )}
=  \theta \left( m-(2n -1)\pi \right)
2\pi (-1)^{n-1} \,e^{\frac{i}{g}[m+ (2n-1)\pi i]^{2}} .
\label{eq:Z2}
\end{align}
Note that this decomposition is well defined for almost all values of $m$
in the sense that it is apparently ambiguous on the Stokes lines
as we will discuss later.
Here $Z_{\rm pt}$ is the perturbative contribution 
while $Z_{\rm np}^{(n )}$ is the nonperturbative contribution.
By expanding the perturbative part $Z_{\rm pt}$ with respect to $t$ and looking into coefficients, 
we obtain the asymptotic form of the perturbative contribution 
as
\begin{align}
Z_{\rm pt}\,
&=\, \frac{\sqrt{i g}}{2}\sum_{q=0}^{\infty} 
\frac{\Gamma(q+1/2)}{\Gamma(q+1)}
\partial_{t}^{q}\left.\left(
\frac{1}{\cosh{\frac{\sqrt{it}-m}{2}}} +\frac{1}{\cosh{\frac{\sqrt{it}+m}{2}}}
\right)\right|_{t=0}\,g^{q}
\nonumber\\
\, &=\,
\frac{\sqrt{i g}}{2} \sum_{q=0}^{\infty} \sum_{a=0}^{\infty}  
\frac{E_{2(q+a)} \Gamma(q+1/2)}
  {2^{2(q+a)}\Gamma(2q+1)\Gamma(2a +1)} m^{2a}\,(i g)^{q}\, ,
\label{eq:coeffP}
\end{align}
where $E_n$ is the Euler number 
\footnote{We used $\frac{1}{\cosh{x}}=\sum_{n=0}^\infty 
\frac{E_{2n}}{(2n)!}x^{2n}$. 
Appendix.~\ref{QEDpert} describes
a rederivation of this result by the standard way.
}.
It is notable that this asymptotic series is 
Borel-summable along $\mathbb{R}_+$ for $m\not=(2n-1)\pi$ 
while it is not for $m=(2n-1)\pi$.
However, even for $m\not=(2n-1)\pi$, 
the Borel resummation of the perturbative series along $\mathbb{R}_+$
does not give an exact result for $m>\pi$. 
These are consistent with the argument on the Stokes phenomena
mentioned above. 
Now we are ready to write down the full transseries expansion of $Z$:
\begin{\eq}
Z
= C_0 \sum_{q =0}^\infty c_q^{(0)} g^{\frac{1}{2}+q}
+\sum_{n=1}^\infty C_n e^{-\frac{S_n}{g}} \sum_{q =0}^\infty c_q^{(n)} g^q .
\label{eq:transU1}
\end{\eq}
Comparing this with the above data,
we identify the above parameters with
\begin{\eqa}
&& C_0 =1 ,\quad
c_q^{(0)}
=\frac{i^{\frac{1}{2}+q}\Gamma(q+1/2)}{2\Gamma(2q +1)}
  \sum_{a=0}^{\infty}  
\frac{E_{2(q +a)} } {2^{2(q +a)}\Gamma(2a +1)} m^{2a}, \NN\\
&& C_n=  \theta \left( m-(2n -1)\pi \right) ,\quad
S_n = -i [m+ (2n-1)\pi i]^{2} ,\quad
c_q^{(n)} = 2\pi (-1)^{n-1}  \delta_{q 0}   .
\end{\eqa}

For $m=(2n-1)\pi$, we need to take a handle with care.
This is 
because $Z_{\rm pt}$ and $Z_{\rm np}^{(n)}$ are ambiguous
due to the non-Borel summability along $\mathbb{R}_+$
and the step function behavior of the transseries parameter $C_n$,
respectively,
while the other non-perturbative corrections are unambiguous at this point.
Their ambiguities are indeed canceled as follows.
In the context of resurgence theory,
the Borel ambiguity is usually estimated by the difference of the lateral Borel resummations as
\begin{align}
({\mathcal S}_{0^+} -{\mathcal S}_{0^-} ) Z(g,m) .
\end{align}
Instead
let us estimate the ambiguities of perturbative and nonperturbative contributions by 
\[
Z_{\rm pt}(g,m=(2n-1)\pi + 0_+ ) - Z_{\rm pt}(g,m=(2n-1)\pi + 0_- )  ,
\]
and
\[
Z_{\rm np}^{(\ell)}(g,m=(2n-1)\pi + 0_+) -  Z_{\rm np}^{(\ell)}(g, m=(2n-1)\pi + 0_- ) .
\]
Noting
\begin{align}
 Z_{\rm pt}(m=(2n-1)\pi + 0_\pm )
= P \int_{0}^\infty dt\ e^{-\frac{t}{g}} \mathcal{B}Z(t) 
 \mp \frac{1}{2}{\rm Res}_{t=t_{n}^{*}}\Bigl[ e^{-\frac{t}{g}} \mathcal{B}Z(t) \Bigr] ,
\end{align}
the Borel ambiguity in the perturbative sector is
\begin{align}
 Z_{\rm pt}(m=(2n-1)\pi + 0_+ )
- Z_{\rm pt}(m=(2n-1)\pi + 0_- )
= -{\rm Res}_{t=t_{n}^{*}}\Bigl[ e^{-\frac{t}{g}} \mathcal{B}Z(t) \Bigr] ,
\end{align}
while the non-perturbative ones are
\begin{align}
 Z_{\rm np}^{(\ell)}(m=(2n-1)\pi + 0_+)
-  Z_{\rm np}^{(\ell)}(m=(2n-1)\pi + 0_- )
=\begin{cases}
 0 & {\rm for}\ \ell\neq n \cr
 +{\rm Res}_{t=t_{n}^{*}}\Bigl[ e^{-\frac{t}{g}} \mathcal{B}Z(t) \Bigr] & {\rm for}\ \ell=n
\end{cases} .
\end{align}
${\rm Res}_{t=t_{n}^{*}}[...]$ stands for the residue at the singularity on the positive real axis denoted as $t_{n}^{*}$.
Thus the ambiguities are canceled
and the whole transseries \eqref{eq:transU1} gives 
the unambiguous result which is equivalent to the exact result.

To sum up, the Borel ambiguity in the perturbative sector for $m=(2n-1)\pi$
is canceled by that of the Stokes coefficient in the $n$-th non-perturbative contribution $Z_{\rm np}^{(n)}$.
Therefore cancellations of ambiguities occur only between perturbative and non-perturbative sectors in this theory
while there is no cancellation among different non-perturbative sectors (namely $Z_{\rm np}^{(\ell )}$'s with different $\ell$'s). 
This is reflected by the fact that the perturbative series in the non-perturbative sectors are truncated at finite orders 
\footnote{
More precisely, at leading orders.
} 
and do not have Borel ambiguities although the Stokes coefficients have ambiguities.
In this sense, the resurgent structure in our example is simpler than ones in typical examples of resurgence
where non-perturbative parts also have Borel ambiguities eventually canceled by those in other non-perturbative parts. 
There is another special property in our example.
In general resurgence relates ambiguous parts of perturbative series around different saddle points
and it is not sufficient to determine whole non-perturbative (perturbative) contribution from perturbative (non-perturbative) one.
In contrast, the perturbative series in our example somehow knows everything on the nonperturbative effects as a result
since the $n$-th nonperturbative part consists only of a contribution which is ambiguous at $m=(2n-1)\pi$
and can be simply determined by the perturbative Borel transformation via the cancellation of the ambiguities
\footnote{To avoid confusion, 
we note that the $n$-th non-perturbative part is not ambiguous for $m\not=(2n-1)\pi$.
This is unambiguously zero for $m< (2n-1)\pi$ while nonzero for $m > (2n-1)\pi$.
}.
This is reflected by the fact that
the exact result is equivalent to the Borel resummation of the perturbative part along the imaginary contour
as in \eqref{eq:U1BorelT}.
Of course we cannot find the above property just by looking at the perturbative part.
To find this, we need some extra information 
such as \eqref{eq:U1BorelT}, direct computation of the nonpertubative parts or something else
since the nonperturbative parts could also have unambiguous contributions in principle.
The above properties are common in our class of examples
since the exact results are always the same as the pertubative Borel resummation along the imaginary contour \cite{Honda:2016vmv}.

We also note that the importance of the Borel singularities at the first and fourth quadrants on the perturbative Borel plane has been stressed in \cite{Grassi:2014cla} 
in the context of $1/N$-expansion of 3d $\mathcal{N}=6$ superconformal field theory
as their residues give exponentially suppressed corrections. 
There, it is argued that these singularities correspond to nonperturbative contributions, thus they should be taken into account even if the perturbative series is Borel-summable. 
The present case is one of the examples consistent with this argument.

We also show the results for a generic number of flavors $N_{f}\geq 1$ with degenerate mass $m_{a}=m$:
\begin{align}
Z_{\rm pt} 
&= \frac{\sqrt{i g}}{2} \sum_{\{q_{a}\}=0}^{\infty}  \sum_{\{l_{a}\}=0}^{\infty}
\frac{\Gamma( \bar{q} + 1/2)}{2^{2(\bar{q}+\bar{l})} } 
\Biggl[ \prod_{a=1}^{N_f} \frac{ E_{2(q_a+l_a)} }{\Gamma(2 q_a + 1) \Gamma(2 l_a + 1)} m^{2 l_a}  \Biggr]
(i g)^{\bar{q}}  \,, \\
Z_{\rm np}&= \frac{\pi i }{2^{N_f-1} \Gamma(N_f)}  \sum_{\ell=1}^{n} 
\lim_{z \rightarrow z_\ell^{*}} \Biggl[ \frac{\pd^{N_f-1}}{\pd z^{N_f-1}} \frac{(z-z^{*}_{\ell})^{N_f} }{\left( \cosh \frac{z-m}{2} \right)^{N_f}}  
e^{\frac{i  z^2}{g}}  \Biggr] , 
\quad\quad\quad{\rm for}\,\,\, (2n-1)\pi<m<(2n+1)\pi
\end{align}
where $\bar{q} = \sum_{a=1}^{N_f} q_a$, $\bar{l} = \sum_{a=1}^{N_f} l_a$ and $z_\ell^{*}=m+ (2\ell-1)\pi i$.
We note that $Z_{\rm np}=0$ for $n=0$.
We again emphasize that the order of poles of Borel singularities is $N_{f}$ when the masses of the flavors are degenerate.
For these cases with the degenerate mass, 
we still have the exact result as the full transseries, where the Stokes phenomena occur
at the special values of the real mass $m=(2n-1)\pi$.
This is regarded as the resurgent structure 
beyond the argument with the standard ``resurgent function" with simple poles or branch cuts \cite{Ec1}.

We end this subsection by a comment on uniqueness of the decomposition of the exact partition function into perturbative and nonperturbative parts in Eq.~(\ref{eq:Z2}). 
We have defined perturbative part as the Borel resummation 
that is obtained by integration of the Borel transform along $\mathbb{R}_+$, 
and have decomposed the exact result into the perturbative and 
nonperturbative parts. 
Provided we have a perturbative series for a certain quantity, 
its Borel resummation just gives one of analytic functions, 
whose asymptotic expansion becomes the perturbative series.
Thus the Borel resummation is not a unique definition of the 
perturbative contribution. 
In addtion, provided we have an exact result of the quantity, 
its decomposition into perturbative and nonperturbative parts 
is not unique. 
This point will be discussed again when we study the thimble 
decomposition of the exact result in the next subsection.

\subsection{Thimble decomposition}
\label{sec:SQED_thimble}
Here we decompose 
the Coulomb branch localization formula (\ref{eq:Zexact}) 
into Lefshetz thimbles (steepest descents)
and compare the result with the transseries expression in the previous subsection.
A brief review on the Lefschetz thimble decomposition is given in Appendix.~\ref{LT}.
Here we concentrate on the $N_f=1$ case for simplicity.
Generalization to multi-flavors is straightforward.
First we rewrite (\ref{eq:Zexact}) as
\begin{\eq}
Z=\int_{-\infty}^\infty d\sigma\ e^{-S[\sigma ]} ,
\label{eq:originalU1}
\end{\eq}
where  
\begin{align}
S[\sigma ] \, = \,  -\frac{i}{g} \sigma^2 - \log \frac{1}{ 2 \cosh \frac{\sigma -m}{2}}\,.
\label{eq:actionU1}
\end{align}
We regard $S[\sigma ]$ as ``action" of the Coulomb branch parameter $\sigma$ and
extend $\sigma \in {\mathbb R}$ to a complex value $z \in {\mathbb C}$
since saddle points and the associated Lefschetz thimbles are complex-valued in general. 
The saddle points $z^{c}$ are obtained from the saddle-point equation,
\begin{align}
\left. \frac{\partial S[z]}{\partial z} \right|_{z=z^{c}} \, 
= \, -\frac{2 i}{g} z^{c} + \frac{1}{2} \tanh \frac{z^{c}-m}{2} \, = \, 0\,.
\label{eq:critical}
\end{align}
Let us label the saddle points by $z_I^c$.
Note that
although we have the infinitely many saddle points $\{ z_I^c \}$,
each saddle point may or may not contribute to the integral \eqref{eq:originalU1}.
This is determined by looking at saddle points 
passed by the steepest descent contours 
obtained by deforming the original contour 
without changing the value of the integral.
In general, this depends on the original integral contour,
the parameters $(g,m)$ and properties of the (dual) Lefschetz thimbles
as explained below. 

The Lefschetz thimble or the steepest descent contour $\mathcal{J}_I$
associated with the saddle point $z_I^c$ 
is obtained by solving the differential equation called the flow equation,
\begin{align}
\left. \frac{d z}{ds} \right|_{\mathcal{J}_I} & 
=  \, \overline{\frac{\partial S[z]}{\partial z}} \nonumber \\
  & = \, + \frac{2 i }{g} \bar{z} + \frac{1}{2} \tanh \frac{\bar{z}-m}{2}\,,
\end{align}
with the initial condition 
\begin{align}
  \lim_{s \rightarrow -\infty} z(s) = z_I^{\rm c} ,
\label{eq:ini_cond_thim}
\end{align}
with $s$ being the flow parameter.
Using the flow equation, we can easily prove
\begin{align}
 \left. \frac{d {\rm Re} S[z(s)]}{ds}\right|_{\mathcal{J}_I} \, \ge \, 0 
  \quad{\rm and}\quad
\left.  \frac{d {\rm Im} S[z(s)]}{ds}\right|_{\mathcal{J}_I} \, = \, 0 ,
\end{align}
which indicate that
integrals along Lefschetz thimbles are rapidly convergent and non-oscillating.
We can express the original contour ${\cal C}_{\mathbb R}$
as the linear combination of the thimbles 
\begin{align}
{\cal C}_{\mathbb R} \,=\, \sum_{ I \in {\rm saddles}} n_I {\cal J}_I . 
\end{align}
When $n_I$ is nonzero, the saddle point $z_I^c$ and its associated thimble contribute to the integral
while we have no contributions from saddle points with $n_I =0$.
It is known that the expansion coefficient $n_I$ is an integer
because $n_I$ is the same as an intersection number 
between the original contour ${\cal C}_{\mathbb R}$
and the dual thimble (steepest ascent contour) $\mathcal{K}_I$ associated with $z_I^c$
defined by
\begin{align}
\left. \frac{d z}{ds} \right|_{\mathcal{K}_I}  
= \overline{\frac{\partial S[z]}{\partial z}} \quad\quad
{\rm with}\   \lim_{s \rightarrow +\infty} z(s) = z_I^{\rm c} . 
\end{align}
In general $n_I$ depends on $(g,m)$ but its dependence is not continuous
since $n_I$ is integer.
Typically $n_I$ is a constant or a step function and
the latter case leads us to a Stokes phenomenon.

Let us analyze the structures of the Lefschetz thimbles in the present example.
The saddle point equation \eqref{eq:critical} implies that
the critical points $z^c$ are complicated functions of $(g,m)$
and it is hard to compute them and their thimbles analytically.
Therefore we exhibit the critical points
and solve the associated flow equations numerically for finite $g$. 
Before showing the numerical results,
we discuss the weak coupling limit analytically
to get an intuitive understanding on the thimble structures.

\subsubsection{Analytical results for small $g$}
\begin{figure}[t]
\begin{center}
\includegraphics[width=7.2cm]{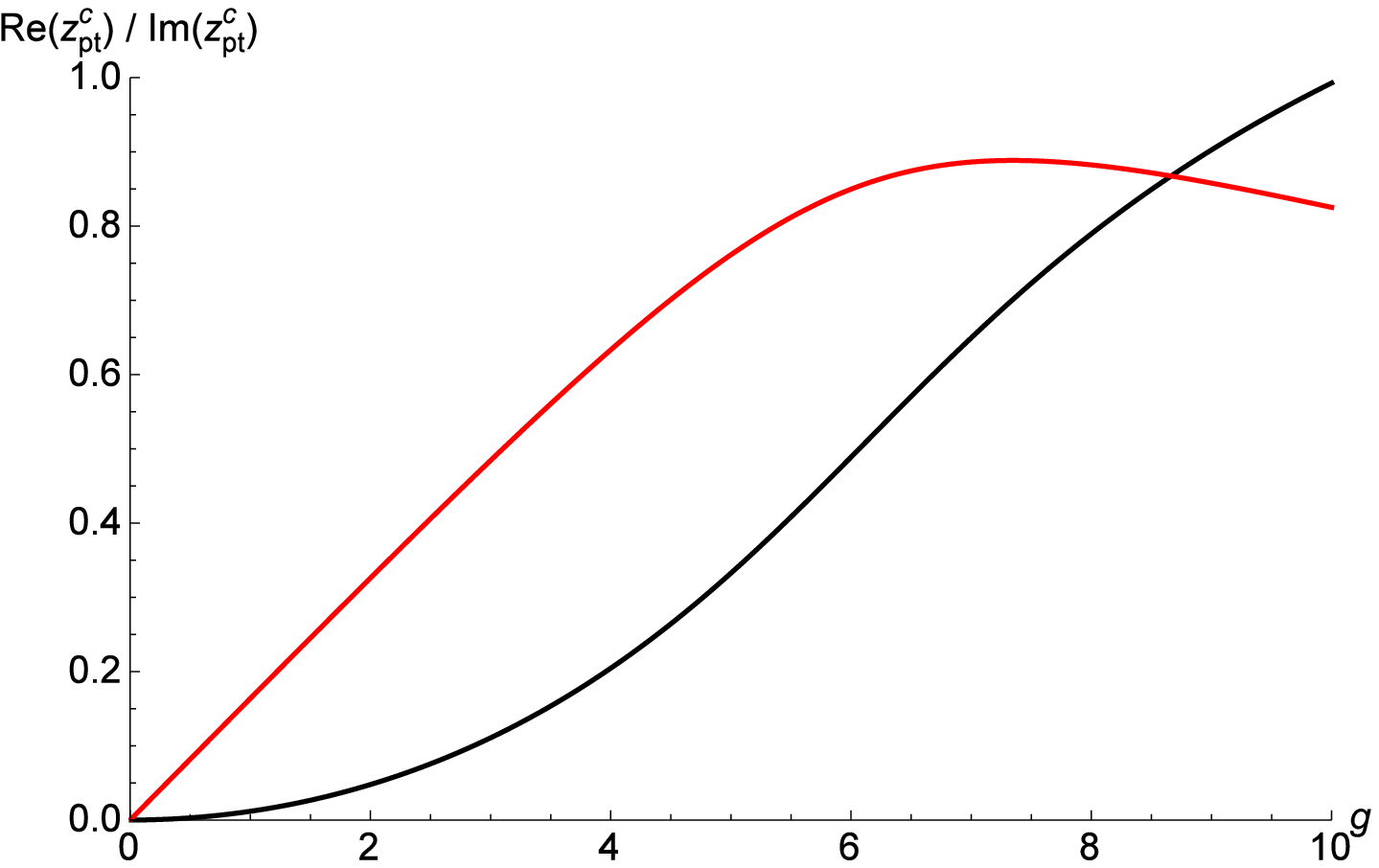}
\includegraphics[width=7.2cm]{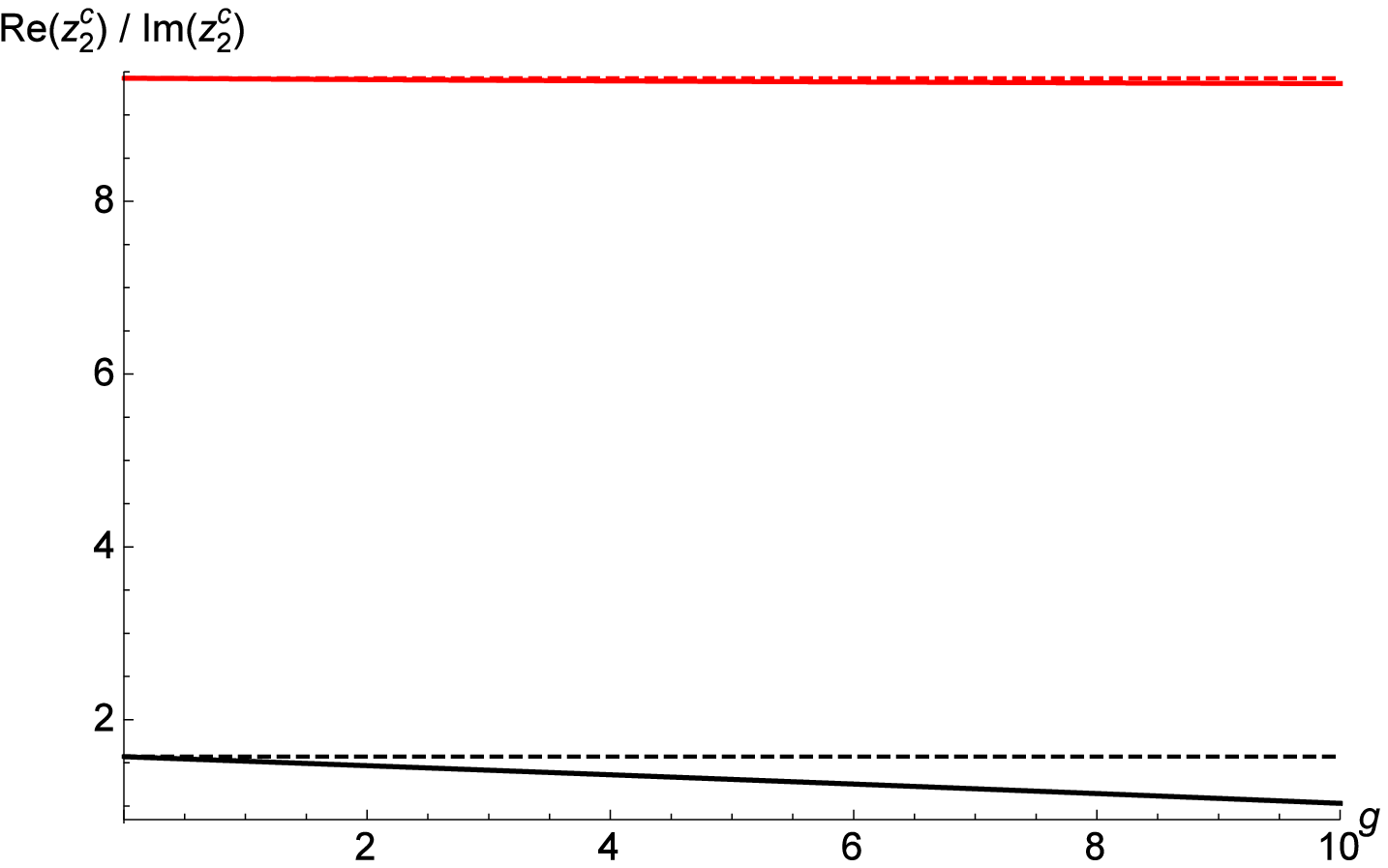}\\
\includegraphics[width=7.2cm]{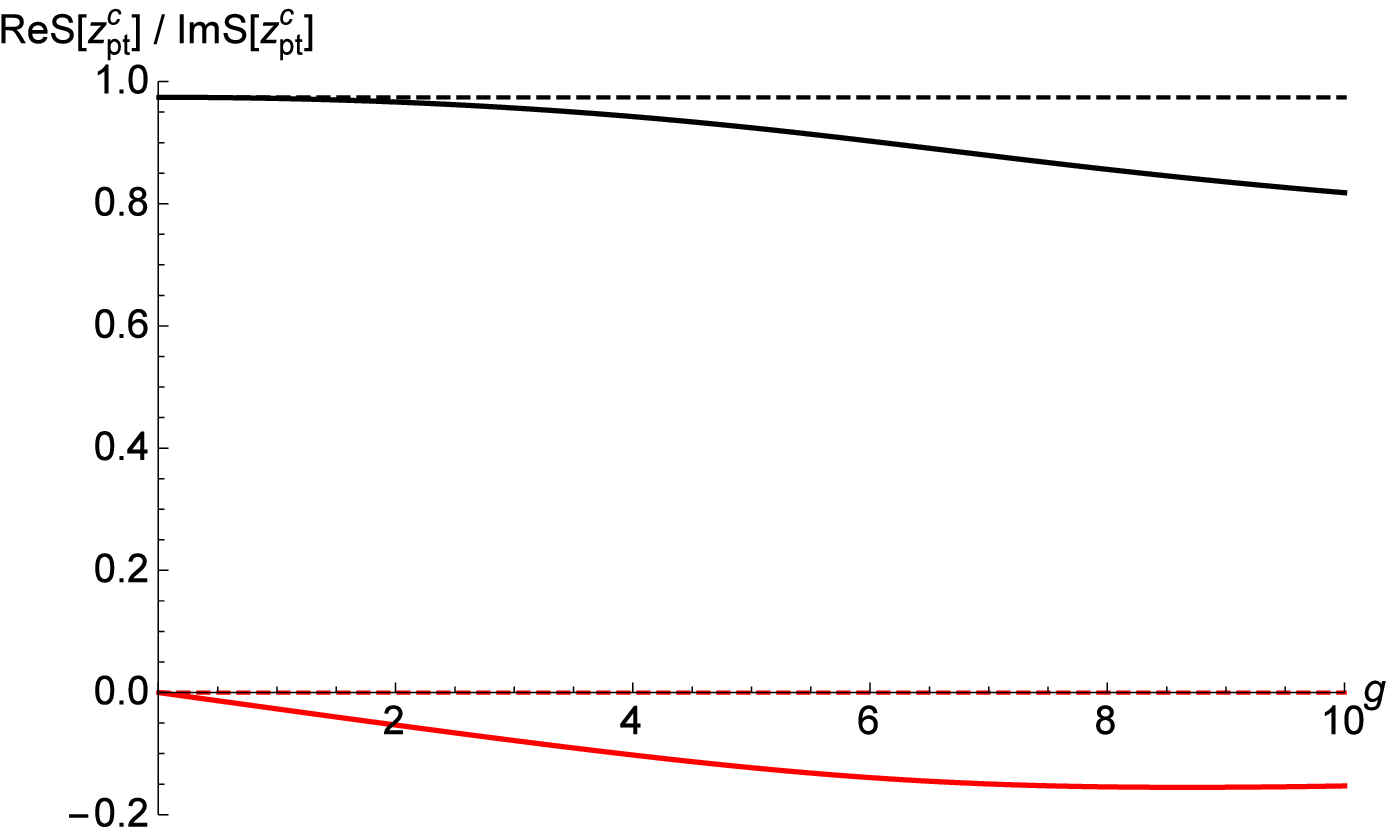}
\includegraphics[width=7.2cm]{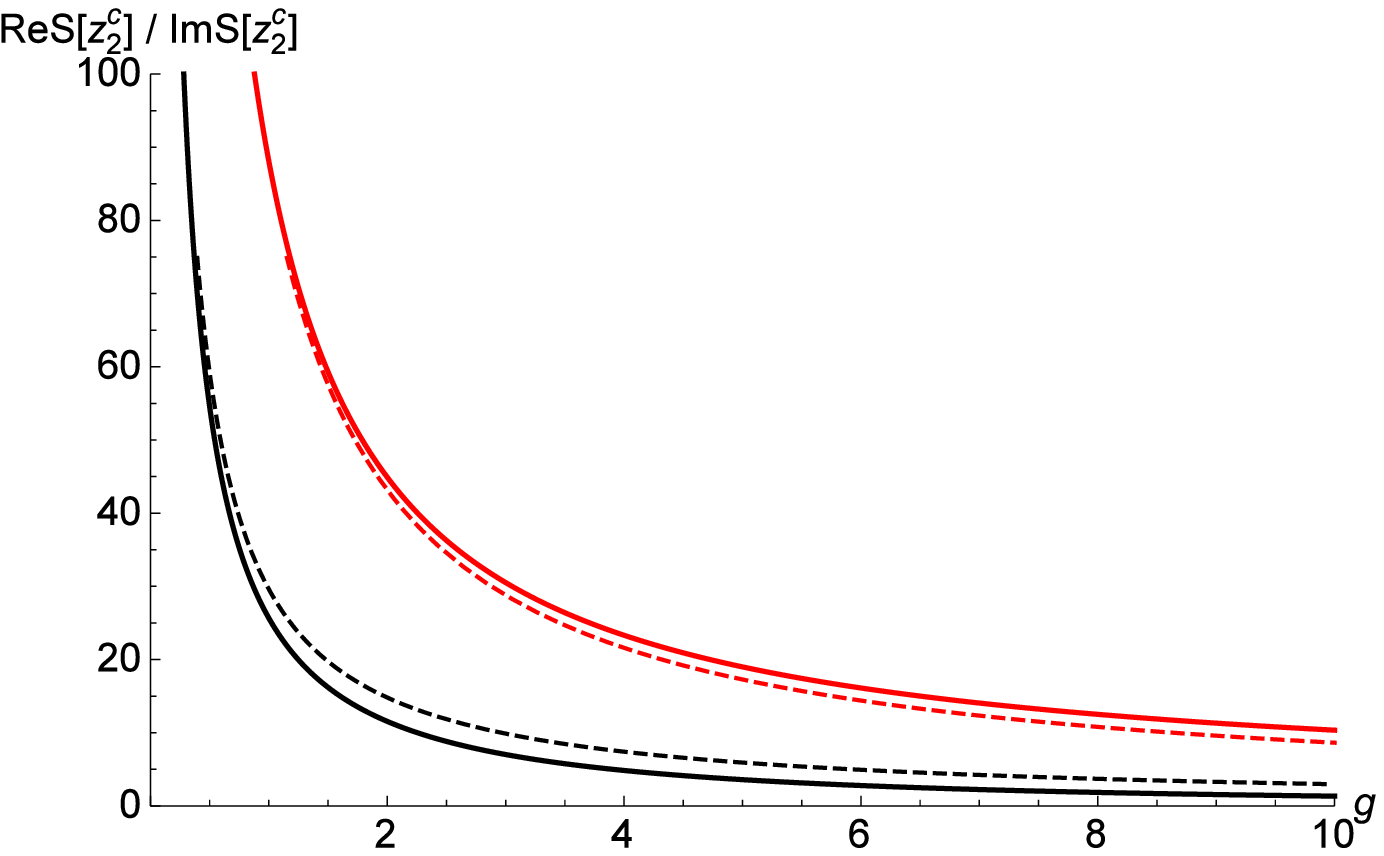}
 \caption{
Locations of critical points and their actions as functions 
of $g$ for $m=\pi /2$,
which are computed numerically.
[Left Top] ${\rm Re}z_{\rm pt}^c$ (black solid) and 
${\rm Im}z_{\rm pt}^c$ (red solid).  
[Right Top] ${\rm Re}z_2^c$ (black solid) and ${\rm Im}z_2^c$ (red solid)
compared with  ${\rm Re}z_2^\ast =m$ (black dotted) 
and ${\rm Im}z_2^\ast =3\pi$ (red dotted).
[Left Bottom] ${\rm Re}S[z_{\rm pt}^c]$ (black solid) and 
${\rm Im}S[z_{\rm pt}^c]$ (red solid)
compared with ${\rm Re}S[0]=\log{2\cosh{\frac{m}{2}}}$ (black dotted) 
and ${\rm Im}S[0]=0$ (red dotted). 
[Right Bottom] ${\rm Re}S[z_2^c]$ (black solid) and 
${\rm Im}S[z_2^c]$ (red solid)
compared with ${\rm Re}\Bigl[ \frac{i}{g}(m+3\pi i)^2 \Bigr]$ 
(black dotted) 
and ${\rm Im}\Bigl[ \frac{i}{g}(m+3\pi i)^2 \Bigr]$ (red dotted). 
}
 \label{fig:actions}
\end{center}
\end{figure}

The saddle point equation \eqref{eq:critical} is simplified in the weak-coupling limit $g \rightarrow 0$.
To see this, we multiply \eqref{eq:critical} by $g\cosh \frac{z^{c}-m}{2}$: 
\begin{align}
 \, -2i z^{c}\cosh \frac{z^{c}-m}{2} + \frac{g}{2} \sinh \frac{z^{c}-m}{2} \, = \, 0\,.
\end{align}
In the limit $g \rightarrow 0$,
we can ignore the second term on the LHS
and the critical points are determined by
\begin{\eq}
z^c (g,m)  \cosh{\frac{z^c (g,m) -m}{2}} =0 \quad\quad {\rm for}\ g\rightarrow 0 ,
\end{\eq}
in which we obtain $z^c (0,m)=0$, $m+(2\ell -1)\pi i$ with $\ell \in {\mathbb Z}$.
Therefore we have an infinite number of critical points approaching these values
in $g\rightarrow 0$.
Let us denote as $z_{\rm pt}^c$ and $z_\ell^c$ the critical points satisfying
\begin{align}
\lim_{g\rightarrow 0}  z_{\rm pt}^{\rm c} (g,m) =0 ,\quad
\lim_{g\rightarrow 0}  z_\ell^{\rm c} (g,m) =m  + (2\ell-1)\pi i .
 \label{eq:crit_U1}
\end{align}
As $g$ increases,
the critical points go away from \eqref{eq:crit_U1}
as shown in the top panels of Fig.~\ref{fig:actions}.
The critical points
$z^{\rm c}_{\rm pt}$ and $z^{\rm c}_\ell$ 
approximately correspond to the saddle points 
for the perturbative and nonperturbative contributions respectively
since, for $g\rightarrow 0$,
the action at $z^{\rm c}_{\rm pt}$ behaves as $\mathcal{O}(1)$
while the one at $z^{\rm c}_\ell$ behaves 
as $-\frac{i}{g} \left( m + (2\ell-1)\pi i \right)^2$ $+\mathcal{O}(1)$
as illustrated in the bottom panels of Fig.~\ref{fig:actions}.
This behavior precisely matches 
with the exponent of the nonperturbative corrections
appearing in our resurgent transseries \eqref{eq:Z2}.
Moreover ${\rm Im}S$ of $z_{\rm pt}^c (0,m)$ and $z_\ell^c (0,m)$
coincide at the special values:
$m=\pm (2\ell -1)\pi$.
This is expected from
the fact that
the Stokes phenomena of the transseries \eqref{eq:Z2} occurs at $m=(2\ell -1)\pi$.

We can easily compute 
the thimble flowing from $z^{\rm c}_{\rm pt}$ in the $g\rightarrow 0$ limit.
The flow equation for $g\to 0$ is given by
\begin{align}
\frac{d z}{ds} \, =  \, + \frac{2 i }{g} \bar{z} \quad\quad\quad {\rm as} \quad g \rightarrow 0\, ,
\end{align}
which is solved by
\begin{align}
\lim_{g\rightarrow 0} z_{\rm pt}(g,m;s) \, 
= \, \epsilon \exp \left( {\frac{2}{g}s + \frac{\pi i }{4}} \right)\, ,
\end{align}
with a parameter $\epsilon \in {\mathbb R}$ for the initial condition.
Note that this thimble corresponds to 
the integration $Z_{\rm pt}$ in (\ref{eq:Z2}), or the perturbative contribution.
For the non-perturbative one $z^{\rm c}_\ell$,
it is hard to analytically solve the flow equation globally
even in the $g\rightarrow 0$ limit.

Note that $z_\ell^c$ for $g\rightarrow 0$ given by \eqref{eq:crit_U1}
is precisely the same as the location of the poles of the integrand $e^{-S[x]}$, 
which are zeroes of $\cosh \frac{z-m}{2}$ in the denominator and given by
\begin{align}
z^{*}_{\ell} (m) \, = \, m + (2\ell-1) \pi i 
\quad {\rm with}\ \ell \in {\mathbb Z} .
\label{eq:sing_U1}
\end{align}
This always happens when we study the following type of integral:
\begin{\eq}
\int \frac{dx}{f(x)} e^{-\frac{1}{g} h(x)} ,
\end{\eq}
where $f(x)$ is a function without poles but may have zeroes.
The critical points for this integrand are determined by
\begin{\eq}
 \frac{\pd h}{\pd x} + g \frac{1}{f}  \frac{\pd f}{\pd x} = 0 .
\end{\eq}
By examining the limiting behavior of critical points as $g\to 0$, 
we find that at least one of critical points inevitably
goes to each zero of $f(x)$ in the limit.
In summary, the asymptotic values of the critical points in the $g\to 0$ limit satisfy
\begin{\eq}
 \frac{\pd h(x)}{\pd x} \cdot f(x) = 0.
\end{\eq}
This fact has important implications for structures of (dual) thimbles.
Since the actions at the poles are $-\infty$,
dual thimbles can end on the poles 
while thimbles cannot pass through the poles.
In other words,
the poles play a role of source of the dual thimble.
Therefore,
noting that the critical points for finite $g$ are located near the poles,
the dual thimble associated with one of the critical points
goes from the pole to another region with ${\rm Re}S \rightarrow -\infty$
via the critical point.
On the other hand,
the thimble associated with the same critical point
connects two regions with ${\rm Re}S \rightarrow +\infty$
via the critical point
but circumvents the poles.
As we will see below,
the thimble integrals associated with the critical points near the poles
are equivalent to their residues.

Next we take into account a small $g$ correction
by taking the ansatz $z^c (g,m) =z^{c,0} (m) +g z^{c,1}(m) +\mathcal{O}(g^2 )$
to see explicitly what would be going on for nonzero $g$.
Matching $\mathcal{O}(1)$ terms in \eqref{eq:critical} gives
\begin{\eq}
 z_{\rm pt}^c (g,m) = \frac{i g}{4} \tanh{\frac{m}{2}} +\mathcal{O}(g^2 ) ,\quad
 z_\ell^c (g,m) = z_\ell^\ast  +\frac{g}{2i z_\ell^\ast } +\mathcal{O}(g^2 ) ,
\label{eq:criticalOg}
\end{\eq}
which have the actions
\begin{\eqa}
&& S[ z_{\rm pt}^c (g,m)]
 = \log{\left( 2\cosh{\frac{m}{2}}\right)}  +\mathcal{O}(g ) ,\NN\\ 
&& S[ z_\ell^c (g,m)]
 = -\frac{i}{g}z_\ell^{\ast 2} +\log{g} 
   +\left(-1+\log{\frac{(-1)^{\ell-1} }{2z_\ell^\ast}}\right) +\mathcal{O}(g ) .
\label{eq:actionOg}
\end{\eqa}
From these actions,
a necessary condition for having Stokes phenomenon is 
\footnote{We are assuming ${\rm arg}(g)=0$.}
\begin{\eqa}
0&=&{\rm Im}\, S [z_\ell^c (g,m)]-{\rm Im} \,S[ z_{\rm pt}^c (g,m)] \NN\\
 &=& -\frac{1}{g}{\rm Re}\left[z_\ell^{\ast 2}  \right] 
   +{\rm arg}\frac{(-1)^{\ell-1}}{z_\ell^\ast} +\mathcal{O}(g ) .
\end{\eqa}
Note that this condition is not satisfied by $m=(2\ell -1)\pi $,
which was a solution in the $g\rightarrow 0$ limit.
This implies that the Stokes phenomenon in the thimble 
decomposition for nonzero $g$ occurs at a different point 
$m=\tilde{m}_\ell (g)$ from those of the transseries 
and they coincide in the weak coupling limit:
\begin{\eq}
\lim_{g\rightarrow 0}\tilde{m}_\ell (g) =(2\ell -1)\pi .
\end{\eq}
Consequently for finite $g$ we need to distinguish 
Stokes phenomena in the sense of thimble decomposition and in 
the sense of transseries.
This happens in general when coupling is not multiplicative 
to the whole action or when we include operators
(with no or different coupling dependence) as a 
part of the effective action. 
We will readily see this effect by a numerical analysis for finite 
$g$ performed below and discuss relation to the resurgent transseries.

\subsubsection{Numerical results for finite $g$ and comparison with resurgent transseries}
\begin{figure}[t]
  \begin{center}
    \begin{tabular}{ccc}
      \begin{minipage}{0.33\hsize}
        \begin{center}
          \includegraphics[clip, width=50mm]{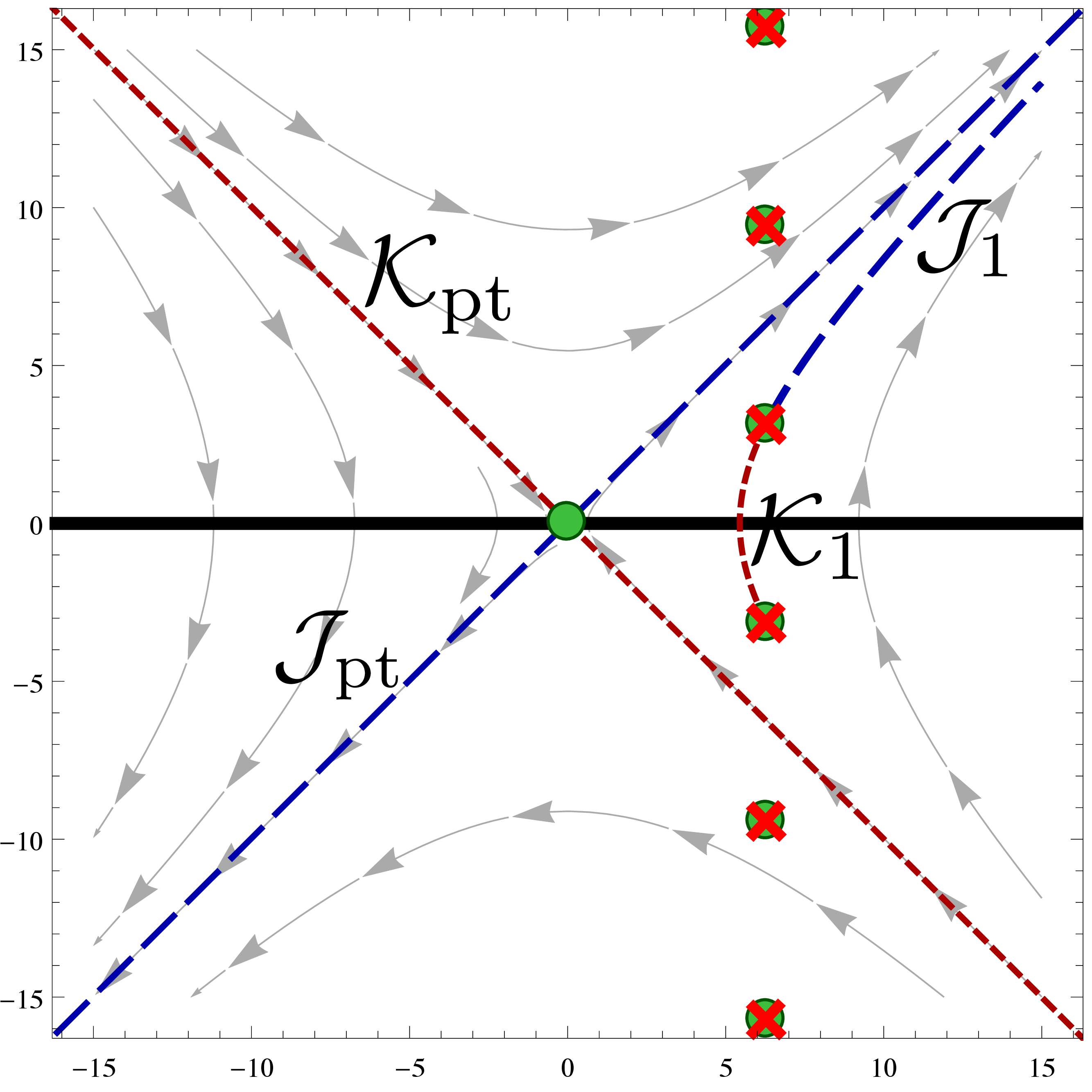}
          \hspace{1.6cm} (a) $m = 2 \pi$ 
        \end{center}
      \end{minipage}
      \begin{minipage}{0.33\hsize}
        \begin{center}
          \includegraphics[clip, width=50mm]{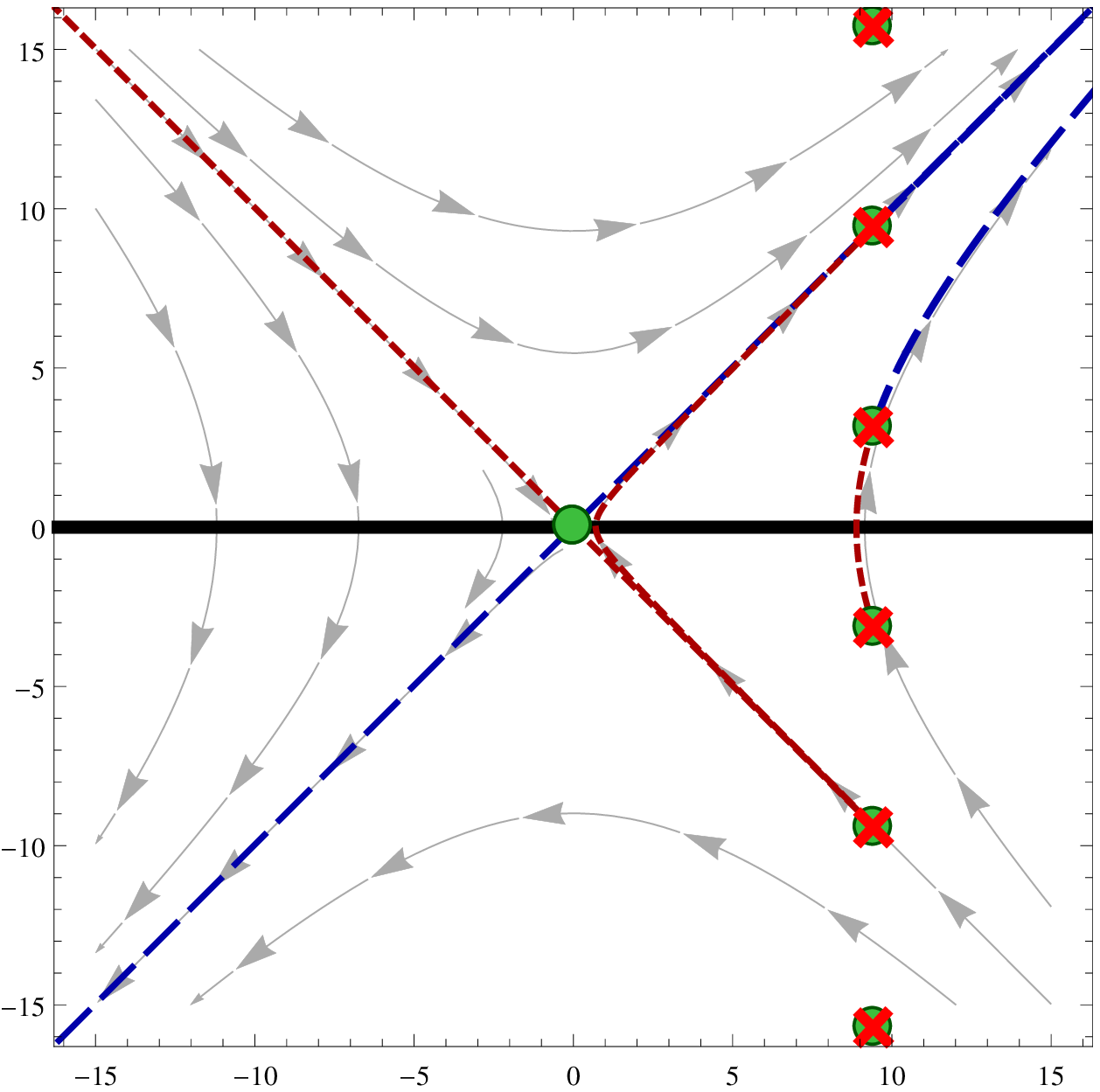}
          \hspace{1.6cm} (b) $m = 3 \pi$  
        \end{center}
      \end{minipage}
      \begin{minipage}{0.33\hsize}
        \begin{center}
          \includegraphics[clip, width=50mm]{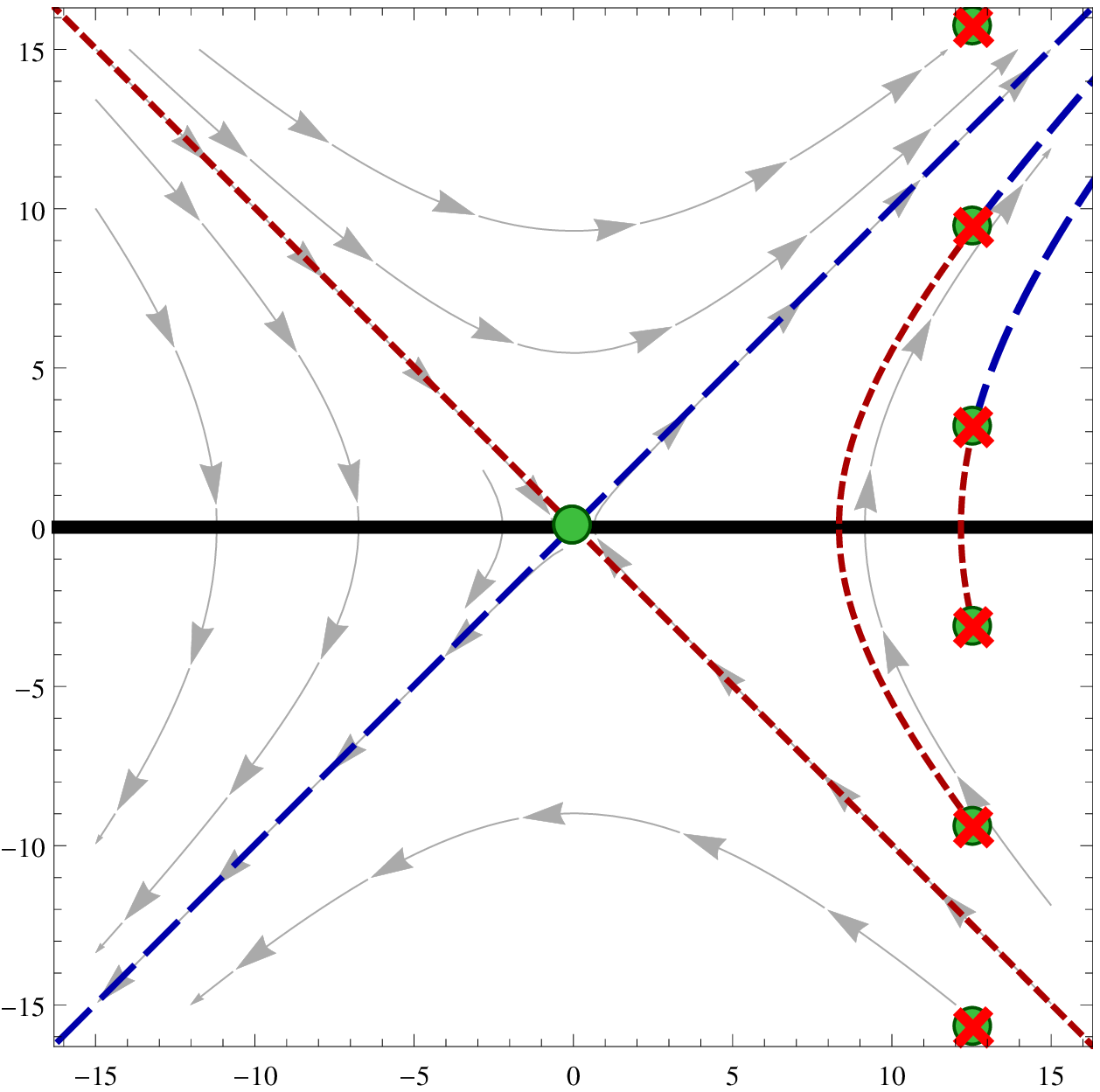}
          \hspace{1.6cm} (c) $m = 4 \pi$ 
        \end{center}
      \end{minipage}
    \end{tabular}
    \caption{Thimble structure of partition function of 
${\mathcal N}=2$ CS SQED with $N_{f}=1$ hyper multiplet for 
$k=100$ ($g=4\pi/k$ $\approx$ $0.126$) on ${\rm Re}z$-${\rm Im}z$ plane.
      The green points and red crosses stand for critical points and 
singularities, respectively.
      The green points (saddle points) are hidden by the red 
crosses (singularities) except at the origin. 
      The red dotted lines stand for dual 
thimbles with nonzero intersection numbers, whereas the blue 
dashed lines for corresponding thimbles.
The arrows represent flow lines for increasing flow parameter $s$.  
}
    \label{fig:thim_CS_U1_g0.1}
  \end{center}
\end{figure}
\begin{figure}[t]
        \begin{center}
          \includegraphics[clip, width=45mm]{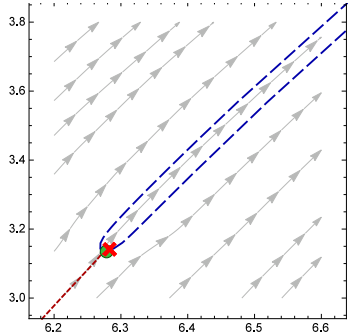}
        \end{center}
    \caption{
    Zoom-up of Fig.~\ref{fig:thim_CS_U1_g0.1} (a) 
    around the singularity $z=z_1^\ast =m+\pi i$.
    The thimble integral associated with $z_1^c$ is equivalent to
    the residue around  $z=z_1^\ast$.
    }
    \label{fig:zoom}
\end{figure}

\begin{figure}[t]
  \begin{center}
    \begin{tabular}{ccc}
      \begin{minipage}{0.33\hsize}
        \begin{center}
          \includegraphics[clip, width=50mm]{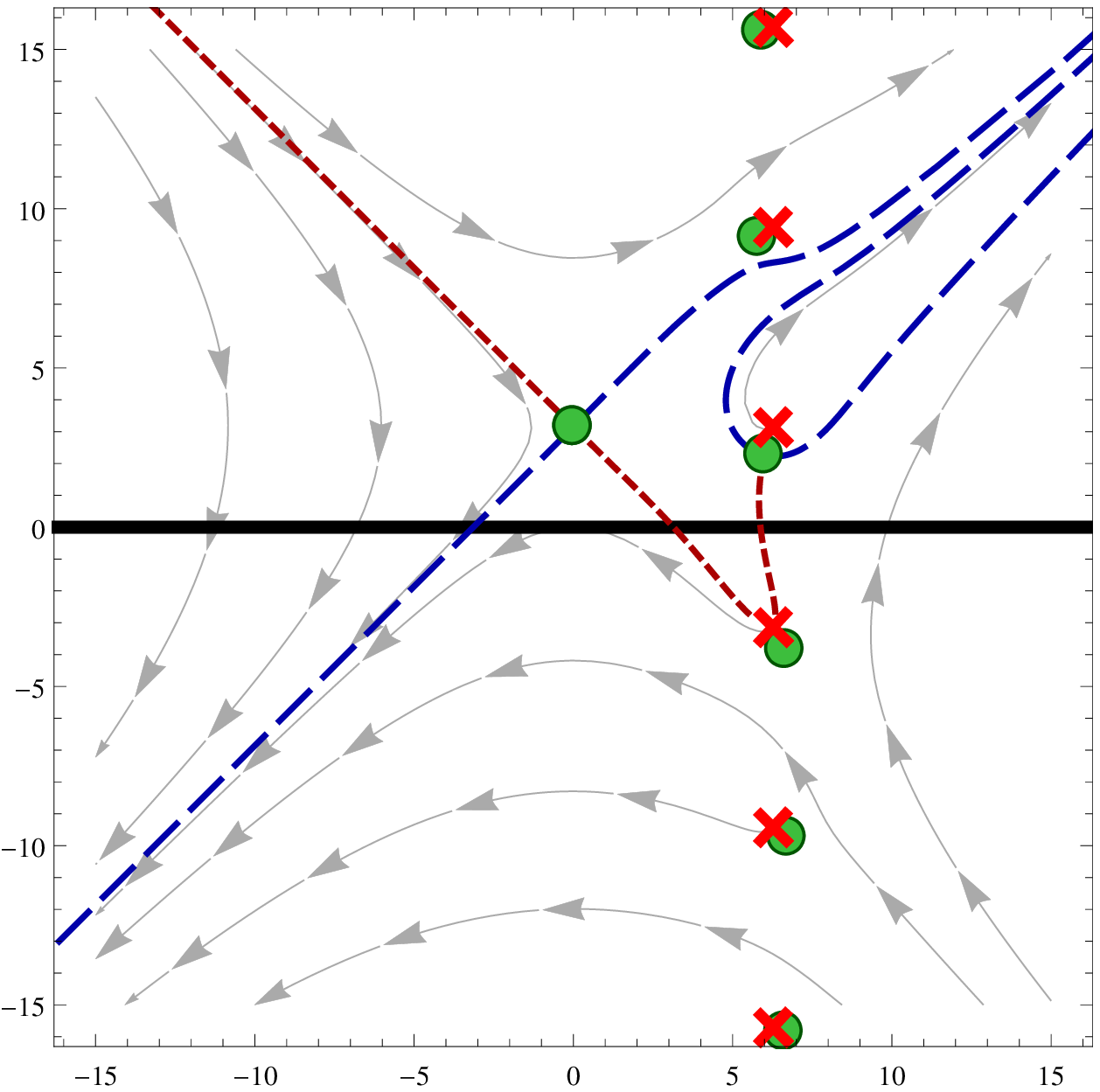}
          \hspace{1.6cm} (a) $m = 2 \pi$ 
        \end{center}
      \end{minipage}
      \begin{minipage}{0.33\hsize}
        \begin{center}
          \includegraphics[clip, width=50mm]{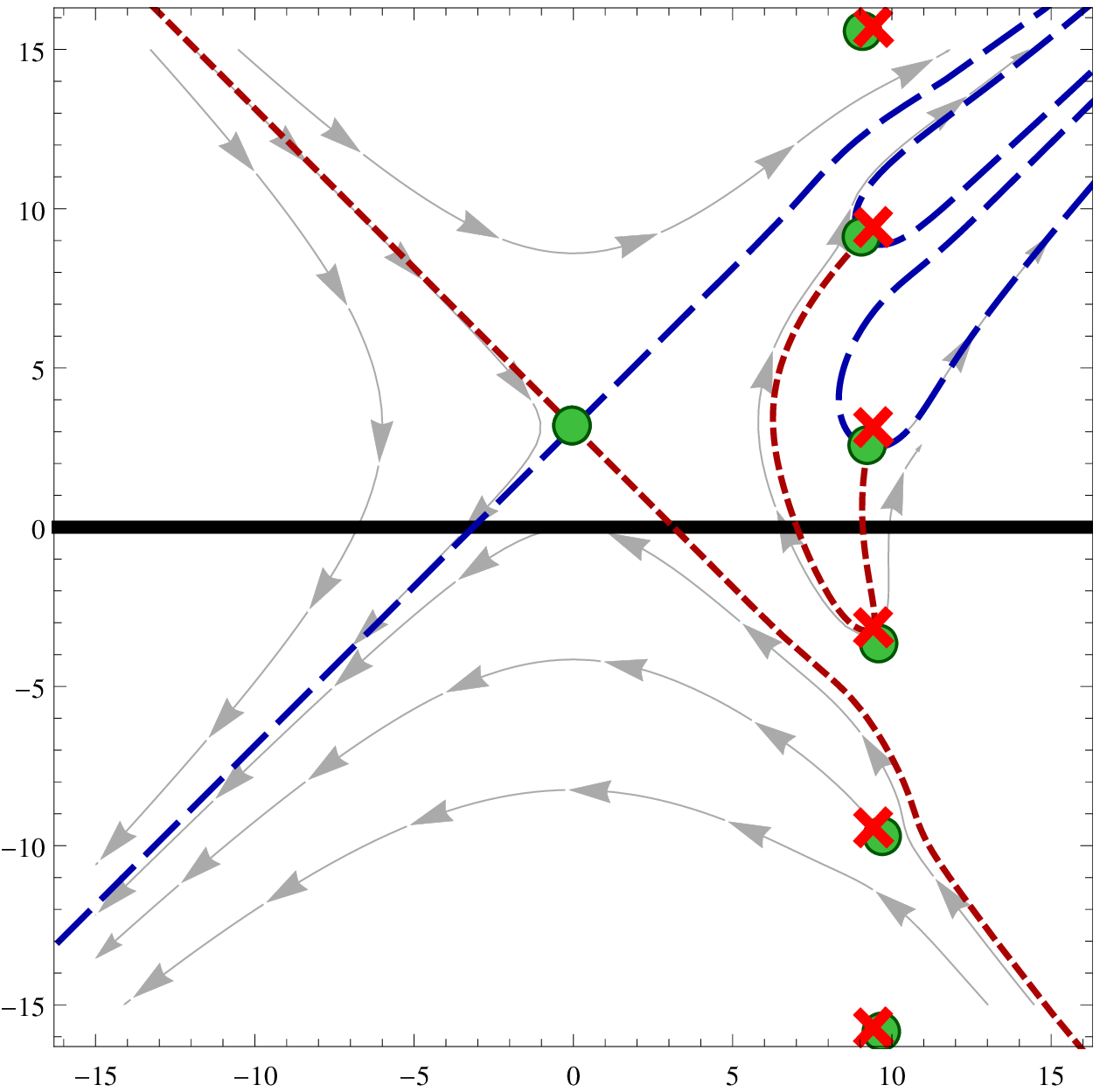}
          \hspace{1.6cm} (b) $m = 3 \pi$  
        \end{center}
      \end{minipage}
      \begin{minipage}{0.33\hsize}
        \begin{center}
          \includegraphics[clip, width=50mm]{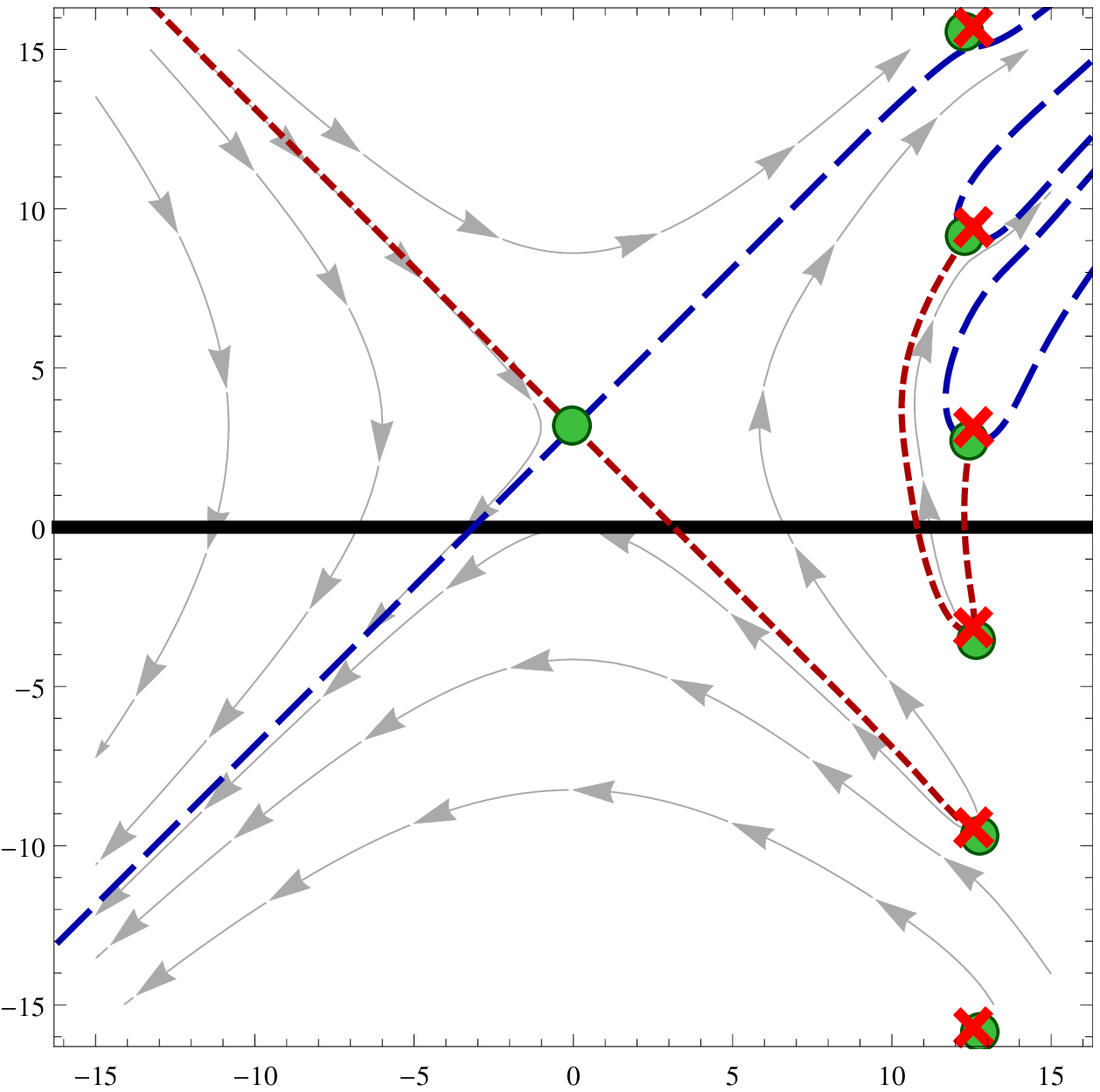}
          \hspace{1.6cm} (c) $m = 4 \pi$ 
        \end{center}
      \end{minipage}
    \end{tabular}
    \caption{Thimble structure of partition function of ${\mathcal N}=2$ 
CS SQED with $N_{f}=1$ hyper multiplet 
for $k=1$ ($g$ $\approx$ $12.6$) on ${\rm Re}\, z$-${\rm Im}\, z$ plane.
    The green points and red crosses stand for critical points and 
singularities, respectively.
      The red dotted lines stand for dual 
thimbles with nonzero intersection numbers, whereas the blue 
dashed lines for corresponding thimbles.
The arrows represent the flow lines. 
}
    \label{fig:thim_CS_U1_g10}
  \end{center}
\end{figure}

Now let us turn to the finite $g$ case.
As we already illustrated in Fig.~\ref{fig:actions}, 
$z_{\rm pt}^c$ and $z_\ell^c$ are distinct from $z=0$ and 
$z=m+(2\ell -1)\pi i$
respectively
and their actions receive finite $g$ corrections.
We have numerically solved the flow equation and obtained 
the thimbles and dual thimbles for the saddle points for finite $g$, 
where we figure out the structure of thimble decomposition 
for several choices of the real mass $m$ as follows. 
Figs.~\ref{fig:thim_CS_U1_g0.1} and \ref{fig:thim_CS_U1_g10} summarize 
the thimble structure for $g=\frac{4\pi}{100}$ $\approx 0.126$ ($k=100$) 
and $g=4\pi$ $\approx 12.56$ ($k=1$) with $m=2\pi,3\pi,4\pi$ 
in complexified $\sigma$ plane ($z$ plane) respectively
\footnote{Although the results could include small numerical errors,
the main arguments in the following are not affected by the details.
}.
For smaller $g$, we see in Fig.~\ref{fig:thim_CS_U1_g0.1} that
the nonperturbative saddle points (green points) and the singularities (red crosses) 
are almost degenerate, 
while they are slightly more separated in Fig.~\ref{fig:thim_CS_U1_g10} for larger $g$.
We term a saddle point near the origin as a ``perturbative" one and others as ``nonperturbative" ones.
As we will see later, this naming gets precisely appropriate only for the $g\to 0$ limit.

We first discuss the case with small $g$ in Fig.~\ref{fig:thim_CS_U1_g0.1}, 
which can be seen as an approximate example of the $g\to0$ limit.
The main results in Fig.~\ref{fig:thim_CS_U1_g0.1} are summarized as follows:
For $m=2\pi$, two of the saddle points contribute to the partition function: 
a thimble associated with the perturbative saddle $z_{\rm pt}^c$ near the origin 
and another one associated with $z_1^c$ near $z=m+\pi i$.
By Cauchy's theorem,
the integral along $\mathcal{J}_{\rm pt}$ is equivalent to the 
one along $e^{\frac{\pi i}{4}} \mathbb{R}$, namely $Z_{\rm pt}$, 
while the integral along $\mathcal{J}_0$ (the first nonperturbative 
thimble) corresponds to the residue at $z=z_1^\ast$ (see Fig.~\ref{fig:zoom}):
\begin{align}
&Z(g,m)=\int_{\mathcal{J}_{\rm pt} +\mathcal{J}_1 }  dz e^{-S[z]} ,\quad
\\
&\int_{\mathcal{J}_{\rm pt}} dz e^{-S[z]} =Z_{\rm pt}(g,m),\quad 
\int_{\mathcal{J}_1} dz e^{-S[z]} = {\rm Res}_{z=z_1^\ast } \Bigl[ e^{-S[z]} \Bigr]  \quad {\rm at}\ m= 2\pi  .
\end{align}
For $m=3\pi$, there are two important changes.
First, the dual thimble associated with $z_2^c$ (another nonperturbative saddle) intersects the real axis.
Second, the thimble associated with $z_{\rm pt}^c$
seems to pass $z_2^c$.
More precisely, this does not pass $z_2^c$ in a rigorous sense 
but almost passes $z_2^c$.
These facts imply that
Stokes phenomena in the sense of the thimble decomposition occur
at a certain point $m=\tilde{m}(g)$ which is slightly below $m=3\pi$
as expected from the subleading small-$g$ correction.
\begin{align}
&Z(g,m)= \int_{\mathcal{J}_{\rm pt} +\mathcal{J}_1  +\mathcal{J}_2} dz e^{-S[z]} ,\quad
\\
&\int_{\mathcal{J}_{\rm pt}} dz e^{-S[z]} =Z_{\rm pt}(g,m+0_{+}),\quad 
\int_{\mathcal{J}_{1,2}} dz e^{-S[z]} = {\rm Res}_{z=z_{1,2}^\ast } \Bigl[ e^{-S[z]} \Bigr]  \quad {\rm at}\ m= 3\pi  .
\end{align}
For $m=4\pi$, 
three of saddle points contribute to the partition function: 
a thimble associated with the perturbative saddle (near the origin) 
and two thimbles associated with the nonperturbative saddles.
\begin{align}
&Z(g,m)= \int_{\mathcal{J}_{\rm pt} +\mathcal{J}_1 +\mathcal{J}_2}  dz e^{-S[z]} ,\quad
\\
&\int_{\mathcal{J}_{\rm pt}} dz e^{-S[z]} =Z_{\rm pt}(g,m), \quad 
\int_{\mathcal{J}_{1,2}} dz e^{-S[z]} = {\rm Res}_{z=z_{1,2}^\ast } \Bigl[ e^{-S[z]} \Bigr]  \quad {\rm at}\ m= 4\pi  .
\end{align}

\begin{figure}[t]
\begin{center}
\includegraphics[width=8.0cm]{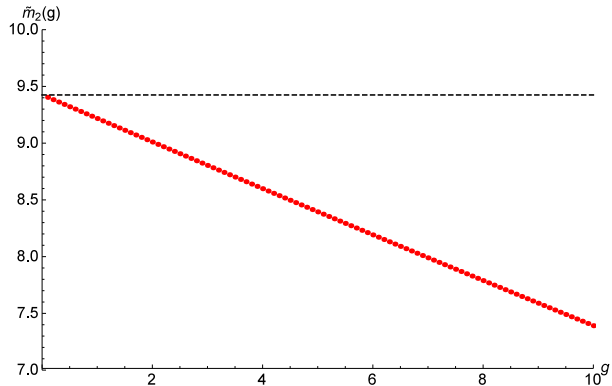}
 \caption{The red dotted line denotes
$\tilde{m}_2 (g)$ as a function of $g$.
The black dotted line denotes $3\pi$.
}
 \label{fig:tildem}
\end{center}
\end{figure}

For larger $g$,
the differences from the weak coupling case are more explicit
as illustrated in Fig.~\ref{fig:thim_CS_U1_g10}.
First the critical points $z_{\rm pt}^c$ and $z_\ell^c$ are clearly separated 
from the origin and singularities respectively.
The thimble structures at $m=2\pi$, $3\pi$ and $4\pi$
are the same as the ones of $g\approx0.126$
but the value of $\tilde{m}_2 (g)$ clearly deviates from $m=3\pi$.
We here give a short explanation on the thimble structure before the detailed discussion:
For $\pi<m<\tilde{m}_2$ the perturbative contribution $Z_{\rm pt}$ is only composed of the thimble associated with the perturbative saddle point near the origin, 
while $Z_{\rm pt}$ gets composed of the perturbative thimble and one more thimble associated with the nonperturbative saddle for $\tilde{m}_2 <m<3\pi$. 
This nonperturbative thimble comes to contribute as the ``genuine" nonperturbative contribution at the Stokes line $m=3\pi$.
In Fig.~\ref{fig:tildem}
we plot $\tilde{m}_2 (g)$ as a function of $g$.
We immediately see that
$\tilde{m}_2 (g)$ deviates from $3\pi$ for strong coupling.

\begin{figure}[t]
\begin{center}
\includegraphics[clip, width=180mm]{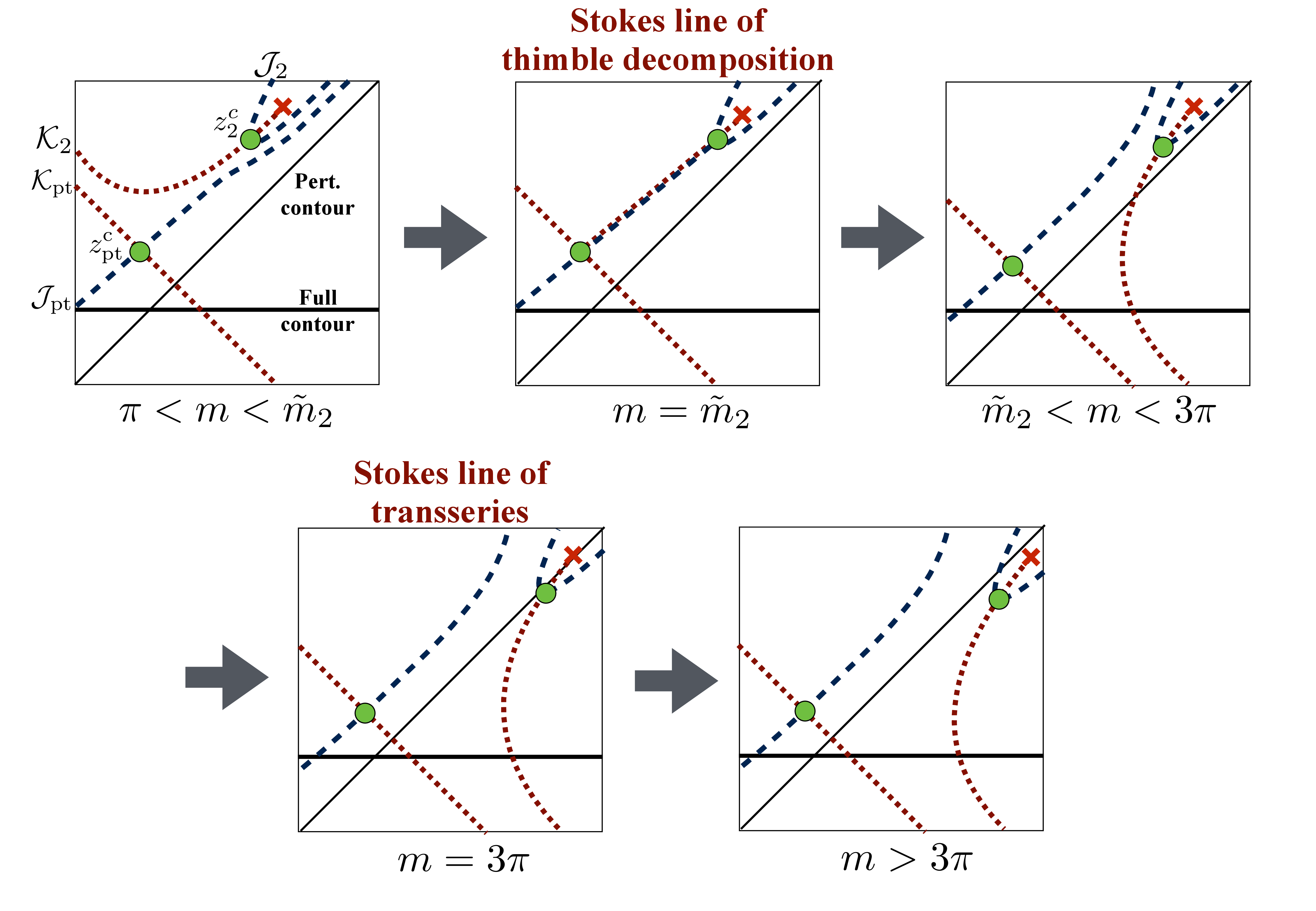}
\caption{Schematic expanded figures for Fig.~\ref{fig:thim_CS_U1_g10} 
around $m\approx3\pi$. We only show two saddle points 
$z^{\rm c}_{\rm pt}$ and $z^{\rm c}_{2}$ (green points) to 
discuss the Stokes phenomena around $m=3\pi$. We also exhibit 
the associated thimbles ${\mathcal J}_{\rm pt}$, ${\mathcal J}_{2}$ 
(blue dashed lines) and dual thimbles 
${\mathcal K}_{\rm pt}$, ${\mathcal K}_{2}$ (red dotted 
lines). One can figure out the Stokes lines by looking into 
the intersection of the dual thimbles with ``Full contour"(black 
bold line) and``Pert. contour"(black solid line).
}
    \label{fig:stokes}
  \end{center}
\end{figure}

We have analyzed the cases for generic values of $(g,m)$ and
summarized the thimble structures related to $z_{\rm pt}^c$ and $z_2^c$
in Fig.~\ref{fig:stokes},
which is the schematic expanded version of Fig.~\ref{fig:thim_CS_U1_g10}. 
In the figure, 
we only show two saddle points $z^{\rm c}_{\rm pt}$, $z^{\rm c}_{2}$ and one singularity $z^{*}_{2}$ to discuss the Stokes phenomena just around $m=3\pi$. 
We manifest their associated thimbles ${\mathcal J}_{\rm pt}, {\mathcal J}_{2}$ 
and dual thimbles ${\mathcal K}_{\rm pt}, {\mathcal K}_{2}$. 
We now look into the intersection of the dual thimbles 
with ``Full contour": $\mathbb{R}$ 
and ``Perturbative contour": $e^{\frac{\pi i}{4}}\mathbb{R}$.
The full contour stands for the integration contour 
giving the exact partition function $Z$ 
while the perturbative contour is the one giving the perturbative part $Z_{\rm pt}$
of the transseries,
which is the Borel resummation along $\mathbb{R}_+$.
The results of Fig.~\ref{fig:stokes} is summarized as follows:

\begin{itemize}
\item For $m<\tilde{m}_2$, the dual thimble ${\mathcal K}_{\rm pt}$ intersects 
with both the full and perturbative contours.
It indicates that 
the perturbative thimble ${\mathcal J}_{\rm pt}$ yields the perturbative contribution $Z_{\rm pt}$ in the full transseries of the partition function:
\begin{\eq}
Z(g,m)=\int_{\mathcal{J}_{\rm pt} +\mathcal{J}_1 }  dz e^{-S[z]} ,\quad
\int_{\mathcal{J}_{\rm pt}} dz e^{-S[z]} =Z_{\rm pt}(g,m),\quad\quad 
 {\rm for}\ \pi < m<\tilde{m}_2  .
\end{\eq}

\item For $m=\tilde{m}_2$, the two saddle points $z^{\rm c}_{\rm pt}$ and $z^{\rm c}_{2}$ are connected by the thimble 
${\mathcal J}_{\rm pt}$ and the dual thimble ${\mathcal K}_{2}$.
This indicates that
our thimble decomposition has the Stokes phenomenon
and is apparently ambiguous at $m=\tilde{m}_2$.

\item For $\tilde{m}_2 <m<3\pi$, 
the dual thimble ${\mathcal K}_{2}$ intersects 
with both the full and perturbative contours,
which means that the nonperturbative thimble ${\mathcal J}_{2}$ contributes, but just as part of the perturbative contribution $Z_{\rm pt}$:
\begin{\eq}
\int_{\mathcal{J}_{\rm pt}+\mathcal{J}_2} dz e^{-S[z]} =Z_{\rm pt}(g,m),\quad\quad 
 {\rm for}\ \tilde{m}_2 <m<3\pi  .
\end{\eq}
Therefore we can express the exact result in this regime as
\begin{\eq}
Z(g,m)
=\int_{\mathcal{J}_{\rm pt} +\mathcal{J}_1 +\mathcal{J}_2} dz e^{-S[z]} 
=Z_{\rm pt}(g,m) +{\rm Res}_{z=z_1^\ast } \Bigl[ e^{-S[z]} \Bigr] ,\quad\quad
 {\rm for}\ \tilde{m}_2 <m<3\pi  ,
\end{\eq}
which agrees the transseries representation.

\item At $m=3\pi$, the integral along the perturbative contour is ill-defined
due to the pole $z_2^\ast$
but the integral along $\mathcal{J}_2$ 
is still related to $Z_{\rm pt}$ as the ambiguous part:
\begin{\eq}
\int_{\mathcal{J}_{\rm pt}} dz e^{-S[z]} =Z_{\rm pt}(g,m+0_+)
= P \int_{0}^\infty dt\ e^{-\frac{t}{g}} \mathcal{B}Z(t) 
 -\frac{1}{2}{\rm Res}_{z=z_2^\ast}\Bigl[ e^{-S[z]} \Bigr],
 \quad\quad  {\rm at}\ m=3\pi  .
\end{\eq}
The Lefschetz thimble decomposition of the exact result 
is well-defined at this point
and the exact result is expressed as
\begin{align}
Z(g,m)
=\int_{\mathcal{J}_{\rm pt} +\mathcal{J}_1 +\mathcal{J}_2} dz e^{-S[z]}
= Z_{\rm pt}(g,m+0_+) +{\rm Res}_{z=z_1^\ast } \Bigl[ e^{-S[z]} \Bigr] 
+{\rm Res}_{z=z_2^\ast } \Bigl[ e^{-S[z]} \Bigr] 
\nonumber\\
{\rm at}\ m=3\pi ,
\end{align}
which is equivalent to the transseries expression at $m=3\pi +0_+$.
Note that using
\[
Z_{\rm pt}(g,m+0_-)
= P \int_{0}^\infty dt\ e^{-\frac{t}{g}} \mathcal{B}Z(t) 
 +\frac{1}{2}{\rm Res}_{z=z_2^\ast}\Bigl[ e^{-S[z]} \Bigr],
\]
we can also write the exact result as
\begin{\eq}
Z(g,m)
= Z_{\rm pt}(g,m+0_-) +{\rm Res}_{z=z_1^\ast } \Bigl[ e^{-S[z]} \Bigr] 
\quad\quad  {\rm at}\ m=3\pi ,
\end{\eq}
which is the transseries representation at $m=3\pi +0_-$.
This is what we expect from the resurgence analysis.
Namely we have manifested that
the transseries has the Stokes phenomena at $m=3\pi$
and the well-defined thimble decomposition of the exact result at $m=3\pi$
coincides with the unambiguous answer obtained by the resurgence:
\begin{\eq}
Z(g,m)
= P \int_{0}^\infty dt\ e^{-\frac{t}{g}} \mathcal{B}Z(t) 
 +{\rm Res}_{z=z_1^\ast}\Bigl[ e^{-S[z]} \Bigr]
 +\frac{1}{2}{\rm Res}_{z=z_2^\ast}\Bigl[ e^{-S[z]} \Bigr]
\quad  {\rm at}\ m=3\pi .
\end{\eq}

\item For $m> 3\pi$, 
let us take $m$ to be smaller than the next Stokes line
to keep that another Stokes phenomena with $z_{\ell\geq 3}^c$ does not matter,
namely$3\pi <m<\tilde{m}_3 (g)$
\footnote{
Strictly speaking, 
we expect $3\pi <\tilde{m}_3 (g)<5\pi$ 
up to some large value of $g$ 
as suggested by fig.~\ref{fig:thim_CS_U1_g10} (c) for $g\approx 12.6$,
but we do not know whether or not this is still true for very large $g$.
For small $g$, we can show $\tilde{m}_\ell (g)=(2\ell -1)\pi -\frac{7\pi /4 -{\rm arg}(-1)^{\ell -1}}{2(2\ell -1)\pi}g +\mathcal{O}(g^2 )$,
which indicates $(2\ell -3)\pi <\tilde{m}_\ell (g)<(2\ell -1)\pi$ up to $\mathcal{O}(g^2 )$
if we take into account ``branch cut problem" commented in the last of this subsection.
Note that we do not need to know thimble structures for large-$g$
in order to compare with the resurgence structures.
}
In this regime
the dual thimble ${\mathcal K}_{2}$ does not intersect with the perturbative contour 
while it still intersects with the full contour. 
It indicates that 
the nonperturbative thimble ${\mathcal J}_{2}$ comes to contribute as the nonperturbative contribution, 
not as part of the perturbative contribution:
\begin{\eq}
\int_{\mathcal{J}_{\rm pt}} dz e^{-S[z]} =Z_{\rm pt}(g,m),\quad\quad 
 {\rm for}\ 3\pi <m<\tilde{m}_2 (g)  ,
\end{\eq}
which leads us to
\begin{align}
Z(g,m)
=\int_{\mathcal{J}_{\rm pt} +\mathcal{J}_1 +\mathcal{J}_2} dz e^{-S[z]} 
=Z_{\rm pt}(g,m) 
+{\rm Res}_{z=z_1^\ast } \Bigl[ e^{-S[z]} \Bigr] 
+{\rm Res}_{z=z_2^\ast } \Bigl[ e^{-S[z]} \Bigr]
\nonumber\\
 {\rm for}\ 3\pi <m<\tilde{m}_3 (g) .
\end{align}
\end{itemize}

If we further increase $m$,
then we encounter Stokes phenomenon with other 
critical points in similar ways.
We conclude that
the Lefschetz thimble decomposition for any $m$ is
\begin{\eq}
Z(g,m)
=\int_{\mathcal{J}_{\rm pt}} dz e^{-S[z]} 
+\sum_{\ell =1}^\infty \theta (m-\tilde{m}_\ell (g)) 
  \int_{\mathcal{J}_\ell} dz e^{-S[z]}  .
\label{eq:decompositionU1}
\end{\eq}
This shows that
we have the decomposition
\be
{\mathcal C}_{\mathbb R}= n_{\rm pt} {\mathcal J}_{\rm pt} + \sum_{\ell} n_{\ell}{\mathcal J}_{\ell}\,,
\ee
with the intersection numbers 
\begin{\eq}
n_{\rm pt}=1 ,\quad n_\ell =\theta \left( m-\tilde{m}_\ell (g)\right) .
\end{\eq}
The thimble integral along $\mathcal{J}_\ell$ is equivalent to the residue of the Borel singularities,
which is the nonperturbative exponential part other than the step function in $Z^{n}_{\rm np}$ of the transseries (\ref{eq:Z2}).
$Z_{\rm pt}(g,m)$ is related to the thimble integrals 
in a complicated way due to the intersection number between $e^{\frac{\pi i}{4}}\mathbb{R}$
and the dual thimbles.
In terms of the Boxcar function $\Pi_{a,b}(x)$
\begin{\eq}
\Pi_{a,b}(x) =\theta (x-a) -\theta (x-b)
=\begin{cases} 
0 & {\rm for}\ x<a \cr 1 & {\rm for}\ a<x<b \cr 0 & {\rm for}\ x>b
\end{cases} ,
\end{\eq}
$Z_{\rm pt}(g,m)$ is decomposed as
\begin{\eq}
Z_{\rm pt}(g,m)
=\int_{\mathcal{J}_{\rm pt}} dz e^{-S[z]} 
+\sum_{\ell =1}^\infty \Pi_{\tilde{m}_\ell (g), (2\ell -1)\pi }(m) 
  \int_{\mathcal{J}_\ell} dz e^{-S[z]}  .
\end{\eq}
This decomposition is ambiguous at $m=\tilde{m}_\ell (g)$ and $(2\ell -1)\pi$.
At $m=(2\ell -1)\pi$, the integral along $\mathcal{J}_{\rm pt}$ is 
related to $Z_{\rm pt}$ by
\begin{\eq}
\int_{\mathcal{J}_{\rm pt}} dz e^{-S[z]} =Z_{\rm pt}(g,m+0_+)
= P \int_{0}^\infty dt\ e^{-\frac{t}{g}} \mathcal{B}Z(t) 
 -\frac{1}{2}{\rm Res}_{z=z_{\ell}^\ast}\Bigl[ e^{-S[z]} \Bigr],
 \quad\quad  {\rm at}\ m=(2\ell -1)\pi  .
\end{\eq}
Thus, at $m=(2\ell -1)\pi$,
we can rewrite the exact result  as 
\begin{align}
Z(g,m)
&= Z_{\rm pt}(g,m+0_+ )
+\sum_{\ell' =1}^{\ell}  {\rm Res}_{z=z_{\ell'}^\ast } \Bigl[ e^{-S[z]} \Bigr] 
= Z_{\rm pt}(g,m+0_- )
+\sum_{\ell' =1}^{\ell-1}  {\rm Res}_{z=z_{\ell'}^\ast } \Bigl[ e^{-S[z]} \Bigr] \NN\\
&= P \int_{0}^\infty dt\ e^{-\frac{t}{g}} \mathcal{B}Z(t) 
 +\sum_{\ell' =1}^{\ell-1} {\rm Res}_{z=z_{\ell'}^\ast}\Bigl[ e^{-S[z]} \Bigr]
 +\frac{1}{2}{\rm Res}_{z=z_\ell^\ast}\Bigl[ e^{-S[z]} \Bigr]\quad\quad
{\rm at}\ m=(2\ell -1)\pi ,
\nonumber\\
\end{align}
which is the same as the unambiguous answer obtained in the resurgent transseries.
We can easily derive 
the resurgent transseries from the thimble decomposition
by considering small-$g$ expansion of the expression \eqref{eq:decompositionU1}.
Noting $\tilde{m}_\ell (g)=(2\ell -1)\pi +\mathcal{O}(g)$,
we can replace $\theta (m-\tilde{m}_\ell (g))$ by $\theta (m-(2\ell -1)\pi)$
and arrive at
\begin{\eq}
Z(g,m)
=Z_{\rm pt}(g,m)
+\sum_{\ell =1}^\infty \theta \left( m-(2\ell -1)\pi \right) 
  {\rm Res}_{z=z_{\ell}^\ast } \Bigl[ e^{-S[z]} \Bigr] , 
\end{\eq}
which is nothing but the resurgent transseries representation.

Now we comment on the definition of the perturbative contribution.
As we mentioned in the end of the previous subsection, the definition of the perturbative contribution
based on the Borel resummation is just one of definitions.
In our work, we define the perturbative part as the Borel resummation of the perturbative series and decompose
the exact result into the perturbative and nonperturbative parts.
We may be able to propose another feasible definition of the perturbative contribution: the thimble integral associated with the perturbative saddle $z^{\rm c}_{\rm pt}$ is regarded as the perturbative contribution while the nonperturbative contributions are defined as the thimble integral associated with the nonperturbative saddles $z^{\rm c}_{\ell}$.
In this alternative definition, the Stokes phenomenon of thimble decomposition at $m=m^{*}$ becomes a Stokes phenomenon of transseries while $m=(2n-1)\pi$ is no longer a Stokes line.
We emphasize that the two definitions get equivalent in the $g\to 0$ limit.

\begin{figure}[t]
\begin{center}
\includegraphics[width=8.0cm]{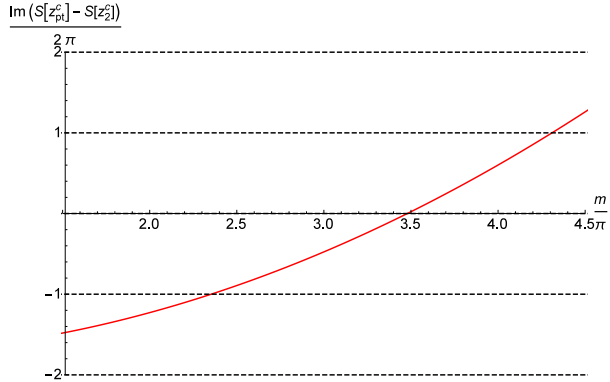}
 \caption{
The difference of ${\rm Im}S$ at $z_{\rm pt}^c$ and $z_2^c$
normalized by $2\pi$ as a function of $m/\pi$
for $g=10$ in the notation ``$\log{1}$"$=0$.
}
\label{fig:ImS}
\end{center}
\end{figure}

Finally we mention a technical subtlety 
of ${\rm Im}S$ for different thimbles.
As well-known,
a necessary condition for having a Stokes phenomenon 
between two thimbles $\mathcal{J}$ and $\mathcal{J}'$ is
to have the same imaginary part of action:
$\left. {\rm Im}S \right|_{\mathcal{J}} =\left. {\rm Im}S \right|_{\mathcal{J}'}$.
However, we have to be careful in evaluating ${\rm Im}S$
when the action has branch cuts as noted in \cite{Fujii:2015bua}.
For our case,
we have infinitely many logarithmic branch cuts
extended from the poles of the integrand,
which generate ambiguities in specifying ``$\log{1}$"$=2\pi i\mathbb{Z}$.
Thus, the necessary condition for the Stokes phenomenon can be modified as
\be
\left. {\rm Im} \, S \right|_{\cal J} = \left. {\rm Im} \, S \right|_{\cal J^\prime} + 2 \pi n' , \quad n' \in {\mathbb Z},
\ee
and one can determine $n'$ of each thimble by looking into the Stokes phenomena in details.
For example,
we present $\frac{{\rm Im}(S[z_{\rm pt}^c]-S[z_2^c])}{2\pi}$
as a function of $m/\pi$ for $g=10$ in Fig.~\ref{fig:ImS}.
We take the notation ``$\log{1}$"$=0$ in computing ${\rm Im}S$.
Here, the Stokes line in the thimble decomposition is given as $\tilde{m}_2 (g=10)\simeq 2.355\pi$.
For this value of $m$, we have $\frac{{\rm Im}(S[z_{\rm pt}^c]-S[z_2^c])}{2\pi}\approx -1$, which means
$\left. {\rm Im}S \right|_{\mathcal{J}_{\rm pt}}$ 
$=\left. {\rm Im}S \right|_{\mathcal{J}_2}-2\pi$ at $m=\tilde{m}_2 (g)$.
This is a clear example where we need to take care of the branch cuts to consider thimble decompositions.

\subsection{Stokes phenomena in terms of ${\rm arg}(g)$}
\label{sec:arg}
So far we have discussed
the Stokes phenomena and the resurgent structure
by changing the real mass parameter
while we have fixed the coupling $g$ to be real positive.
It would be also interesting to change ${\rm arg}(g)$ with fixed $m$
in the integral \eqref{eq:Zexact}
as in the usual analyses of the resurgence theory.
Note that it is unclear 
whether or not \eqref{eq:Zexact} for complex $g$ can be interpreted 
as $S^3$ partition function of the theory with complex $g$
except ${\rm arg}(g)=0,\pi$ 
\footnote{More precisely, except $k\in \mathbb{Z}$.}
since the localization procedure requires gauge invariance naively.
In order to see the relation in a more precise manner,
we need to perform analogue of the reference \cite{Witten:2010zr} 
for 3d $\mathcal{N}=2$ CS matter theory
but we do not discuss this in the present work.
This subsection is motivated 
by technical comparison with the standard resurgence analyses. 
We take $N_f =1$ for simplicity in this subsection.

\subsubsection{Resurgent transseries}
\begin{figure}[t]
  \begin{center}
    \begin{tabular}{cc}
      \begin{minipage}{0.5\hsize}
        \begin{center}
          \includegraphics[clip, width=80mm]{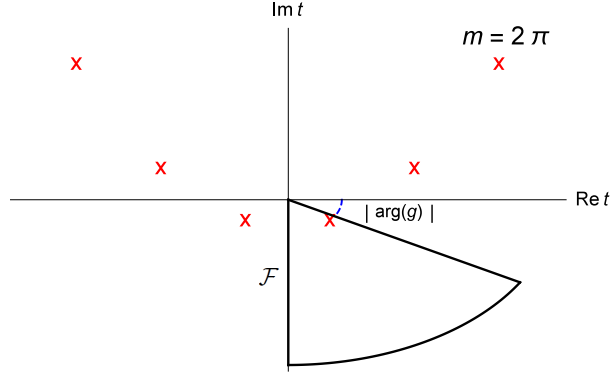}
          \hspace{1.6cm}
        \end{center}
      \end{minipage} 
    \end{tabular}
    \caption{
 Borel singularities and the fan $\mathcal{F}$
 in the case of $m=2\pi$ and ${\rm arg}(g)=-\pi /6$.
Whether the singularities are inside $\mathcal{F}$ depends on ${\rm arg}(g)$
while their locations are independent.
    }
  \label{fig:Borel_complex_g}
  \end{center}
\end{figure}

Let us take complex $g$ in the integral \eqref{eq:Zexact}.
In order to keep the integral finite,
we restrict ourselves to 
\begin{\eq}
-\pi \leq {\rm arg} (g) \leq 0 \quad  (0 \leq {\rm arg}  (k)\leq \pi ) .
\end{\eq}
Repeating the argument of \cite{Honda:2016vmv},
we can easily show that
the exact result for nonzero ${\rm arg}(g)$ can be still written as
\begin{\eq}
Z =\int_{0}^{-i \infty } dt\ e^{-\frac{t}{g}} \mathcal{B}Z(t)\,,
\end{\eq}
where the Borel transformation $\mathcal{B}Z(t)$
is given by \eqref{eq:U1BorelT}.
The main difference from the ${\rm arg}(g)=0$ case is that
the ``standard direction" of the Borel resummation
is $\varphi ={\rm arg}(g)$ rather than $\varphi =0$,
which is equivalent to 
${\rm arg} (\sigma) =\frac{\pi}{4}+\frac{{\rm arg}(g)}{2}$
in the language of $\sigma$.
Therefore considering a contour integral along the fan 
$\mathcal{F}$ connecting $-i\mathbb{R}_+$ and
$e^{i {\rm arg}(g)}\mathbb{R}_+$
(see Fig.~\ref{fig:Borel_complex_g}),
we find 
\begin{\eq}
Z =
\int_{0}^{e^{i{\rm arg} (g)} \infty } dt\
  e^{-\frac{t}{g}} \mathcal{B}Z(t)
+\sum_{{\rm poles}\in \mathcal{F} } 
{\rm Res}\Bigl[ e^{-\frac{t}{g}} \mathcal{B}Z(t) \Bigr] ,
\end{\eq}
which is the extension of \eqref{eq:decompose} to general ${\rm arg}(g)$.
As in the ${\rm arg}(g)=0$ case,
we identify the first and second terms 
with perturbative and non-perturbative contributions respectively.
On the Borel plane, the non-perturbative corrections are given 
by the residues around the Borel singularities 
$t_n^\ast =-i[m+(2n-1)\pi i]^2 $
satisfying $\varphi={\rm arg}(g) >{\rm arg}(t_n^\ast )$.
As changing ${\rm arg}(g)$ from $0$ to $-\pi$, 
the Stokes line rotates clockwise 
but the locations of the singularities are unchanged
since they are independent of $g$ with fixed $m$. 
Then, except for the Stokes lines with respect to ${\rm arg}(g)$,
we can write the partition function as
\footnote{We assume that $m>0$ as with the case ${\rm arg}(g)=0$.}
\begin{align}
& Z =Z_{\rm pt}+\sum_{n} Z_{\rm np}^{(n )}\,,  \NN\\
& Z_{\rm pt}
=  \int_{0}^{e^{i {\rm arg}(g)} \infty} dt\ e^{-\frac{t}{g}} 
\mathcal{B}Z(t) ,\quad
 Z_{\rm np}^{(n )}
=  \theta \left( {\rm arg}(g) -{\rm arg}(t_n^\ast )    \right)
2\pi (-1)^{n-1} \,e^{\frac{i}{g}[m+ (2n-1)\pi i]^{2}}, 
\label{eq:Z_np_g_comp}
\end{align}
where
\begin{\eq}
{\rm arg}(t_n^\ast ) 
= -\frac{\pi}{2} +2\arctan{\frac{(2n-1)\pi}{m}}
= -\arctan{\left( \frac{m^2 -(2n-1)^2 \pi^2}{2m(2n-1)\pi}\right)} .
\end{\eq}
Note that the only differences from the ${\rm arg}(g)=0$ case are
the change of the contour of the perturbative Borel resummation
and the step function in $Z_{\rm np}^{(n )}$.
Namely the perturbative series in every sector is unchanged
and only the transseries parameter is changed.
We emphasize that we are changing ${\rm arg}(g)$ rather than $m$.
This is why the variable in the step function is not $m$ but ${\rm arg}(g)$.
We can see from (\ref{eq:Z_np_g_comp}) that
for $-\pi/2 < {\rm arg}(g) \le 0$, 
the total partition function has 
the non-perturbative part coming only from the Borel singularities $t_n^\ast =-i[m+(2n-1)\pi i]^2 $
with the positive $n$
and the fan $\mathcal{F}$ becomes narrower 
for smaller ${\rm arg}(g)$ (larger $|{\rm arg}(g)|$) in this regime.
In particular, for ${\rm arg}(g)=-\pi /2$,
the fan coincides with $-i\mathbb{R}_+$ and the exact result has only the perturbative part.
For $-\pi \le {\rm arg}(g) < -\pi/2$,
the partition function receives non-perturbative corrections 
from $n\in\mathbb{Z}_{\leq 0}$.

For ${\rm arg}(g)={\rm arg}(t_n^\ast)$,
$Z_{\rm pt}$ and $Z_{\rm np}^{(n)}$ are apparently ambiguous
since the integral in $Z_{\rm pt}$ hits the singularity at $t=t_n^\ast$
and the step function in $Z_{\rm np}^{(n)}$ is ambiguous.
The ambiguities are indeed canceled as in Sec.~\ref{sec:SQED_trans}.
Let us estimate
the Borel ambiguity by
\begin{align}
\left( 
{\mathcal S}_{{\rm arg}(t_n^\ast ) +0^+} -{\mathcal S}_{{\rm arg}(t_n^\ast ) +0^-}
\right) Z(g,m) ,
\end{align}
as usual.
Noting
\begin{align}
\left. Z_{\rm pt}(g, m) \right|_{{\rm arg}(g)={\rm arg}(t_n^\ast ) +0^\pm}
= P \int_{0}^{e^{i{\rm arg}(g)}\infty} dt\ e^{-\frac{t}{g}} \mathcal{B}Z(t) 
 \mp \frac{1}{2}{\rm Res}_{t=t_{n}^{*}}\Bigl[ e^{-\frac{t}{g}} \mathcal{B}Z(t) \Bigr] ,
\end{align}
the Borel ambiguity in the perturbative sector is
\begin{align}
\left. Z_{\rm pt}(g, m) \right|_{{\rm arg}(g)={\rm arg}(t_n^\ast ) +0^+}
-\left. Z_{\rm pt}(g, m) \right|_{{\rm arg}(g)={\rm arg}(t_n^\ast ) +0^-}
= -{\rm Res}_{t=t_{n}^{*}}\Bigl[ e^{-\frac{t}{g}} \mathcal{B}Z(t) \Bigr] ,
\end{align}
while the non-perturbative ones are
\begin{align}
\left. Z_{\rm np}^{(\ell)} \right|_{{\rm arg}(g)={\rm arg}(t_n^\ast ) +0^+}
-\left. Z_{\rm np}^{(\ell)} \right|_{{\rm arg}(g)={\rm arg}(t_n^\ast ) +0^-}
=\begin{cases}
 0 & {\rm for}\ \ell\neq n \cr
 +{\rm Res}_{t=t_{n}^{*}}\Bigl[ e^{-\frac{t}{g}} \mathcal{B}Z(t) \Bigr] & {\rm for}\ \ell=n
\end{cases} .
\end{align}
Therefore the ambiguities are canceled
and the whole transseries gives the unambiguous answer,
which agrees with the exact result.

\subsubsection{Thimble decomposition}
Let us decompose the exact result into Lefschetz thimble contributions.
First, we discuss a small-$g$ regime analytically.
As with the ${\rm arg}(g) =0$ case,
the critical points up to $\mathcal{O}(g^2 )$ are given by \eqref{eq:criticalOg}
\[
 z_{\rm pt}^c (g,m) = \frac{i g}{4} \tanh{\frac{m}{2}} +\mathcal{O}(g^2 ) ,\quad
 z_\ell^c (g,m) = z_\ell^\ast  +\frac{g}{2i z_\ell^\ast } +\mathcal{O}(g^2 ) ,
\]
which approach the origin and the positions of poles of the integrand 
in the $g\rightarrow 0$ limit respectively.
The perturbative thimble in the $g\rightarrow 0$ limit is given by
\begin{align}
\lim_{g\rightarrow 0} z_{\rm pt}(g,m;s) \, 
= \, \epsilon \exp\Biggl[ {\frac{2s}{|g|}
 +i\left(\frac{\pi}{4} - \frac{{\rm arg}(g)}{2} \right) } \Biggr]\, ,
\end{align}
with 
a parameter $\epsilon \in {\mathbb R}$ for the initial condition.
The actions at the critical points are still given by \eqref{eq:actionOg}, but the necessary condition for having Stokes phenomenon is slightly modified as
\begin{\eq}
0= -{\rm Re}\left[ \frac{z_\ell^{\ast 2}}{g}  \right] 
 +{\rm arg}\frac{(-1)^{\ell-1} g}{z_\ell^\ast} +\mathcal{O}(g ) ,
\end{\eq}
or equivalently
\begin{\eq}
0= -\frac{|z_\ell^\ast|^2}{|g|} 
\cos{\left( 2{\rm arg}(z_\ell^\ast ) -{\rm arg}(g) \right)}
 +{\rm arg}\frac{(-1)^{\ell-1} g}{z_\ell^\ast} +\mathcal{O}(g ) .
\label{eq:ImS_complex_g}
\end{\eq}
In the $|g| \rightarrow 0$ limit,
one of the solutions of this condition is 
${\rm arg}(g)=-\pi /2 +2{\rm arg}(z_n^\ast)={\rm arg}(t_n^\ast )$.
This is consistent 
with the Stokes phenomena of the transseries \eqref{eq:Z_np_g_comp}
at ${\rm arg}(g)={\rm arg}(t_n^\ast )$,
which we encountered above.
Note that ${\rm arg}(g)={\rm arg}(t_n^\ast )$ 
is no longer solution of \eqref{eq:ImS_complex_g} for nonzero $|g|$.
This indicates that for nonzero $|g|$,
we have the Stokes phenomena of the thimble decomposition
at a different point ${\rm arg}(g)={\rm arg}(\tilde{g}_n )(|g|,m)$
which approaches ${\rm arg}(t_n^\ast )$ in the weak coupling limit:
\begin{\eq}
\lim_{|g|\rightarrow 0} {\rm arg}(\tilde{g}_n )(|g|,m)
={\rm arg}(t_n^\ast ) ,
\end{\eq}
which is the counter part of $\tilde{m}_n (g)$ 
in the case of ${\rm arg}(g)=0$ with varying $m$. 
It is worth to note that
the Stokes line in the $g$-plane is curved rather than straight for given $m$
since ${\rm arg}(\tilde{g}_n )$ depends also on $|g|$.

\begin{figure}[t]
  \begin{center}
    \begin{tabular}{ccc}
      \begin{minipage}{0.33\hsize}
        \begin{center}
          \includegraphics[clip, width=50mm]{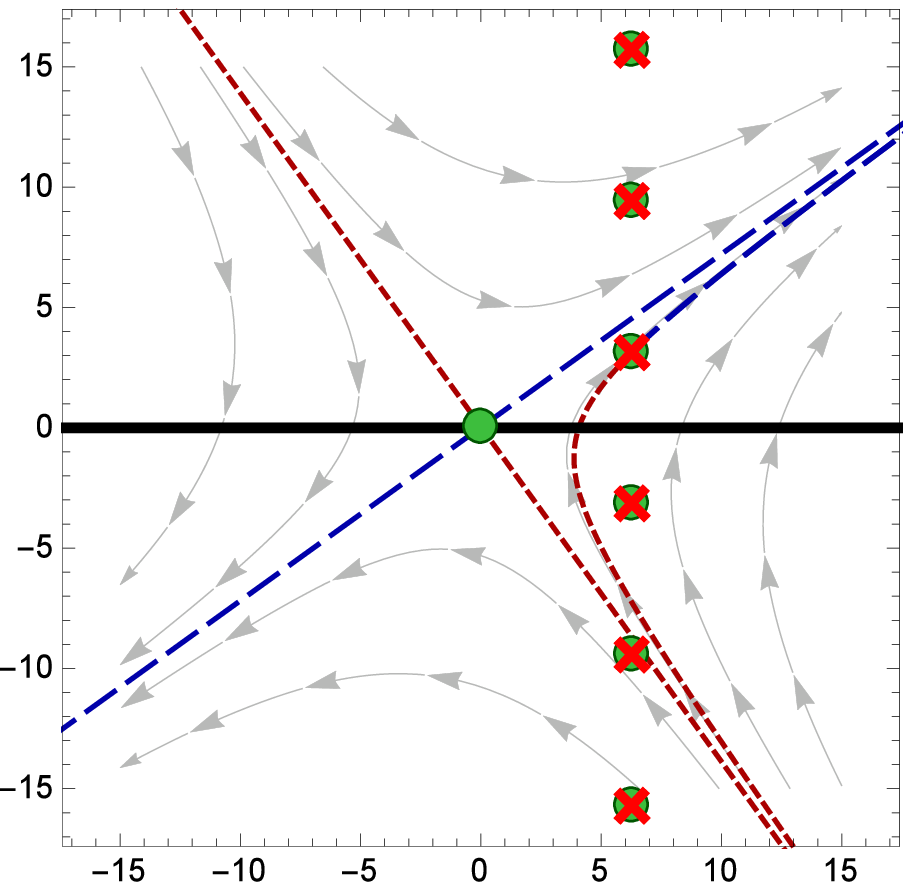}
          \hspace{1.6cm} 
          (a) ${\rm arg}(g)=\frac{{\rm arg}(t_1^\ast)}{2}\simeq -0.32$ 
        \end{center}
      \end{minipage}
      \begin{minipage}{0.33\hsize}
        \begin{center}
          \includegraphics[clip, width=50mm]{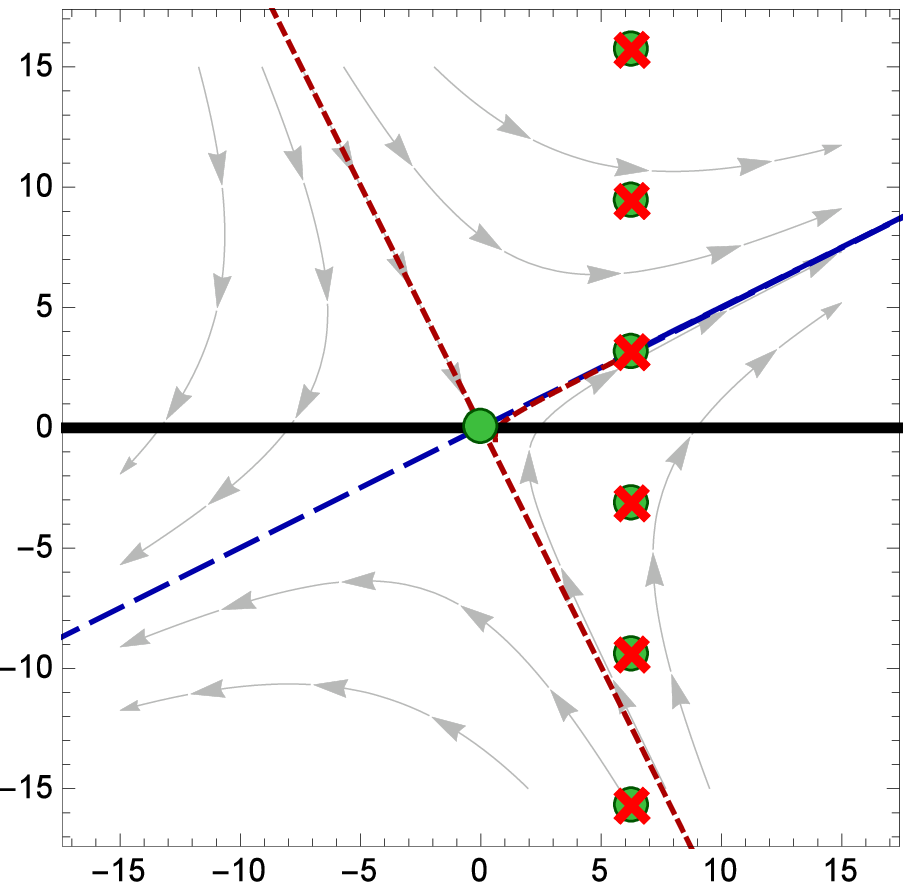}
          \hspace{1.6cm}
          (b) ${\rm arg}(g)={\rm arg}(t_1^\ast)\simeq -0.64$
        \end{center}
      \end{minipage}
      \begin{minipage}{0.33\hsize}
        \begin{center}
          \includegraphics[clip, width=50mm]{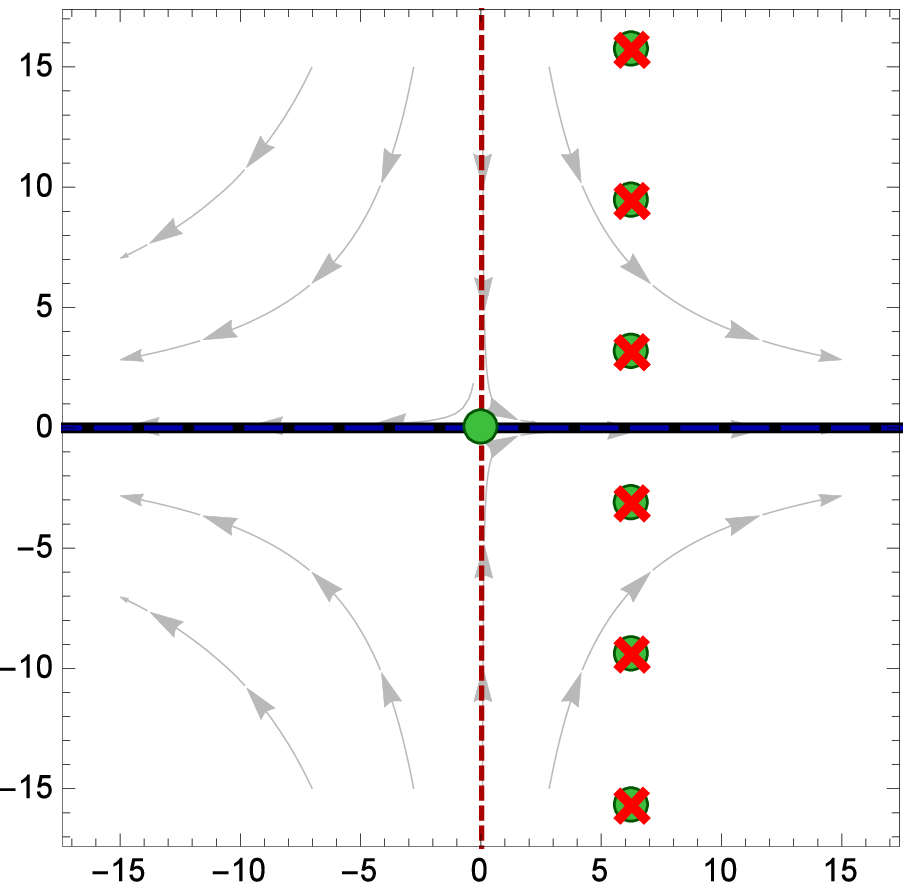}
          \hspace{1.6cm} 
          (c) ${\rm arg}(g) = -\frac{\pi}{2}\simeq -1.57$ 
        \end{center}
      \end{minipage}
    \end{tabular}\\ \vspace{1.5em}
        \begin{tabular}{cc}
          \begin{minipage}{0.33\hsize}
            \begin{center}
              \includegraphics[clip, width=50mm]{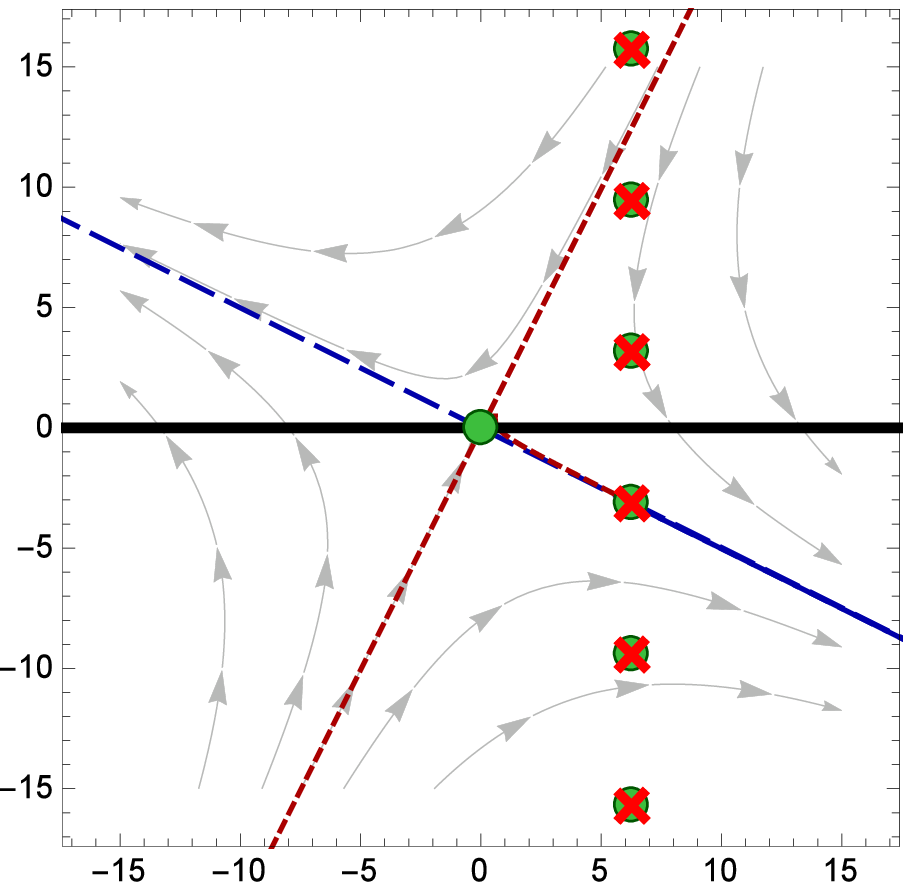}
              \hspace{1.6cm} 
              (d) ${\rm arg}(g)={\rm arg}(t_{0}^\ast)\simeq -2.50$ 
               \end{center}
          \end{minipage}
          \begin{minipage}{0.33\hsize}
            \begin{center}
              \includegraphics[clip, width=50mm]{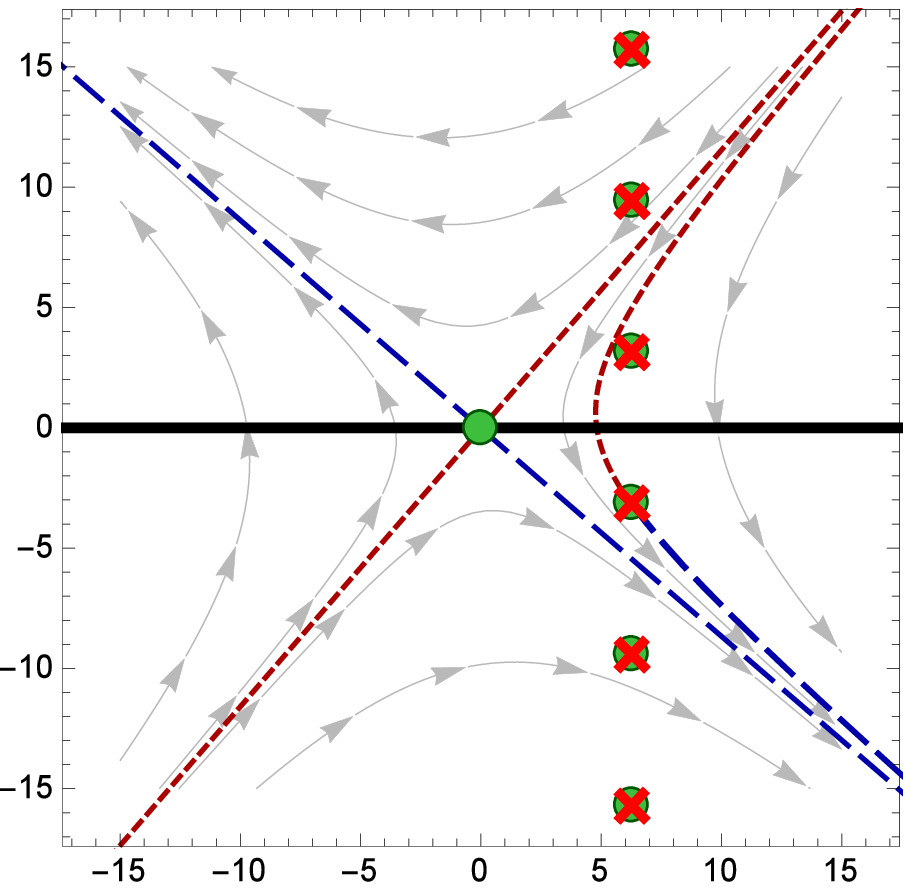}
              \hspace{1.6cm}
              (e) ${\rm arg}(g)=1.2\cdot {\rm arg}(t_{0}^\ast)\simeq -3.00$
            \end{center}
          \end{minipage}
        \end{tabular}  
    \caption{
    Thimble structures for $|g|=\frac{4\pi}{100}$, $m=2\pi$ 
    with varying ${\rm arg}(g)$.
    Values of ${\rm arg}(t_1^\ast )$ and ${\rm arg}(t_{0}^\ast )$ for $m=2\pi$
    are $ -\arctan{\frac{3}{4}}$ and $-\frac{\pi}{2} -2\arctan{\frac{1}{2}}$ respectively.
    }
    \label{fig:thimble_complex_g}
  \end{center}
\end{figure}

In Fig.~\ref{fig:thimble_complex_g},
we show numerical plots of the thimble structures
for $|g|=\frac{4\pi}{100}$, $m=2\pi$ with varying ${\rm arg}(g)$.
Since $|g|$ is small, we expect that Stokes phenomena occur 
around the Stokes lines of the transseries, namely ${\rm arg}(g) ={\rm arg}(t_n^\ast )$.
One can check this expectation 
by looking at Fig.~\ref{fig:thimble_complex_g} (b) with ${\rm arg}(g)={\rm arg}(t_1^\ast )$ and (d) with ${\rm arg}(g)={\rm arg}(t_{0}^\ast )$ 
at which the transseries has the Stokes phenomena. 
We easily see from these figures that the perturbative thimbles approximately pass the two critical points.
Furthermore 
Fig.~\ref{fig:thimble_complex_g} (a), (c) and (e) show that
the number of contributing critical points is changed when we cross ${\rm arg}(g)\simeq {\rm arg}(t_1^\ast )$ and ${\rm arg}(g)\simeq {\rm arg}(t_{0}^\ast )$.
In summary, for $\pi <m <3\pi $ and $|g| \ll 1$,
we have the following pictures:
\begin{itemize}
\item For $0\geq {\rm arg}(g) > {\rm arg}(\tilde{g}_1) \simeq {\rm arg}(t_1^\ast )$,
we have contributions from $z_{\rm pt}^c$ and $z_1^c$.

\item For ${\rm arg}(\tilde{g}_1) > {\rm arg}(g) > {\rm arg}(\tilde{g}_{0}) \simeq {\rm arg}(t_{0}^\ast )$,
only the perturbative critical point $z_{\rm pt}^c$ contributes.
Especially, the perturbative Lefschetz thimble for ${\rm arg}(g)=-\pi /2$ is almost the same as the original integral contour.

\item For ${\rm arg}(t_1^\ast ) > {\rm arg}(g) \geq  -\pi$,
we have contributions from $z_{\rm pt}^c$ and $z_{0}^c$.
\end{itemize}

As $|g|$ increases,  ${\rm arg}(\tilde{g}_n )$ becomes typically further from ${\rm arg}(t_n^\ast )$.
In other regimes of $m$, the number of ${\rm arg}(t_n^\ast )$'s satisfying $0\geq {\rm arg}(t_n^\ast )\geq -\pi$ is different which determines the number of times we encounter the Stokes phenomena.

\subsection{``Mirror" description}
\label{sec:mirror}
The CS SQED has another description, which is connected to the original description by 3d mirror symmetry \cite{Intriligator:1996ex}.
The $S^3$ partition function has a different integral representation but turns out to take the same value.
In this subsection we briefly study thimble structures of the mirror integral.
To derive the mirror description,
it is convenient to use the Fourier transformation \cite{Kapustin:2010xq}:
\begin{\eq}
\frac{1}{2\cosh{\frac{x}{2}}} 
=\frac{1}{2\pi}\int dp \frac{e^{\frac{i}{2\pi}px}}{2\cosh{\frac{p}{2}}} ,
\end{\eq}
which leads us to
\begin{\eq}
Z 
= \frac{1}{2\pi}\int_{-\infty}^\infty d\sigma 
\int_{-\infty}^\infty d\tilde{\sigma}\   
\frac{e^{\frac{ik}{4\pi}\sigma^2
+\frac{i}{2\pi}(\sigma -m)\tilde{\sigma}}}{2\cosh{\frac{\tilde{\sigma}}{2}}} = \sqrt{\frac{i}{k}}\int_{-\infty}^\infty d\tilde{\sigma} 
\frac{e^{-\frac{i}{4\pi k}\tilde{\sigma}^2
 -\frac{i}{2\pi}m\tilde{\sigma}}}{2\cosh{\frac{\tilde{\sigma}}{2}}} .
\end{\eq}
This is formally the same as the Coulomb branch localization formula for the $S^3$ partition function
of $U(1)$ Chern-Simons theory coupled to charge-$1$ hyper multiplet
with level $-1/k$ and FI-parameter $-m/2\pi$.

Let us perform thimble decomposition in this integral representation.
\begin{\eq}
\tilde{Z} 
= \int_{-\infty}^\infty d\tilde{\sigma}\ e^{-S_{\rm mirror}[\tilde{\sigma }]}
\end{\eq}
where
\begin{\eq}
S_{\rm mirror}[\tilde{z}]
=\frac{ig}{16\pi^2}\tilde{z}^2 +\frac{i}{2\pi}m\tilde{z}
  +\log{\left( 2\cosh{\frac{\tilde{z}}{2}}\right)} .
\end{\eq}
Note that the action becomes large for $g\rightarrow 0$
since weak coupling in the original theory corresponds to strong coupling
in the mirror theory and vice versa.
Therefore it is much easier to analyze Lefschetz thimble 
for $g\rightarrow \infty$.
In this limit, the saddle point $\tilde{z}^c (g,m)$
is approximately determined by
\begin{\eq}
\tilde{z}^c (g,m) \cosh{\frac{\tilde{z}^c (g,m)}{2}} =0\quad\quad
{\rm for}\ g\rightarrow \infty ,
\end{\eq}
which leads us to
$\lim_{g\rightarrow\infty} \tilde{z}^c (g,m)$
$=$ $0$, $(2\ell -1)\pi i$ with $\ell\in\mathbb{Z}$.
We denote as $\tilde{z}_{\rm origin}^c$ and $\tilde{z}_\ell^c$
the critical points satisfying
\begin{\eq}
\lim_{g\rightarrow\infty} \tilde{z}_{\rm origin}^c (g,m) = 0,  \quad\quad
\lim_{g\rightarrow\infty} \tilde{z}_\ell^c (g,m) =(2\ell -1)\pi i .
\end{\eq}
Note their roles in the transseries are unclear just from this information
in contrast to $z_{\rm pt}^c$ and $z_\ell^c$ in the original theory.
In other words, we do not have one-to-one correspondences
between the critical points in the original and mirror theories 
although their final results are the same.
In the large-$g$ expansion,
$\tilde{z}_{\rm origin}^c (g,m)$ and $\tilde{z}_\ell^c$
correspond to perturbative and non-perturbative critical points
of $1/g$-expansion.
We can easily solve the thimble associated with $\tilde{z}_{\rm origin}^c$
in the $g\rightarrow \infty$ limit by
\begin{\eq}
\lim_{g\rightarrow\infty} \tilde{z}_{\rm origin}(g,m;s)
= \epsilon \exp\left( \frac{g}{16\pi^2}s  -\frac{\pi i}{4}\right) .
\end{\eq}
For the other critical points,
it is hard to solve the flow equation analytically as in the original theory.

\begin{figure}[t]
  \begin{center}
        \begin{tabular}{ccc}
          \begin{minipage}{0.33\hsize}
            \begin{center}
              \includegraphics[clip, width=50mm]{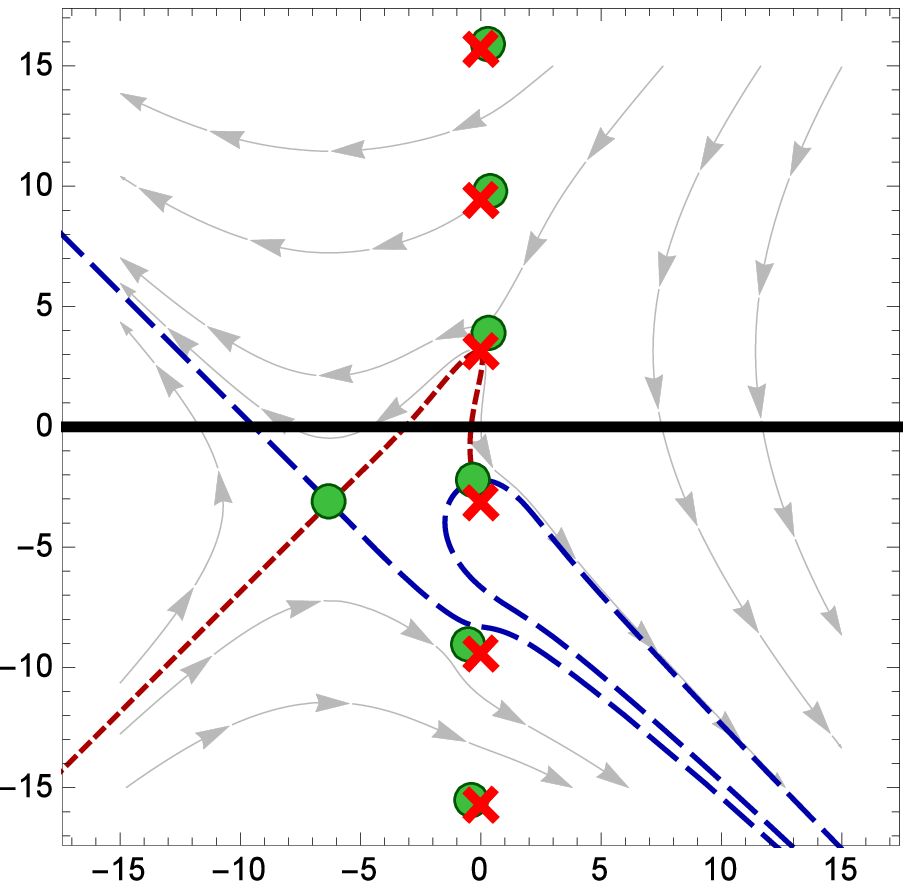}
              \hspace{1.6cm} (a) $g=4\pi$, $m = 2 \pi$ 
            \end{center}
          \end{minipage}
          \begin{minipage}{0.33\hsize}
            \begin{center}
              \includegraphics[clip, width=50mm]{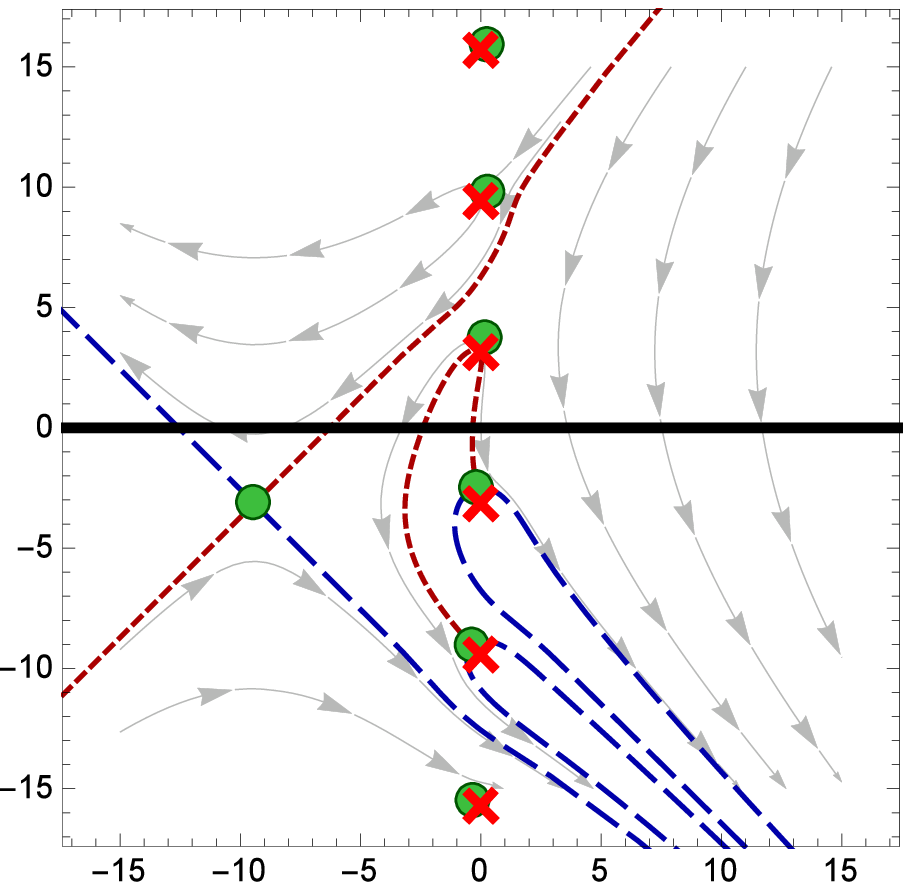}
              \hspace{1.6cm} (b) $g=4\pi$, $m = 3 \pi$ 
            \end{center}
          \end{minipage}
          \begin{minipage}{0.33\hsize}
            \begin{center}
              \includegraphics[clip, width=50mm]{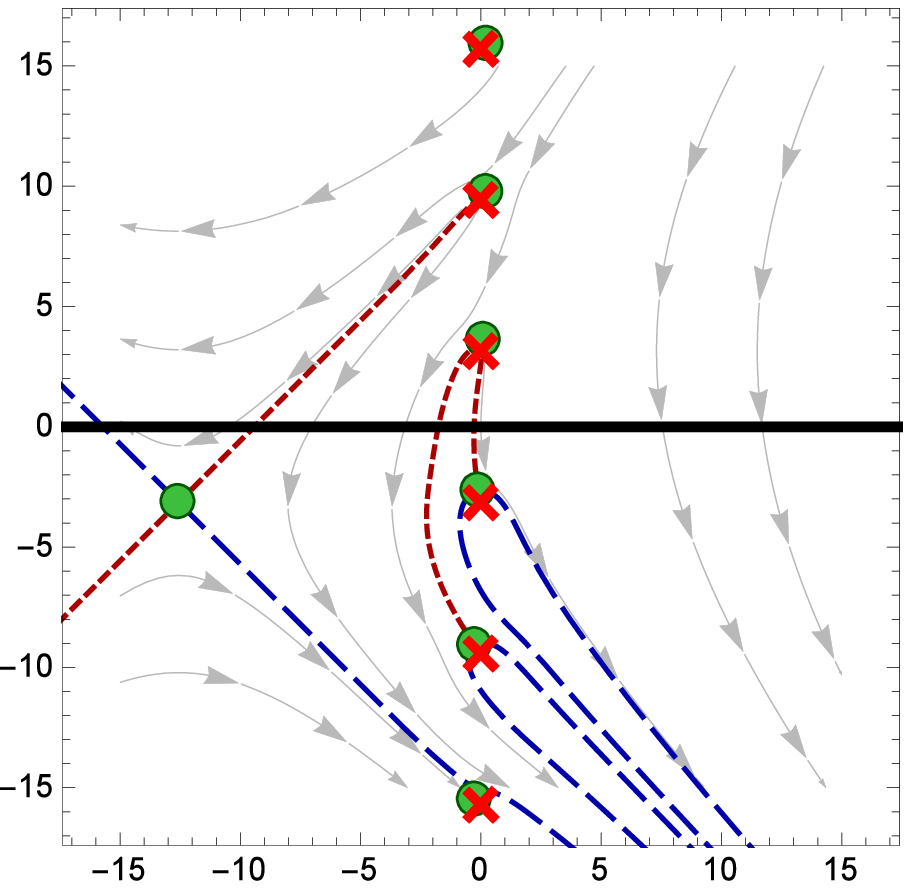}
              \hspace{1.6cm} (c) $g=4\pi$, $m = 4 \pi$ 
            \end{center}
          \end{minipage}
        \end{tabular}\\ \vspace{1em}
                \begin{tabular}{cc}
                  \begin{minipage}{0.33\hsize}
                    \begin{center}
                      \includegraphics[clip, width=50mm]{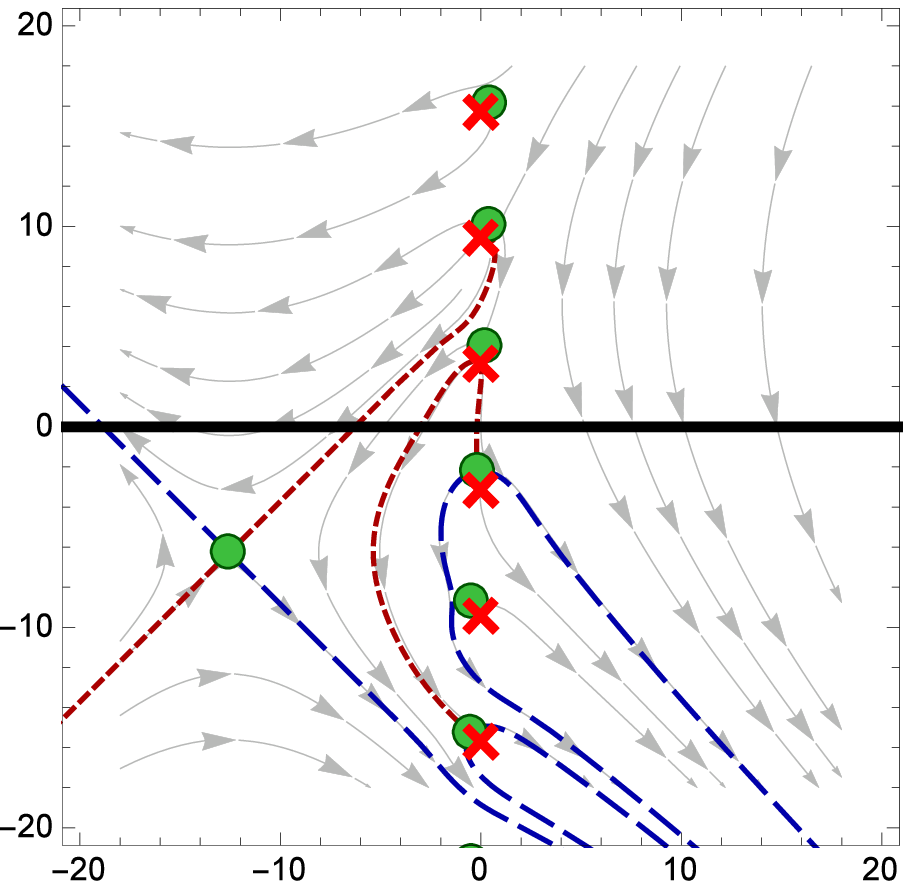}
                      \hspace{1.6cm} (d) $g=2\pi$, $m = 2 \pi$ 
                    \end{center}
                  \end{minipage}
                  \begin{minipage}{0.33\hsize}
                    \begin{center}
                      \includegraphics[clip, width=50mm]{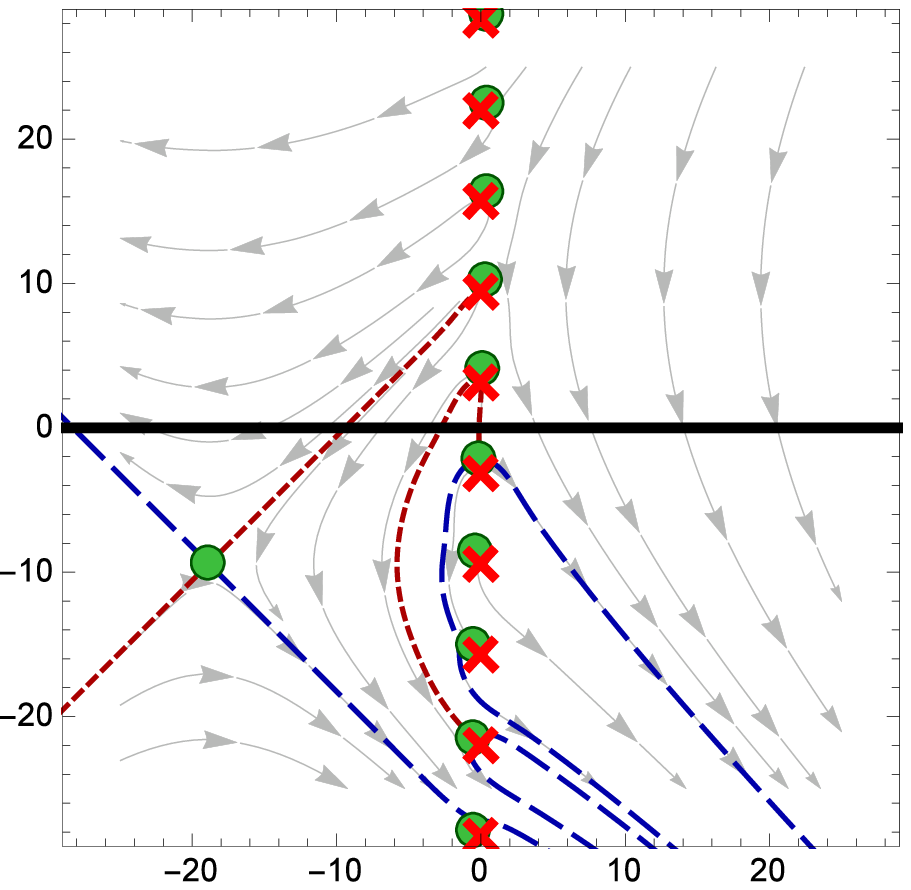}
                      \hspace{1.6cm} (e) $g=\frac{4\pi}{3}$, $m = 2\pi$ 
                    \end{center}
                  \end{minipage}
                \end{tabular}
   \caption{
    Thimble structures in the mirror theory.
}
    \label{fig:mirror_thimble}
  \end{center}
\end{figure}

In Fig.~\ref{fig:mirror_thimble}
we present numerical plots for the thimble structures in the mirror theory.
We take $(g,m)$ as parameters in Fig.~\ref{fig:mirror_thimble} (a)-(c)
as in Fig.~\ref{fig:thim_CS_U1_g10}.
For $(g,m)=(4\pi ,2\pi )$,
contributing critical points are $\tilde{z}_{\rm origin}^c$ and $\tilde{z}_{0}^c$,
and the thimble integral associated with $\tilde{z}_{0}^c$ 
is equivalent to the residue around the pole, which is here denoted as $\tilde{z} = \tilde{z}_{0}^\ast$.
For $m=3\pi $ and $4\pi$ with $g=4\pi$,
another critical point $\tilde{z}_{-1}^c$ also contributes 
and the thimble associated with $\tilde{z}_{\rm origin}^c$
passes between $ \tilde{z}_{-1}^\ast$ and $ \tilde{z}_{-2}^\ast$
in contrast to the $m=2\pi$ case.
We have more complicated structures for smaller $g$:
Fig.~\ref{fig:mirror_thimble} (d) shows that
the contributing critical points are 
$\tilde{z}_{\rm origin}^c$, $\tilde{z}_{0}^c$ and $\tilde{z}_{-2}^c$
for $(g,m)=(2\pi ,2\pi )$.
For this case, the thimble integral associated with $\tilde{z}_{0}^c$
is equivalent to the residues 
around the two poles $\tilde{z} = \tilde{z}_{0}^\ast$ and $\tilde{z}_{-1}^\ast$.
Similarly, for $(g,m)=(4\pi /3 ,2\pi )$,
we have contributions from
$\tilde{z}_{\rm origin}^c$, $\tilde{z}_{0}^c$ and $\tilde{z}_{-3}^c$,
and the thimble integral associated with $\tilde{z}_{0}^c$
is the same as the residues 
around the three poles $\tilde{z} = \tilde{z}_{0}^\ast$, $\tilde{z}_{-1}^\ast$ and $\tilde{z}_{-2}^\ast$.
These results clearly show that
the thimble decomposition in the mirror theory has the Stokes phenomena.
While the sum of the thimble integrals over the contributing critical points
is the same as the exact result by construction, 
we have not found precise understanding 
on a connection between the thimble structure and the resurgent structure in the mirror theory.
For large-$g$, the mirror theory becomes weak coupling
and we expect that the mirror integral for strong coupling has similar thimble structures to the one in the original theory for small-$g$.
This should be useful to understand resurgence structures for the large-$g$ expansion in the original theory.
It would be interesting to study the above problems in more details in the future.

\section{$\mathcal{N}=3$ $SU(2)$ Chern-Simons SQCD}
\label{SQCD}
We next investigate 
the $S^3$ partition function of 
the 3D $\mathcal{N}=3$ $SU(2)_k$ CS theory
with $N_f$ fundamental hyper multiplets and real masses $m_a$,
which we call CS SQCD
\footnote{
In 3D $\mathcal{N}=2$ language,
this theory consists of $SU(2)_k$ vector multiplet,
adjoint chiral multiplet with $U(1)_R$-charge 1 and
$N_f$ pairs of fundamental chiral multiplets with $U(1)_R$-charge $1/2$
and the real masses. 
}.
We rewrite the exact partition function obtained 
by the Coulomb branch localization 
into the full transseries with nonperturbative exponential contributions.
We also discuss the thimble decomposition and the Stokes phenomena in a manner parallel to
the case of CS SQED in the previous section.

\subsection{Exact results as resurgent transseries}
\label{sec:SU2_trans}
The partition function of the $\mathcal{N}=3$ $SU(2)$ CS SQCD
is given by
\footnote{
In 3d $\mathcal{N}=2$ language,
we have $SU(N_f )\times SU(N_f )\times U(1)_B \times U(1)_A$ global symmetry
where $U(1)_B$ is the baryon symmetry.
The diagonal part of $m_a$ corresponds to the real mass associated with $U(1)_B$
while we are turning off the one associated with $U(1)_A$.
}
\begin{\eq}
Z =\int_{-\infty}^\infty d\sigma\ 
e^{\frac{ik}{2\pi}\sigma^2}
\frac{\left( 2\sinh{\sigma} \right)^2}
{\prod_{a=1}^{N_f} 2\cosh{\frac{\sigma -m_a}{2}}\cdot 2\cosh{\frac{\sigma +m_a}{2}}}\,.
\label{eq:Zexact2}
\end{\eq}
We again focus on $k>0$ and $m_{a}>0$ mainly. 
Taking $g=\frac{2\pi}{k}$ and $\sigma =\sqrt{i t}$, 
we rewrite the partition function as
\begin{\eq}
Z =\int_{0}^{-i \infty} dt\ e^{-\frac{t}{g}} \mathcal{B}Z(t) ,
\label{eq:exact_SU2}
\end{\eq}
where $\mathcal{B}Z(t)$ is the Borel transformation of the perturbative series of the CS SQCD \cite{Honda:2016vmv}:
\begin{\eq}
\mathcal{B}Z(t)
=\frac{i \left( 2\sinh{\sqrt{it}} \right)^2}
{\sqrt{it}\prod_{a=1}^{N_f} 
2\cosh{\frac{\sqrt{it}-m_a}{2}}\cdot 2\cosh{\frac{\sqrt{it}+m_a}{2}}}\, .
\label{bt_sqcd}
\end{\eq}
This Borel transformation has simple poles at
$t = -i \left[m_{a}\pm (2n_{a}-1)\pi i \right]^2$ with 
$n_{a}\in {\mathbb N}$.
With $m_a =(2n_{a}-1)\pi$, we have Borel singularities at positive real axis as
$\left. t \right|_{m_a =(2n_{a}-1)\pi} = \pm 2(2n_{a}-1)^2 \pi^2$,
leading to non-Borel-summability of the perturbative series, thus $m_a =(2n_{a}-1)\pi$ is a Stokes line.

As with the case of CS SQED, 
the exact result \eqref{eq:exact_SU2}
is decomposed into the Borel resummation along $\mathbb{R}_+$ ({\it perturbative part}) and
the residue of all the singularities in the fourth quadrant of the Borel plane ({\it non-perturbative part}):
\begin{align}
&Z=Z_{\rm pt}+Z_{\rm np}\,,  
\label{tsQCD}
\\
& Z_{\rm pt}=\int_{0}^{\infty} dt\ e^{-\frac{t}{g}} \mathcal{B}Z(t) ,\quad
Z_{\rm np} = \sum_{{\rm poles}\in {\rm 4th\ quadrant}} {\rm Res}_{t=t_{\rm pole}}\Bigl[  e^{-\frac{t}{g}} \mathcal{B}Z(t) \Bigr] .
\end{align}
The number of the singularities in the region is $|n_{a}|$ for the real mass $(2n_{a}-1)\pi<m_{a}<(2n_{a}+1)\pi$ for each of flavors, 
thus another singularity comes to contribute to the partition function at $m_{a}=(2n_{a}+1)\pi$, 
leading to ambiguity of the perturbative Borel resummation, 
that is the Stokes phenomenon.
It is also notable that, for the degenerate masses $m=m_{a}$, the singularities are degenerate, where the order of their poles gets equivalent to $N_{f}$.

We first focus on $N_{f}=2$ for simplicity.
By expanding $Z_{\rm pt}$ with respect to $t$ and extracting coefficients,
we obtain an asymptotic series of the perturbative part as
\be
Z_{\rm pt}
&=&\frac{\sqrt{ig}}{8} \sum_{\{s_{b}\}=0}^{\infty}\sum_{\{q_{b}\}=0}^{\infty} \sum_{\{l_{b}\}=0}^{\infty} 2^{-2\bar{q}} 
\left( \prod_{b=1}^2 \frac{1}{\Gamma(2s_b+2)} \right)
\left( \prod_{b=1}^4 \frac{E_{2q_b} }
{\Gamma(2q_b -l_b+1) \Gamma(l_b+1)} \right)
\nonumber\\
&&\times\, \Gamma(\bar{q} - \bar{l}/2 + \bar{s} + 3/2 ) 
\cdot (i g)^{\bar{q} - \bar{l}/2 + \bar{s}+1} 
m_1^{l_1+l_2} m_2^{l_3 +l_4}  \, \delta_{\bar{l} \, {\rm mod}2, 0}.
\ee
with $\bar{q} = \sum_{b=1}^{4} q_b$, $\bar{l} = \sum_{b=1}^{4} l_b$, 
and $\bar{s} = s_1 + s_2$.
This 
perturbative series is Borel-summable 
along $\mathbb{R}_+$ for $m_{a}\not=(2n_{a}-1)\pi$.
However, even if $m_{a}\not=(2n_{a}-1)\pi$, the Borel resummation of the perturbative series does not give an exact result for $m_{a}>\pi$
as in the CS SQED case. 
The nonperturbative part $Z_{\rm np}$ can be calculated by
the residues of the Borel singularities in the fourth quadrant of the Borel plane.
We below show the results of $Z_{\rm np}$ for non-degenerate and degenerate real masses, separately.
For $m_1 \ne m_2$ with $(2n_{1}-1)\pi<m_{1}<(2n_{1}+1)\pi$ and 
$(2n_{2}-1)\pi<m_{2}<(2n_{2}+1)\pi$, 
the nonperturbative part is given by
\be
Z_{{\rm np}} 
= 4 \pi i \left[ \sum_{\ell_1 = 1}^{n_{1}} 
\frac{ e^{\frac{i}{g} [m_{1}+(2\ell_1 -1 ) \pi i ]^2 } \sinh m_1 }{ \cosh m_1 -\cosh m_2} + 
\sum_{\ell_2 = 1}^{n_{2}} 
\frac{ e^{\frac{i}{g}  [m_{2}+(2\ell_2 - 1 ) \pi i  ]^2 } \sinh m_2 }{ \cosh m_2 -\cosh m_1 } \right]\,. 
\ee
For $m=m_1 = m_2$ with $(2n-1)\pi<m_{1}<(2n+1)\pi$,
it is obtained as
\be
Z_{{\rm np}} 
= 4 \pi i  \sum_{\ell = 1}^{n} e^{\frac{i}{g} [m+ (2\ell - 1) \pi i]^2}
 \left( \frac{2i[ m+(2\ell-1) \pi i]}{g} + \frac{1}{\tanh m} \right)\,.
\ee
In these expressions of the full transseries expansion,
each of the nonperturbative parts corresponds to the contribution with the action $S=-i[m_{a}+(2\ell_{a}-1)\pi i]^{2}/g$, 
which is consistent with the position of the singularities in the Borel transform (\ref{bt_sqcd}).

In the case of general $N_f$ with degenerate mass, $Z_{\rm pt}$ and $Z_{\rm np}$ are given by
\be
Z_{\rm pt} 
&=& \frac{\sqrt{i g}}{2^{2N_f-1}} \sum_{\{s_{b}\}=0}^{\infty}\sum_{\{q_{b}\}=0}^{\infty} \sum_{\{l_{b}\}=0}^{\infty} 2^{-2\bar{q}} \left( \prod_{b=1}^{2} \frac{1}{\Gamma(2s_b+2)} \right)
\left( \prod_{b=1}^{2N_f} \frac{E_{2q_b} }{\Gamma(2q_b-l_b+1) \Gamma(l_b+1)} \right)
\nl 
&& \quad \times\, \Gamma(\bar{q} - \bar{\ell}/2 + \bar{s} + 3/2 ) 
\cdot (i g)^{\bar{q} - \bar{\ell}/2 + \bar{s}+1}
m^{\bar{l}}  \, 
 \delta_{\bar{\ell} \, {\rm mod}2, 0}\,,
\\
Z_{\rm np} &=&  \frac{\pi i }{2^{2N_f-3} \Gamma(N_f)}  \sum_{\ell=0}^{n-1} 
\lim_{z \rightarrow z^{*}_\ell} \frac{\pd^{N_f-1}}{\pd z^{N_f-1}} \frac{(z-z_\ell^{*})^{N_f} \sinh^2z \cdot e^{i z^2/g}}{\left( \cosh \frac{z-m}{2} \cosh \frac{z+m}{2}  \right)^{N_f}}\,, \label{eq:SU2_generic}
\ee
with $\bar{q} = \sum_{b=1}^{2N_f} q_b$, $\bar{\ell} = \sum_{b=1}^{2N_f} \ell_b$, $\bar{s}=\sum_{b=1}^{2}s_{b}$ and $z^{*}_\ell=m+(2\ell-1)\pi i$.
Introducing the step function,
the nonperturbtive part for general $(N_f , m_a )$ is expressed as
\begin{\eq}
Z_{\rm np}
=\sum_{a=1}^{N_f} \sum_{\ell_a =1}^\infty \theta (m_a -(2\ell_a -1)\pi ) 
{\rm Res}_{t=-i[m_a +(2\ell_a -1)\pi i]^2} \Bigl[  e^{-\frac{t}{g}} \mathcal{B}Z(t) \Bigr] .
\end{\eq}
As in the CS SQED cases,
the transseries expression is apparently ambiguous for $m_a =(2n_a -1)\pi$ $(n_{a}\in {\mathbb N})$
due to the Borel ambiguities and step function behaviors of the transseries parameters.
The ambiguity in the perturbative part is estimated by
\begin{align}
\left. Z_{\rm pt}(\{ m_b \}  ) \right|_{m_a =(2n_a -1)\pi +0_+}
-\left. Z_{\rm pt}(\{ m_b \}  ) \right|_{m_a =(2n_a -1)\pi +0_-}
= -{\rm Res}_{t=-i[m_a +(2n_a -1)\pi i]^2} \Bigl[ e^{-\frac{t}{g}} \mathcal{B}Z(t) \Bigr] ,
\end{align}
while the non-perturbative ambiguity is
\begin{\eq}
\left. Z_{\rm np} (\{ m_b \}  ) \right|_{m_a =(2n_a -1)\pi +0_+} -\left. Z_{\rm np}(\{ m_b \}  ) \right|_{m_a =(2n_a -1)\pi +0_-} 
= +{\rm Res}_{t=-i[m_a +(2n_a -1)\pi i]^2} \Bigl[ e^{-\frac{t}{g}} \mathcal{B}Z(t) \Bigr] .
\end{\eq}
It is clear that these ambiguities are canceled 
and we obtain the unambiguous result equivalent to the exact result.

\subsection{Thimble decomposition}
The effective action of the present example with respect to $\sigma$ reads as
\be
S[\sigma ] 
= -\frac{i \sigma^2}{g}  - \log \frac{4\sinh^2 \sigma}{\prod_{a=1}^{N_f} 2 \cosh  \frac{\sigma-m_a}{2} \cdot 2 \cosh  \frac{\sigma+m_a}{2}} \,.
\label{eq:action_CS_QCD1_Nf} 
\ee
We consider the complexification $\sigma\to z$ 
and study thimble structures
in a parallel manner to the CS SQED case.
First the saddle point $z^c$ is determined by
\begin{\eq}
\left. \frac{\partial S[z]}{\partial z} \right|_{z=z^c}
=-\frac{2i}{g} z^c -2\coth{z^c}
+\frac{1}{2}\sum_{a=1}^{N_f}\sum_\pm \tanh{\frac{z^c \pm m_a}{2}} =0 .
\label{eq:saddle_SU2}
\end{\eq}
As in the $U(1)$ case,
we can analytically find the saddle points in the $g\rightarrow 0$ limit:
\begin{\eq}
z^c  \sinh^{2}{z^c} \prod_{a=1}^{N_f} \cosh{\frac{z^c -m_a}{2}} \cosh{\frac{z^c +m_a}{2}} =0  \quad
{\rm for}\ g\rightarrow 0 .
\end{\eq}
Note that while the third factor comes from the poles of the integrand,
which we had also in the CS SQED case,
the second factor comes from the zeroes,
which were absent in the $U(1)$ case.
The zeros of the integrand are given by
\be
\sinh^2 z_{\rm zero} = 0 \quad &\Rightarrow& \quad z_{\rm zero} = n \pi i  , \quad n \in {\mathbb Z}.
\ee
The zeroes add qualitatively new features to the thimble structure
because they can be end points of Lefschetz thimbles
and thimbles may terminate at finite $z$ ($=z_{\rm zero}$) 
in contrast to the CS SQED.
This always happens when we analyze the following type of integral
\begin{\eq}
\int dx\ \frac{P(x)}{Q(x)} e^{-\frac{1}{g}h(x)} ,
\end{\eq}
where $P(x)$ and $Q(x)$ are functions without poles.
Saddle points of this integral in the $g\rightarrow 0$ limit are given by
\begin{\eq}
\frac{\partial h(x)}{\partial x} P(x) Q(x) = 0,
\end{\eq}
which indicates that the poles and zeroes coincide with the saddle points in the $g\rightarrow 0$ limit.
In the present example with $g\rightarrow 0$,
the perturbative part of the transseries, 
which is the Borel resummation along $\mathbb{R}_+$,
corresponds to a sum of two thimble integrals
associated with two saddle points around $z=0$
\footnote{
The fact that the number of the saddles around $z=0$ is two 
can be analytically checked 
by considering the expansion $z^c$ $=$ $0$ $+g^{1/2}z^{c,1}$ $+\mathcal{O}(g^{3/2})$ in \eqref{eq:saddle_SU2},
which leads us to $z^{1,c}=\pm e^{\frac{\pi i}{4}}$.
} 
as we will see soon.
Another important feature comes from the fact that the action \eqref{eq:action_CS_QCD1_Nf} is invariant under the ${\mathbb Z}_2$ transformation $z \rightarrow - z$.
This symmetry forces the singularities and saddle points
to be located symmetrically in the complex $z$-plane. 
These facts imply that each of the contributions in the transseries (\ref{tsQCD}) is composed of a pair of two thimble integrals associated with two saddle points even in $g\rightarrow 0$ limit.

\begin{figure}[t]
  \begin{center}
    \begin{tabular}{ccc}
      \begin{minipage}{0.33\hsize}
        \begin{center}
          \includegraphics[clip, width=50mm]{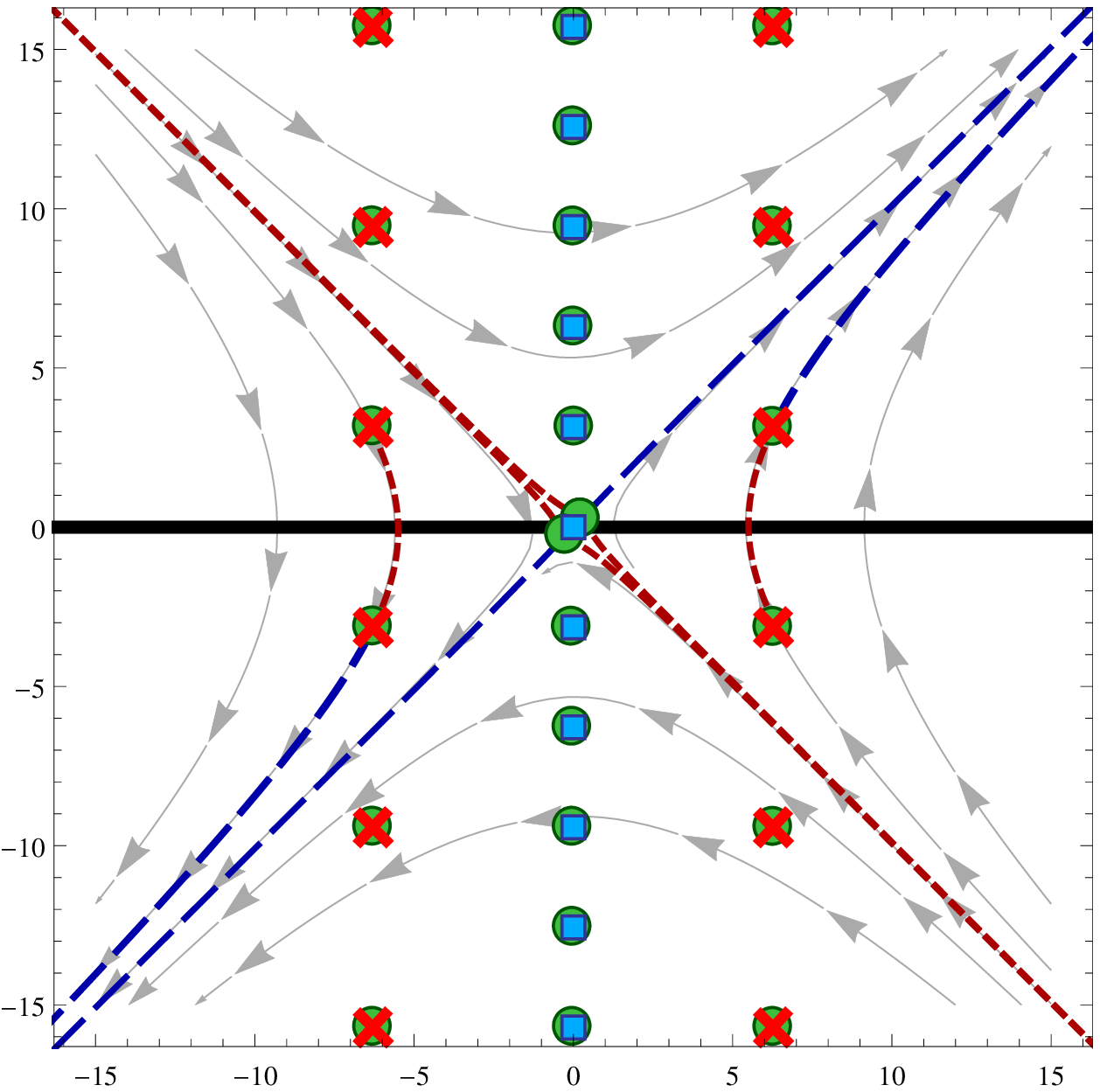}
          \hspace{1.6cm} (a) $m = 2 \pi$ 
        \end{center}
      \end{minipage}
      \begin{minipage}{0.33\hsize}
        \begin{center}
          \includegraphics[clip, width=50mm]{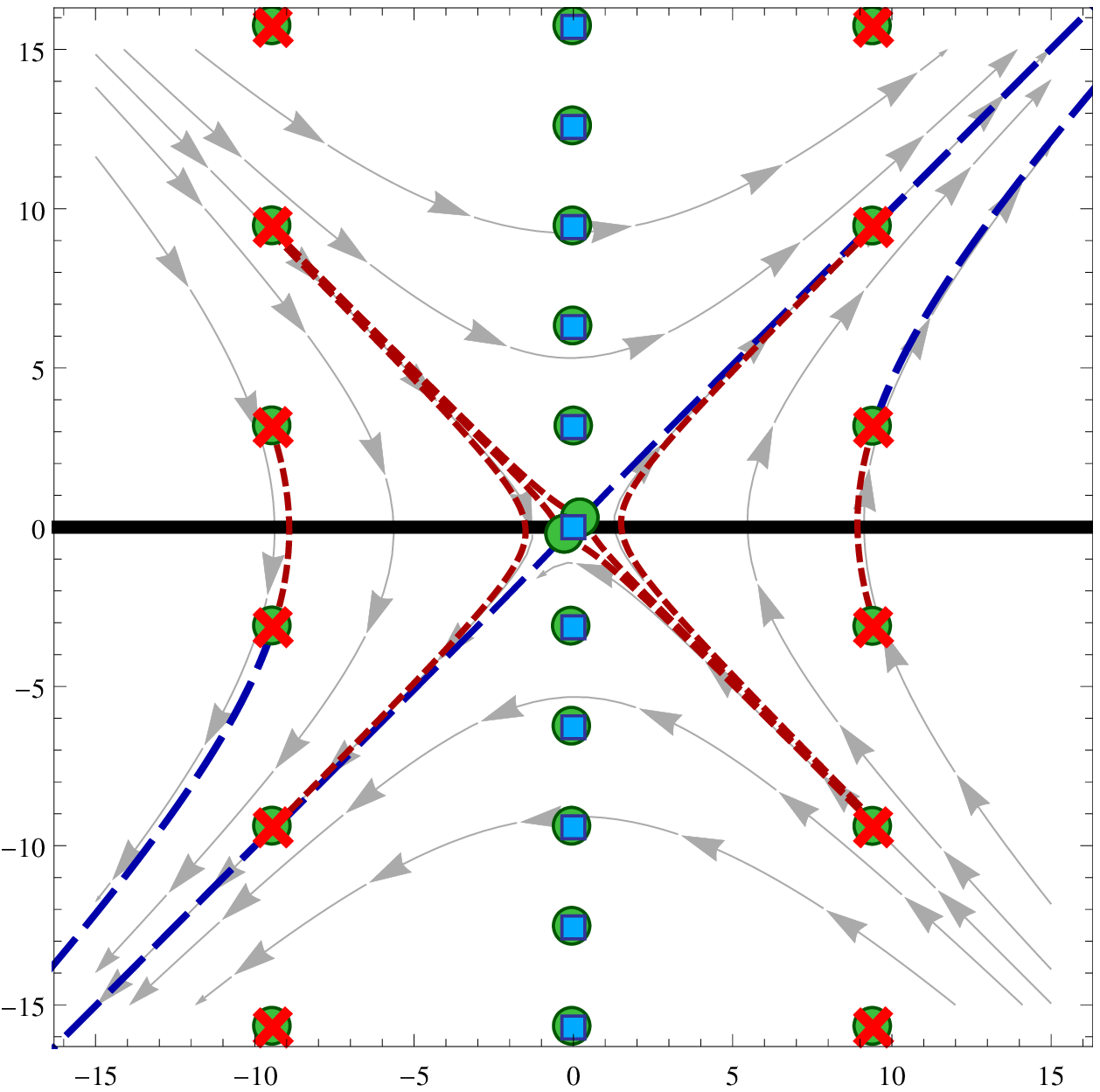}
          \hspace{1.6cm} (b) $m = 3 \pi$  
        \end{center}
      \end{minipage}
      \begin{minipage}{0.33\hsize}
        \begin{center}
          \includegraphics[clip, width=50mm]{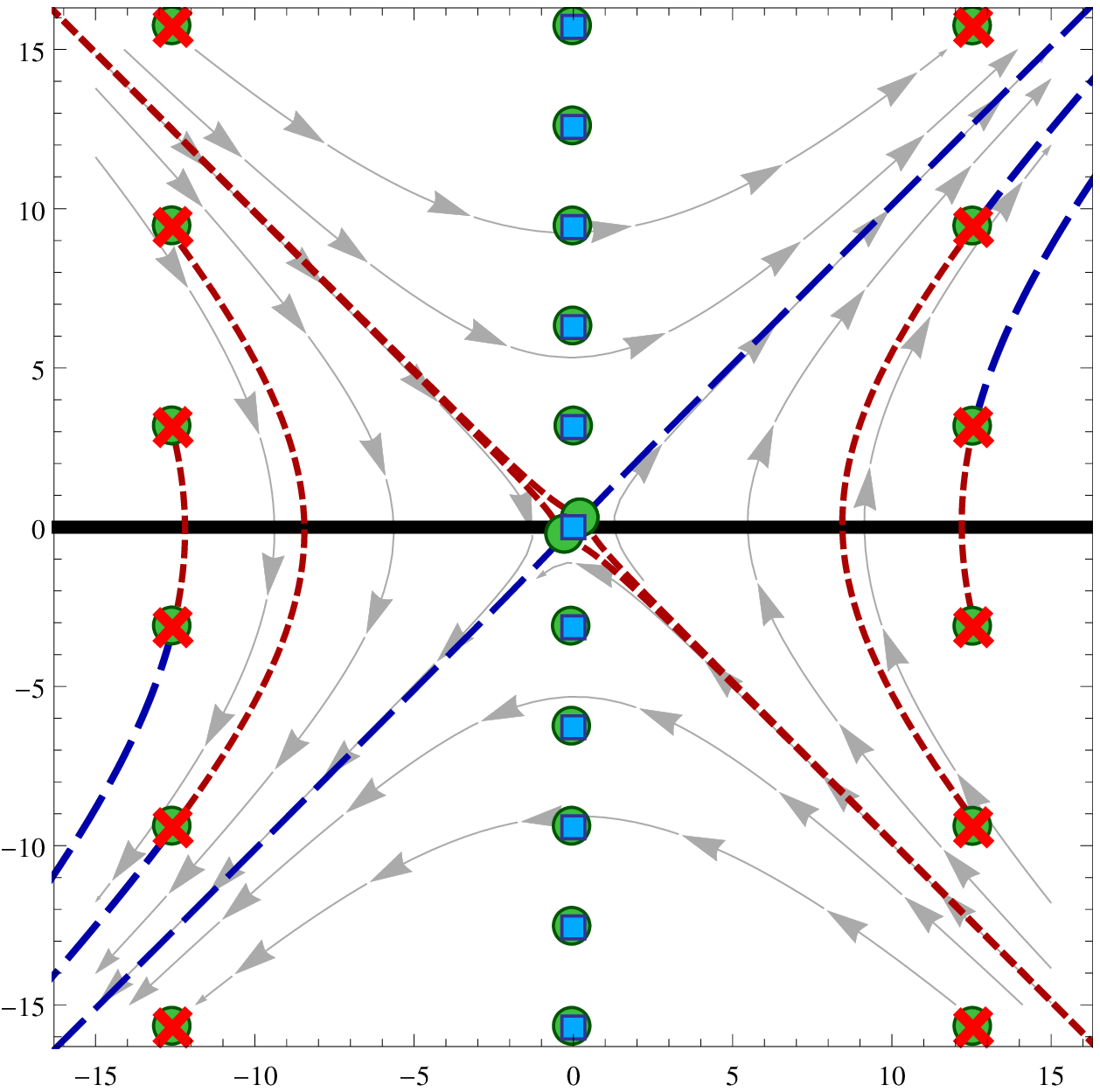}
          \hspace{1.6cm} (c) $m = 4 \pi$ 
        \end{center}
      \end{minipage}
    \end{tabular}
    \caption{Thimble structures of the $SU(2)$ CS SQCD 
    with $N_f =1$ and $g\approx 0.126$ ($k=100$).
      The green points, red crosses, and blue square 
      stand for critical points, poles and zeroes of the integrand, respectively.
      The red dotted lines denote dual thimbles 
      having nonzero intersection numbers with the original integral contour $\mathbb{R}$ 
      while the blue dashed lines for the corresponding thimbles.
The arrows represent flow lines for increasing flow parameter $s$.  }
    \label{fig:thim_CS_SU2_g0.1}
  \end{center}
\end{figure}

Now we present some samples of numerical results. 
Fig.~\ref{fig:thim_CS_SU2_g0.1} depicts 
the thimble structure for $N_f = 1$ in CS SQCD with $g\approx 0.126$ ($k=100$) and $m=2\pi$, $3\pi$, $4\pi$, which can be regarded as approximate cases of the weak-coupling limit.  
We term two saddle points near the origin as ``perturbative" 
ones and others as ``nonperturbative" ones.
Note that the red crosses are almost overlapped with the green circles
since the nonperturbative saddle points (green circles) and the singularities (red crosses) 
are almost degenerate for small $g$.
Each pair of saddle points constituting one sector 
of the transseries is located in a $\mathbb{Z}_{2}$ symmetrical manner.
For $m=2\pi$, four thimbles (two pairs) contribute to the partition 
function: two thimbles associated with the perturbative saddles 
near the origin and the other two thimbles associated with the 
nonperturbative saddles around $z=\pm (m+\pi i)$.
For $m= 3\pi$, 
the perturbative thimbles almost pass the saddles around $z=\pm (m+3\pi i)$.
This reflects the fact that $m=3\pi$ corresponds to the Stokes line of the transseries
and the result starts to receive contributions from the two thimbles associated 
with the saddles around $z=\pm (m+3\pi i)$ as the nonperturbative effects.
Note that, in this limit, the Stokes lines of transseries and thimble decomposition almost coincides.
For $m=4\pi$, the six thimbles (three pairs) contribute to the 
partition function: the two thimbles associated with the perturbative saddles  
and four thimbles associated with the nonperturbative saddles
around $z=\pm (m+\pi i)$ and $\pm (m+3\pi i)$.

\begin{figure}[t]
  \begin{center}
    \begin{tabular}{ccc}
      \begin{minipage}{0.33\hsize}
        \begin{center}
          \includegraphics[clip, width=50mm]{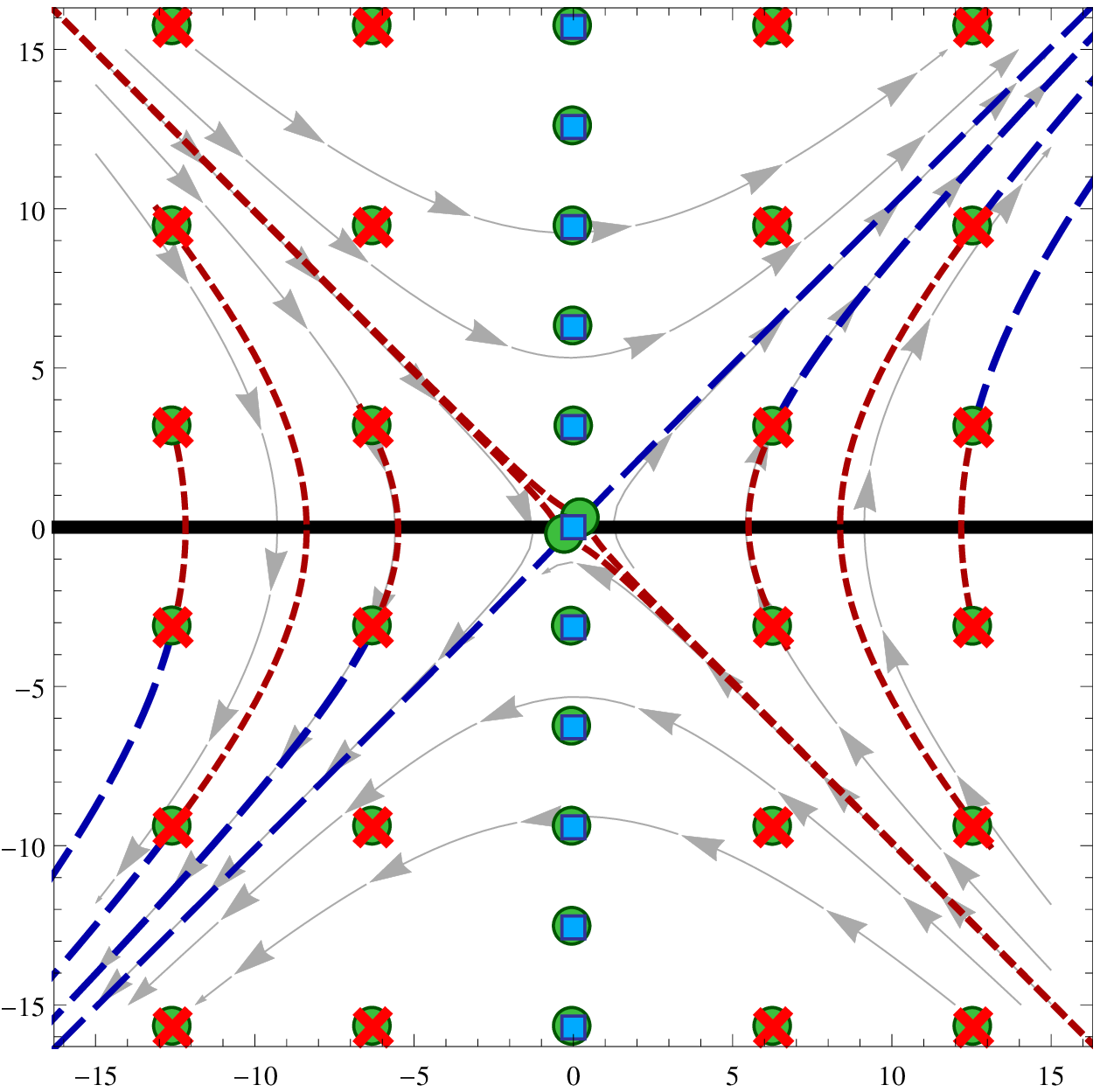}
          \hspace{1.6cm} (a) $m_1 = 2 \pi$ 
        \end{center}
      \end{minipage}
      \begin{minipage}{0.33\hsize}
        \begin{center}
          \includegraphics[clip, width=50mm]{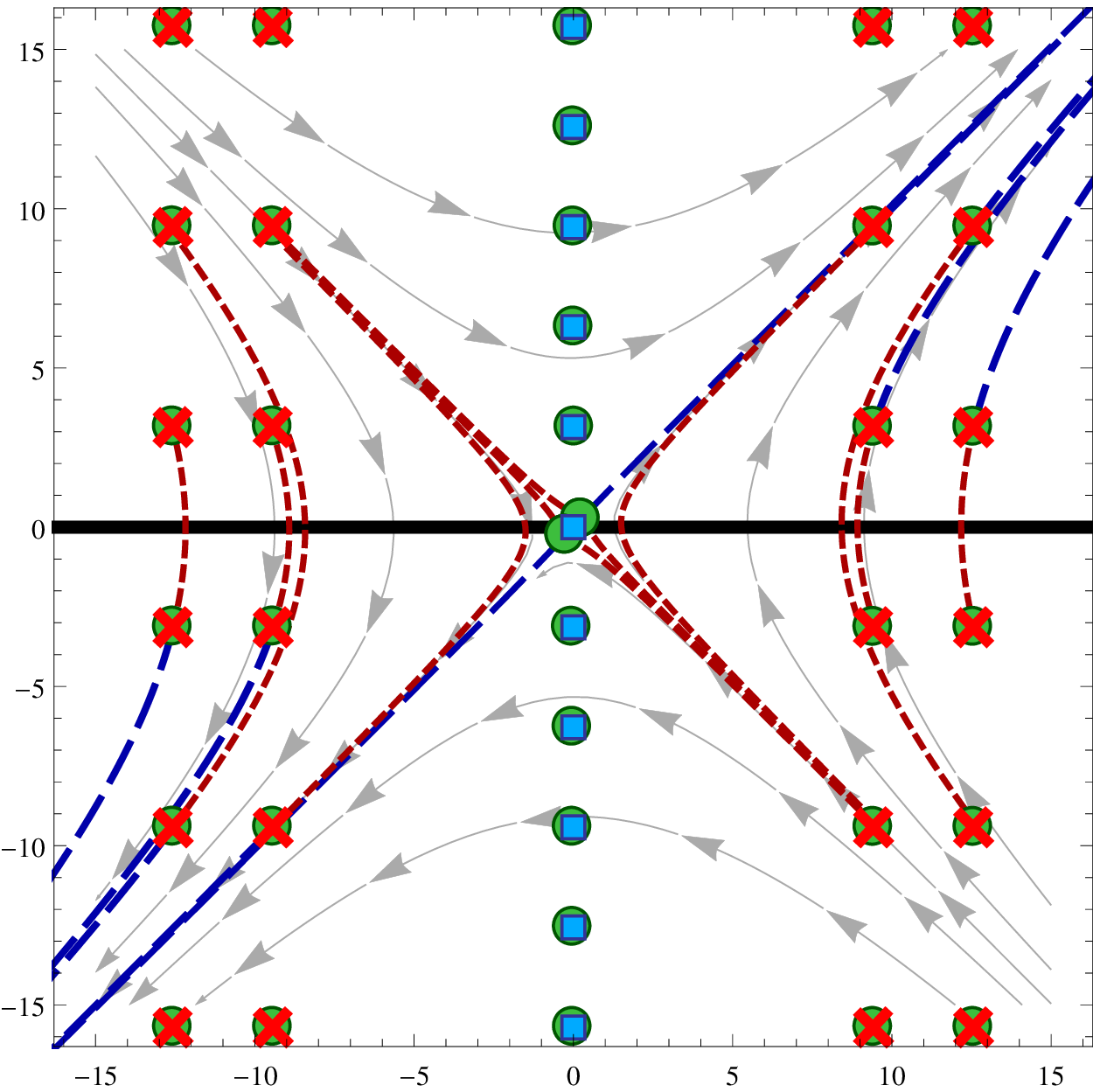}
          \hspace{1.6cm} (b) $m_1 = 3 \pi$  
        \end{center}
      \end{minipage}
      \begin{minipage}{0.33\hsize}
        \begin{center}
          \includegraphics[clip, width=50mm]{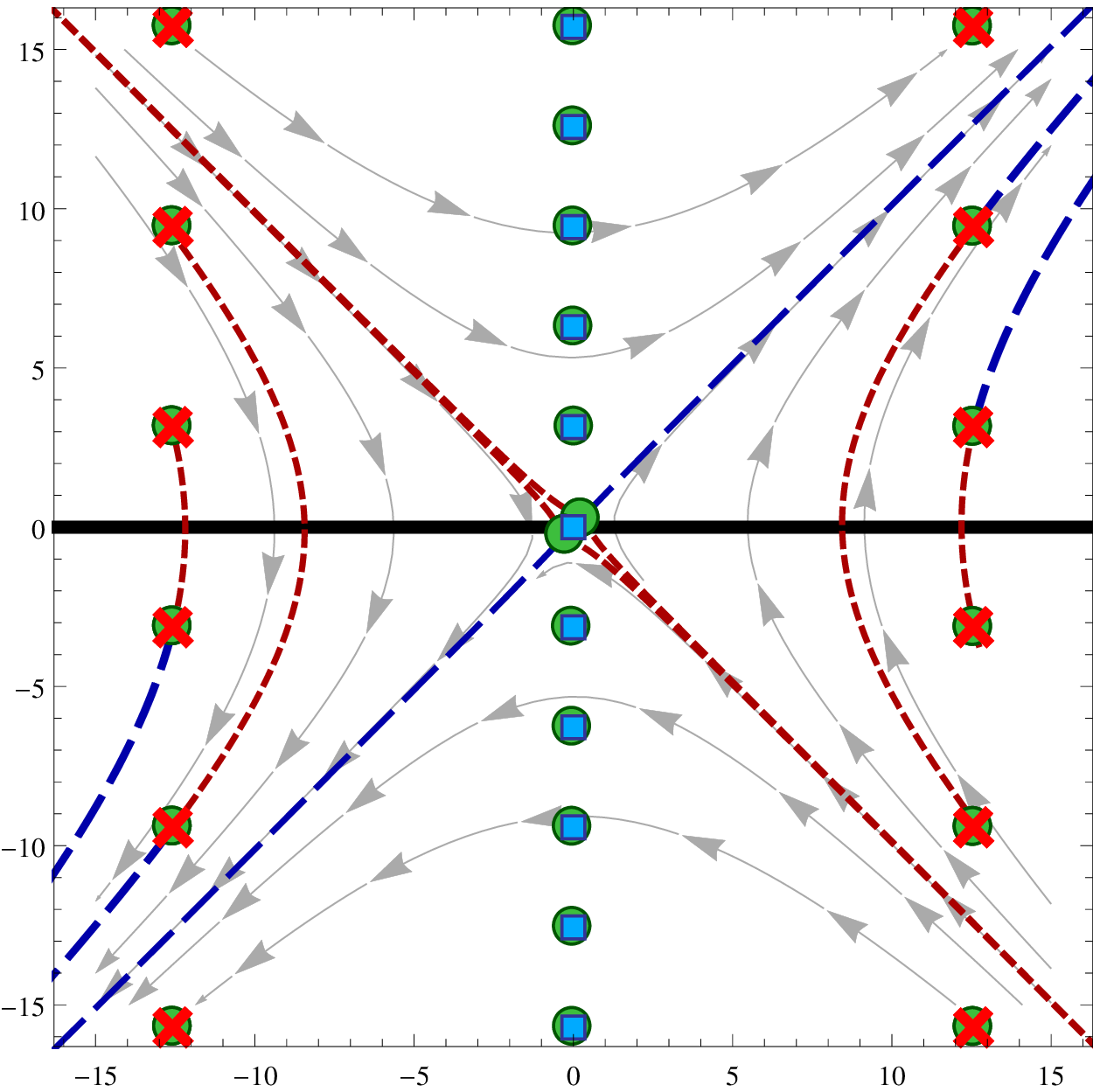}
          \hspace{1.6cm} (c) $m_1 = 4 \pi$ 
        \end{center}
      \end{minipage}
    \end{tabular}
    \caption{Thimble structure of two-flavor $SU(2)$ CS theory 
    with $g\approx 0.126$ ($k=100$) and $m_2 = 4 \pi$.
      The green points, red crosses, and blue square stand for critical points, 
singularities, and zero-points, respectively.
      The red dotted lines stand for dual 
thimbles with nonzero intersection numbers, and the blue dashed
lines for the corresponding thimbles.
The arrows represent the flow lines.
}
    \label{fig:thim_CS_SU2_g0.1_dif}
  \end{center}
\end{figure}

Fig.~\ref{fig:thim_CS_SU2_g0.1_dif} shows the $N_f = 2$ case.
In this case, we depict the thimble structures for 
$m_{1}=2\pi,3\pi,4\pi$ with $g\approx 0.126$ 
and $m_{2}=4\pi$ fixed,
where the nonperturbative saddles (green points) and singularities 
(red crosses) are almost degenerate again.
For $m_{1}=2\pi$, eight thimbles (four pairs) contribute to the 
partition function: two thimbles associated with the perturbative 
saddles (near the origin) 
and the other six thimbles associated 
with the nonperturbative saddles
around $z=\pm (m_2 +\pi i)$, $\pm (m_2 +3\pi i)$ and $\pm (m_1 +\pi i)$.
For $m_{1}= 3\pi$, two more thimbles (one pair) associated with 
the nonperturbative saddles come in as the nonperturbative 
contributions since this parameter is the Stokes line of the transseries.  
For $m_{1}=4\pi$, 
the poles get degenerate and
therefore nonperturbative saddles also become degenerate, 
where the eight nonperturbative saddles 
around $z=\pm (m_{1,2} +\pi i)$ and $\pm (m_{1,2} +3\pi i)$
are merged into the four degenerate saddles. 
We end up with two thimbles (one pair) associated with the 
perturbative saddles near the origin and four thimbles (two 
pairs) associated with the nonperturbative degenerate saddles.

In the $g\to 0$ limit, each of the thimble integrals associated with nonperturbative saddle points
is equivalent to each of the residues of the Borel singularities.
Thus, when the real mass crosses the Stokes line $m_{a}\approx (2n_{a}-1)\pi$, 
the saddle point around the pole starts to contribute to the partition function as the nonperturbative effect.
As in the case of the CS SQED, for finite $g$,
the Stokes phenomena of the thimble decomposition
occur at different points $\tilde{m}$'s from those of the transseries
which approach the same points in the weak coupling limit $g\rightarrow 0$.
Let us consider the $N_f =1$ case for simplicity.
The perturbative contribution $Z_{\rm pt}$ is only composed of a pair of the thimbles associated with the perturbative saddle points near the origin for $(2n-1)\pi<m<\tilde{m}$ ($n\geq1$).
However, it gets composed of these perturbative thimbles and two more thimbles
associated with nonperturbative saddles for $\tilde{m}<m<(2n+1)\pi$. 
The role played by these nonperturbative thimbles changes at $m=(2n+1)\pi$, 
where they come to contribute to the partition function as the ``genuine" nonperturbative contribution.
As in the $U(1)$ case,
if we are not on the Stokes lines,
the result based on the thimble decomposition is in exact agreement with that of the resurgent transseries without subtleties.
On the Stokes lines, they are apparently ambiguous and
we need to take limits from opposite sides as in the $U(1)$ case.
Up to these subtleties, they are equivalent for any $(g,\{m_a \} )$.

\section{Generalization}
\label{sec:generalization}
So far we have analyzed 
the sphere partition functions of the $\mathcal{N}=3$ theories for simplicity.
In this section
we generalize these analyses to more general theories and other observables.

\subsection{General rank-1 $\mathcal{N}=2$ Chern-Simons matter theory}
\label{sec:rank1}
Let us consider general rank-1 $\mathcal{N}=2$ Chern-Simons matter theory,
which is $U(1)_k$ theory coupled to charge-$q_a$ chiral multiplets
with R-charge $\Delta_a$ and real mass $m_a$,
or
$SU(2)_k$ theory coupled to isospin-$j_a$ chiral multiplets
with R-charge $\Delta_a$ and real mass $m_a$.
The localization formula for the sphere partition function is given by
\footnote{
We have rescaled $\sigma$ as $\sigma \rightarrow 2\pi \sigma$
as well as $m_a$.
}
\begin{\eq}
Z
= \int^{\infty}_{-\infty} d\sigma \, 
e^{\frac{i}{g}\sigma^2} Z_{\rm 1-loop}(\sigma )
\label{eq:exact_general}
\end{\eq}
where $g=1/\pi k$ for $U(1)$ and $g=1/2\pi k$ for $SU(2)$.
$Z_{\rm 1loop}$ is given by
\begin{\eq}
Z_{\rm 1loop}(\sigma )
=\begin{cases}
\frac{1}{\prod_{a=1}^{N_f} s_1(q_a \sigma +m_a - i(1-\Delta_a ))} 
& {\rm for}\ U(1)   \cr
\frac{4\sinh^2{(\pi\sigma )}}
{\prod_{a=1}^{N_f}\hat{\prod}_{q_a =-2j_a}^{2j_a} s_1( q_a \cdot \sigma +m_a - i(1-\Delta_a ))} 
& {\rm for}\ SU(2)  
\end{cases} ,
\end{\eq}
where $\hat{\prod}$ denotes $\prod$ with step 2 and
$s_1(x)$ is given by
\be
&& s_1(x) = \prod_{n=1}^{\infty} \left( \frac{n-ix}{n+ix} \right)^n.
\ee
The most important difference from the $\mathcal{N}=3$ theories is that
each matter contribution has both zeroes and poles,
whose degrees are not necessarily one.

\subsubsection{Exact results as resurgent transseries}
We can extend the analyzes in Sec.~\ref{sec:SQED_trans} and \ref{sec:SU2_trans}
straightforwardly.
Taking $\sigma =\sqrt{it}$ again leads us to
\be
Z(g,\{ m_a \} )
= \int_0^{-i\infty} dt\ e^{-\frac{t}{g}} \mathcal{B} Z(t) ,
\ee
where $\mathcal{B} Z(t)$ is the perturbative Borel transformation
\begin{\eq}
\mathcal{B} Z(t)
=\frac{i}{\sqrt{it}} \sum_\pm Z_{\rm 1loop}(\pm \sqrt{it}) .
\end{\eq}
Note that the $a$-th chiral multiplet gives 
poles of $Z_{\rm 1loop}(z)$ with degree $\ell_a$ at
\begin{\eq}
z_{a,\ell}^\ast
= - m_a -i(1-\Delta_a +\ell_a ) ,
\end{\eq}
which gives Borel singularities at
\begin{\eq}
t_{a,\ell}^\ast = -\frac{i}{q_a^2} (m_a +i(1-\Delta_a +\ell_a ))^2 \quad
{\rm with}\ \ell_a \in\mathbb{Z}_+ .
\end{\eq}
Changing the integral contour as in Fig.~\ref{fig:Borel_complex_g}
we decompose the exact result
into the perturbative and nonperturbative parts:
\begin{\eq}
Z (g, \{ m_a \} )
= \int_{0}^\infty dt\ e^{-\frac{t}{g}} \mathcal{B}Z(t)
+\sum_{{\rm poles}\in {\rm 4th\ quadrant}} {\rm Res}_{t=t_{\rm pole}} 
\Bigl[ e^{-\frac{t}{g}} \mathcal{B}Z(t) \Bigr] .
\end{\eq}
Noting that
the poles start to come into 4th quadrant when $m_a =1-\Delta_a +\ell_a$,
we can write the partition function as
\begin{\eq}
Z(g, \{ m_a \} ) = Z_{\rm pt} 
+\sum_{a=1}^{N_f} \sum_{\ell_a =1}^\infty Z_{\rm np}^{(a,\ell_a )} ,
\end{\eq}
where
\begin{\eqa}
 Z_{\rm pt} 
&=&  \int_{0}^\infty dt\ e^{-\frac{t}{g}} \mathcal{B}Z(t)  ,\NN\\
 Z_{\rm np}^{(a, \ell )} 
&=& \theta (m_a -(1-\Delta_a +\ell ) ) \  
{\rm Res}_{t=t_{a,\ell}^\ast}  \Bigl[ e^{-\frac{t}{g}} \mathcal{B}Z(t) \Bigr] .
\end{\eqa}
As in the previous cases,
this decomposition is apparently ambiguous for $m_a =1-\Delta_a +\ell_a$
because of the Borel ambiguities and step function behavior of the transseries parameter.
Indeed the Borel ambiguity in the perturbative sector is
\begin{align}
\left. Z_{\rm pt}(\{ m_b \}  ) \right|_{m_a =1-\Delta_a +\ell_a +0_+}
-\left. Z_{\rm pt}(\{ m_b \}  ) \right|_{m_a =1-\Delta_a +\ell_a +0_-}
= -{\rm Res}_{t=t_{a,\ell}^\ast}\Bigl[ e^{-\frac{t}{g}} \mathcal{B}Z(t) \Bigr] ,
\end{align}
while the non-perturbative ones are
\begin{\eqa}
&& \left. Z_{\rm np}^{(b,\ell )} (\{ m_b \}  ) \right|_{m_a =1-\Delta_a +\ell_a +0_+}
-\left. Z_{\rm np}^{(b,\ell )}(\{ m_b \}  ) \right|_{m_a =1-\Delta_a +\ell_a +0_-} \NN\\
&=& \begin{cases}
 0 &\  {\rm for}\ b\neq a ,\ {\rm or}\ \ell\neq \ell_a \cr
 +{\rm Res}_{t=t_{a,\ell}^\ast} \Bigl[ e^{-\frac{t}{g}} \mathcal{B}Z(t) \Bigr] &\  {\rm for}\ b=a\ {\rm and}\ \ell=\ell_a 
\end{cases} .
\end{\eqa}
Thus the ambiguities are canceled and we find the unambiguous answer consistent with the exact result.

We can also find the resurgent structure
in the situation with fixing $m$ and varying ${\rm arg}(g)$
as in Sec.~\ref{sec:arg}.
By use of similar arguments,
the exact result is decomposed as
\begin{align}
& Z =Z_{\rm pt} +\sum_{a=1}^{N_f} \sum_{\ell_a =1}^\infty 
Z_{\rm np}^{(a,\ell_a )}\,,  \NN\\
& Z_{\rm pt}
=  \int_{0}^{e^{i {\rm arg}(g)} \infty} dt\ e^{-\frac{t}{g}} 
\mathcal{B}Z(t) ,\quad
 Z_{\rm np}^{(a,\ell )}
=  \theta \left( {\rm arg}(g) -{\rm arg}(t_{a,\ell}^\ast )    \right)
{\rm Res}_{t=t_{a,\ell}^\ast}  \Bigl[ e^{-\frac{t}{g}} \mathcal{B}Z(t) \Bigr] .
\end{align}
Although this decomposition apparently has ambiguities 
for ${\rm arg}(g) ={\rm arg}(t_{a,\ell}^\ast )$
estimated by
\begin{align}
( {\mathcal S}_{{\rm arg}(t_{a,\ell}^\ast )  +0^+} 
-{\mathcal S}_{{\rm arg}(t_{a,\ell}^\ast ) +0^-} ) Z(g,\{ m_b \}) ,
\end{align}
they are precisely canceled and the transseries leads us to the unambiguous answer.

\subsubsection{Thimble decomposition}
We discuss thimble decomposition of the integral
\begin{\eq}
Z=\int_{-\infty}^\infty d\sigma\ e^{-S[\sigma ]},\quad
S[z] = -\frac{iz^2}{g} -\log{Z_{\rm 1loop}(z)} .
\end{\eq}
Saddle point equation under this action is given by
\begin{\eq}
0=\begin{cases}
-\frac{2iz}{g} +\pi i \sum_{a=1}^{N_f}  \frac{q_a z +m_a - i(1-\Delta_a )}{\tanh{\left( \pi ( q_a z +m_a - i(1-\Delta_a )) \right)}} 
 & {\rm for}\ U(1) \cr
-\frac{2iz}{g} -\frac{2\pi}{\tanh{(\pi z)}} +\pi i \sum_{a=1}^{N_f} \hat{\sum}_{q_a =-2j_a}^{2j_a} 
\frac{q_a z +m_a - i(1-\Delta_a )}{\tanh{\left( \pi ( q_a z +m_a - i(1-\Delta_a )) \right)}} 
 & {\rm for}\ SU(2)
\end{cases},
\end{\eq}
where we have used the identity
\cite{Jafferis:2010un}
\be
\frac{\pd \log s_1(z) }{\pd z} = \frac{\pi i z}{\tanh (\pi z)} .
\ee

We can analytically solve this equation in weak coupling limit as in the previous cases.
For $g\rightarrow 0$ the saddle points are approximately determined by
\begin{\eq}
0=\begin{cases}
z^c \prod_{a=1}^{N_f}   \frac{\sinh{\left( \pi ( q_a z^c +m_a - i(1-\Delta_a )) \right)}}{q_a z^c +m_a - i(1-\Delta_a )} 
 & {\rm for}\ U(1) \cr
z^c \sinh{(\pi z^c )}  \prod_{a=1}^{N_f} \hat{\prod}_{q_a =-2j_a}^{2j_a} 
\frac{\sinh{\left( \pi ( q_a z^c +m_a - i(1-\Delta_a )) \right)}}{q_a z^c +m_a - i(1-\Delta_a )} 
 & {\rm for}\ SU(2)
\end{cases} ,\quad ( g\rightarrow 0) ,
\end{\eq}
whose solutions are $z=0$, zeros and poles of the integrand in \eqref{eq:exact_general} as expected.
Note that these general cases have much more critical points than the $\mathcal{N}=3$ cases
since each $\mathcal{N}=2$ chiral multiplet gives an infinite number of zeroes as well as poles.
Let $z_{\rm pt}^c$ denoting the critical point satisfying
\begin{\eq}
\lim_{g\rightarrow 0} z_{\rm pt}^c = 0,
\end{\eq}
then we can easily compute the Lefschetz thimble associated with $z_{\rm pt}^c$
in the weak coupling limit:
\begin{\eq}
\lim_{g\rightarrow 0} z_{\rm pt}(s) 
= \epsilon \exp{\left( \frac{2}{g}s  +\frac{\pi i}{4}\right)} .
\label{eq:pert_general}
\end{\eq}

As in the $\mathcal{N}=3$ cases,
it is hard to find critical points analytically for nonzero $g$.
In addition, numerical analysis is also inapplicable without specifying theories.
Therefore we here provides 
expected thimble structures for general case 
based on its resurgent structure in the last subsubsection
and the examples of the thimble structures.
For weak coupling, there are critical points around $z=0$, the zeroes and poles of the integrand.
We identify $z_{\rm pt}^c$ as a ``perturbative critical point" and the ones around the poles as ``nonperturbative critical points".
There are two possibilities of the behavior of the perturbative thimble $z_{\rm pt}(s)$ for finite $g$: 
it would run between $e^{\pm\frac{\pi i}{4}}\infty$ as in \eqref{eq:pert_general}
or it would terminate at a zero of the integrand. 
For the latter case, 
another critical point around the zero contributes and 
its thimble runs from the zero to $e^{\pm\frac{\pi i}{4}}\infty$
so that the thimble combined with the perturbative thimble $z_{\rm pt}(s)$ gets equivalent to \eqref{eq:pert_general} as in Fig.~\ref{fig:thim_CS_SU2_g0.1} 
for the $\mathcal{N}=3$ $SU(2)$ SQCD case.
It is also expected that 
a critical point around the pole $z=z_{a,\ell}^\ast$ starts to contribute around $m_a =1-\Delta_a +\ell $.
Then, there are again two possibilities: the thimble integral associated with this critical point would be equivalent to residue around $z=z_{a,\ell}^\ast$ or would terminate at a zero.
In the latter case, a combination of thimble integrals of the critical points around the pole and the zero 
gets equivalent to the residue.
It is left for the future work to check these expectations explicitly.

\subsection{Other observables}
\label{sec:observables}
So far we have considered only the partition function on a round sphere.
In this subsection we discuss extension of our argument to other observables. 
\subsubsection*{Supersymmetric Wilson loop}
Let us start with the Wilson loop
\begin{\eq}
W_{\mathbf{R}}(C)
={\rm tr}_{\mathbf{R}}
P\exp{\Biggl[ \oint_C ds (iA_\mu \dot{x}^\mu +\sigma |\dot{x}| ) \Biggr]} ,
\end{\eq}
It is known that
this operator preserves two supercharges 
if the contour $C$ is the great circle of $S^3$ \cite{Kapustin:2009kz}. 
Hence we can compute an expectation value of the SUSY Wilson loop by localization:
\begin{\eq}
\langle W_{\mathbf{R}}({\rm Circle})  \rangle
=\langle {\rm tr}_{\mathbf{R}} e^\sigma \rangle_{\rm M.M.} ,
\label{eq:Wilson}
\end{\eq}
where $\langle \cdots \rangle_{\rm M.M.}$ denotes an expectation value in the integral \eqref{eq:Clocalization}.
Note that the difference from the sphere partition function is just insertion of entire function of $\sigma$.
Therefore we can repeat the analyses in the previous sections straightforwardly.
Namely, the SUSY Wilson loop has the same Borel singularities as the sphere partition function
and their resurgent structures are the same
although there are differences in some details such as values of perturbative coefficients and residues around the poles.
The insertion of the Wilson loop changes saddle point equation of the integral and hence thimble structures as well. 
However, since the difference is negligible in the weak coupling limit,
the Wilson loop should not affect the relation between transseries and thimble decomposition,
which we have seen in the sphere partition functions.

\subsubsection*{Bremsstrahrung function in SCFT on $\mathbb{R}^3$}
If we restrict ourselves to superconformal case,
we can also compute Bremsstrahrung function $B$ on $\mathbb{R}^3$ by localization
which determines an energy radiated by accelerating quarks with small velocities
as $E=2\pi B\int dt \dot{v}^2$.
It was conjectured in \cite{Lewkowycz:2013laa} that
the Bremsstrahrung function in 3d $\mathcal{N}=2$ superconformal theory
is given by 
\begin{\eq}
B(g)
=\frac{1}{4\pi^2}\left. \frac{\partial}{\partial b}
\log \langle {\rm tr} e^{b\sigma} \rangle_{\rm M.M.} \right|_{b=1} .
\label{eq:Bremsstrahrung}
\end{\eq}
As in the Wilson loop, the net effect is just insertion of the entire function
and hence we basically arrive at the same conclusion as the Wilson loop.
However, note that we cannot turn on real masses for this case
since we are considering superconformal case.
In other words, we can formally turn on real masses at the level of the integral \eqref{eq:Clocalization}
but its physical interpretation is unclear.
Nevertheless, it is notable that the RHS of \eqref{eq:Bremsstrahrung} with nonzero $m$ shares 
the common resurgent structures with the sphere partition function and Wilson loop.

\subsubsection*{Two-point function of $U(1)$ flavor symmetry currents in SCFT on $\mathbb{R}^3$}
We can also compute 
two-point function of the $U(1)$ flavor symmetry current $j_\mu^a$ for superconformal cases by localization.
It is known that
the two-point function is fixed by the 3d conformal symmetry as
\begin{\eq}
\langle j_a^\mu (x) j_b^\nu (0) \rangle
=\frac{\tau_{ab}}{16\pi^2}
 (\delta^{\mu\nu}\partial^2 -\partial^\mu \partial^\nu ) \frac{1}{x^2} 
+\frac{i\kappa_{ab}}{2\pi}
 \epsilon^{\mu\nu\rho} \partial_\rho \delta^{(3)}(x) ,
\end{\eq}
where $\tau_{ab}(g)$ and $\kappa_{ab}(g)$ are coefficients depending on couplings.
The work \cite{Closset:2012vg} showed that
these coefficients are generated by the sphere partition function with real mass $\{m_a \}$ associated with the $U(1)$ symmetries:
\begin{\eqa}
&& \tau_{ab} (g)
= -\frac{2}{\pi^2}{\rm Re}\left[ 
\frac{1}{Z_{S^3}(g,0)} \frac{\partial^2 Z_{S^3}(g,m) }{\partial m_a \partial m_b}  
  \right]_{ \{m_a \} =0} ,\NN\\
&& \kappa_{ab}(g) = \frac{1}{2\pi} {\rm Im}\left[ 
\frac{1}{Z_{S^3}(g.0)} \frac{\partial^2 Z_{S^3}(g,m) }{\partial m_a \partial m_b}  
 \right]_{ \{m_a \} =0} .
\end{\eqa}
The derivatives by the real masses do not change locations of singularities while their degrees are changed.
This difference, however, does not lead to qualitative change on the resurgent structure and the thimble structures for weak coupling.

\subsubsection*{Partition function and Wilson loop on Squashed $S^3$}
Let us consider partition function on squashed sphere $S^3_b$ with the squashing parameter $b$,
which has a simple relation to supersymmetric Renyi entropy \cite{Nishioka:2013haa}.
The difference from the round sphere partition function in localization formula is just the one-loop determinant \cite{Hama:2011ea}: 
\begin{\eq}
 Z_{\rm 1loop}(\sigma ) 
= \frac{\prod_{\alpha \in {\rm root}_+ }
 4\sinh{(\pi b\alpha \cdot \sigma )} \sinh{(\pi b^{-1}\alpha \cdot \sigma )} }
   {\prod_{a=1}^{N_f} \prod_{\rho_a \in \mathbf{R_a}}  
   s_b \left(  \rho_a \cdot \sigma -\frac{iQ}{2}(1-\Delta_a ) \right)} , 
\label{eq:change_Sb}
\end{\eq} 
where $Q=b+b^{-1}$ and 
\begin{\eq}
 s_b (z)
= \prod_{n_1 =0}^\infty \prod_{n_2 =0}^\infty 
\frac{n_1 b +n_2 b^{-1}+Q/2 -iz}{n_1 b +n_2 b^{-1}+Q/2 +iz} .
\end{\eq}
Note that the round sphere case corresponds to $b=1$.
It was shown in \cite{Honda:2017qdb} that
we can obtain the perturbative Borel transform for general $b$ in a parallel way to the $b=1$ case
and rewrite the $S_b^3$ partition function as the Borel resummation along $\varphi =-\pi /2$.
Therefore the change of the 1-loop determinant \eqref{eq:change_Sb} affects Borel singularities.
Two important differences for us are 
\begin{itemize}

\item Borel singularities associated with each of chiral multiplets become simple poles
and are labeled by two integers.

\item Locations of the singularities depend on $b$.

\end{itemize}
Even for this case, we can still write the partition function as
\begin{\eq}
Z_{S_b^3} (g, \{ m_a \} )
= \int_{0}^\infty dt\ e^{-\frac{t}{g}} \mathcal{B}Z(t)
+\sum_{{\rm poles}\in {\rm 4th\ quadrant}} {\rm Res}_{t=t_{\rm pole}} 
\Bigl[ e^{-\frac{t}{g}} \mathcal{B}Z_{S_b^3} (t) \Bigr] ,
\label{eq:trans_Sb}
\end{\eq}
which is a valid expression except for the Stokes lines.
We regard the first and second terms as perturbative and nonperturbative parts
respectively.
As in the previous cases,
we need to take care of the singularities in the fourth quadrant.
A short calculation shows that
the Borel singularities come on $\mathbb{R}_+$
for $m_a =\frac{n_1 +\Delta}{2}b +\frac{n_1 +\Delta}{2}b^{-1}$
with $n_1,n_2 \in\mathbb{Z}_{\geq 0}$ for $b\in\mathbb{R}_+$
\footnote{
If $b$ is complex, this condition becomes
$ m_a -\frac{n_1 +\Delta}{2}{\rm Im}b -\frac{n_1 +\Delta}{2}{\rm Im}b^{-1}$ 
$=$ $\frac{n_1 +\Delta}{2}{\rm Re}b +\frac{n_1 +\Delta}{2}{\rm Re}b^{-1}$.
}.
For this case, the decomposition \eqref{eq:trans_Sb}
is apparently ambiguous but the ambiguities are canceled between perturbative and nonperturbative parts.

We can also put the supersymmetric Wilson loop on a squashed sphere
constructed in \cite{Tanaka:2012nr}.
Localization formula for the Wilson loop
is insertion of ${\rm tr}_{\mathbf{R}} e^\sigma$ 
or ${\rm tr}_{\mathbf{R}} e^{b^\pm \sigma}$
to the localization formula of the partition function.
Therefore the Wilson loop gives only minor differences
such as values of perturbative coefficients and
details on thimble structures for nonzero $g$.
 
\subsubsection*{Two point function of stress tensor in SCFT on $\mathbb{R}^3$}
For superconformal case,
we can also compute a two-point function of the normalized stress tensor at separate points,
whose expression is determined by conformal symmetry as
\begin{\eq}
\langle T_{\mu\nu}(x) T_{\rho\sigma}(0) \rangle
=\frac{c_T}{64} (P_{\mu\rho}P_{\nu\sigma} +P_{\nu\rho}P_{\mu\sigma} -P_{\mu\nu}P_{\rho\sigma}) \frac{1}{16\pi^2 x^2},
\end{\eq}
where $P_{\mu\nu}=\delta_{\mu\nu}\partial^2 -\partial_\mu \partial_\nu$
\footnote{
We take a normalization such that
$c_T =1$ for single free real scalar. 
}.
The coefficient $c_T (g)$ is generated by $Z_{S_b^3}$ as \cite{Closset:2012ru}  
\begin{\eq}
c_T (g)
= -\frac{32}{\pi^2}{\rm Re} \left[ 
 \frac{1}{Z_{S^3}(g)} \frac{\partial^2 Z_{S_b^3}(g)}{\partial b^2}
 \right]_{b=1} .
\end{\eq}
Although the derivative with respect to $b$ changes degrees of the singularities,
this does not change the resurgent structures so much
as the two-point function of the flavor symmetry currents.

\section{Path integral interpretation of the non-perturbative effects}
\label{Int}
In this section
we discuss possible interpretations of the non-perturbative effects
appearing in the transseries from the path-integral viewpoint.
It is technically obvious that
the non-perturbative effects come from the Borel singularities
or equivalently the poles of the integrand 
of the Coulomb branch localization formula.
In \cite{Honda:2017qdb}, one of the present authors has proposed that
the Borel singularities correspond to complexified SUSY solutions (CSS)
which satisfy SUSY conditions but are not on the original path-integral contour.
The CSS have been constructed for generic 3d $\mathcal{N}=2$ SUSY theory
with Lagrangian and $U(1)_R$ symmetry put on a sphere.
For theories with CS terms,
it has been shown that
classical actions of the CSS are precisely the same 
as the exponents of the residues around Borel singularities \cite{Honda:2017qdb},
which give the non-perturbative corrections appearing in the transseries we have discussed in the present work.
In more detail,
the work \cite{Honda:2017qdb} discussed that
there exist two types of CSS in general:
one has a bosonic parameter while the other has a fermionic one,
which are referred to
as bosonic and fermionic complexified supersymmetric solutions, respectively.
Then it has been proposed that
if there are $n_B$ bosonic and $n_F$ fermionic solutions
with the action $S=S_c /g$,
then 
the Borel transformation includes the following factor 
\begin{\eq}
\mathcal{B}Z (t)
\supset \prod_{\rm CSS} \frac{1}{(t-S_c )^{n_B -n_F}} .
\end{\eq}
For example, in the $\mathcal{N}=3$ $U(1)$ theory discussed in Sec.~\ref{SQED},
there are $\ell$ bosonic and $(\ell -1)$ fermionic solutions 
with the action 
\be
S_{\ell}=-\frac{i}{g}\left[m+(2\ell -1)\pi i\right]^{2}\, ,
\ee
which are precisely the exponentials in the transseries.
Thus it is reasonable to conjecture that
{\it the non-perturbative effects appearing in the transseries correspond 
to the complexified SUSY solutions}.
In this interpretation,
the Borel ambiguity in the perturbative sector
is canceled by ambiguities in the nonperturbative CSS contributions
and the total unambiguous answer obtained in this procedure agrees with the exact result.

However, there are three subtleties in this interpretation.
First,
SUSY solutions are not necessarily saddle points on the curved space
contrary to the flat space.
Indeed the CSS constructed in \cite{Honda:2017qdb}
are saddle points of 3d $\mathcal{N}=2$ SYM coupled to matters
but when we turn on either CS or FI terms,
the CSS do not satisfy saddle-point equations 
while SUSY conditions are still satisfied.
We emphasize that
this does not contradict Lipatov's argument \cite{Lipatov:1977cd}
which states that Borel singularities correspond to saddle points of the theory;
In the localization procedure,
we analyze the following type of path integral
\begin{\eq}
Z(g)
= \int D\Phi \ e^{-\frac{S}{g} -t_{\rm def}QV} ,
\label{eq:localization}
\end{\eq}
where $S$ is the original action, 
$Q$ is supercharge and $V$ is a fermionic functional.
The result is independent of the deformation parameter $t_{\rm def}$,
which is usually taken to be $t_{\rm def}\rightarrow 0$
so that the saddle point analysis becomes exact.
In 3d $\mathcal{N}=2$ theories,
actions of the $\mathcal{N}=2$ SYM theory and chiral multiplets 
can be written in $Q$-exact forms
while the CS and FI terms are $Q$-closed but not $Q$-exact.
In Coulomb branch localization,
we regard 
the SYM and matter actions as the deformation term,
and the CS/FI terms as ``operators" technically.
Therefore the CSS are saddle point of the deformation term
but may not be for the whole action.
Now let us extend the Lipatov's argument to the integral \eqref{eq:localization}.
We can extract $n$-th order perturbative coefficient by
\begin{\eqa}
\frac{1}{2\pi i} \oint \frac{dg}{g^{n+1}} Z(g)
= \frac{1}{2\pi i} \oint dg \int D\Phi  
\exp{\Biggl[ -\frac{1}{g}S[\Phi ] -t_{\rm def}QV -(n +1)\ln{g} \Biggr]} ,
\end{\eqa}
which is independent of $t_{\rm def}$.
For large-$n$, the integral is dominated by the conditions
\begin{\eq}
\left.
\frac{\delta }{\delta \Phi}\left( \frac{S}{g_\ast} +t_{\rm def}QV \right)
\right|_{\Phi =\Phi_\ast} =0 ,\quad\quad
-\frac{1}{g_\ast^2}S[\Phi_\ast ] +\frac{n +1}{g_\ast} =0 ,
\end{\eq}
which leads us to the Borel singularity at $t=S[\Phi_\ast ]$.
Now we use the extra property of the integral \eqref{eq:localization},
namely independence of $t_{\rm def}$.
When $t_{\rm def}$ is very large, 
the first condition approximately becomes
\begin{\eq}
\left. \frac{\delta }{\delta \Phi}QV \right|_{\Phi =\Phi_\ast} =0 ,
\end{\eq}
which is nothing but the condition giving a localization locus.
Therefore, Borel singularities correspond to the localization loci, where the positions of the Borel singularities are given by the actions of the original theory evaluated on the localization loci.
This is similar to the case of the perturbative series for some operators in field theory
in the sense that operators slightly affect saddle point equations
and the original action in localization procedure 
can be technically regarded as an operator.

Second, to verify our conjecture,
we have to check the following two facts:
(1) The Stokes phenomena regarding the nonperturbative contributions we have shown 
should be identified as jumps of intersection numbers 
between the original path integral contour and
dual Lefschetz thimbles associated with the CSS.
(2) The perturbative series in the nonperturbative sector of the transseries 
should agree with perturbative series around the CSS.
Especially the perturbative series should terminate at the one-loop order.
We may be able to check this statement in future works.

Third, to our knowledge,
most of analyses of SUSY localization in the literature 
have not preformed serious saddle-point analysis including complex saddles 
and therefore there is possibility that
we are missing contributions from complex saddles.
In particular, in the Coulomb branch localization formula 
for 3d $\mathcal{N}=2$ theory on $S^3$,
we have picked up only real SUSY solutions which are Coulomb branch solutions,
but we currently know the existence of the CSS,
which may or may not contribute. 
Although we think that this possibility is very unlikely 
since the localization formula has passed many nontrivial tests
such as dualities, AdS/CFT, $F$-theorem and so on,
we have to verify at least that 
the known localization formula is exact by using Lefschetz thimble analysis.

\section{Summary and Discussion}
\label{SD}
We summarize the results obtained in this paper as follows:

(i) We have expressed 
the exact results for the SUSY observables in 3d $\mathcal{N}=2$ Chern-Simons matter theories, which are also seen as $\mathcal{N}=3$ theories,
as the full resurgent transseries composed of 
the perturbative and nonperturbative sectors.
The nonperturbative sectors are given by the residues around the Borel singularities in the fourth quadrant. 
The transseries is also understood from the viewpoint of Lefschetz thimble
associated with saddle points of the effective action with respect to Coulomb branch parameter.

(ii) We have found that,
when the real masses cross the special values,
some of Borel singularities get on the real positive axis and
come to contribute to the partition function as nonperturbative contributions.
It leads to Stokes phenomena, where the perturbative Borel resummation becomes ambiguous.
For example, in the $\mathcal{N}=3$ $U(1)$ CS theory with $N_{f}=1$
for the mass $(2n-1)\pi<m<(2n+1)\pi$,
the $|n|$ Borel singularities contribute to the partition function,
and one more singularity comes to contribute at $m=(2n+1)\pi$.
In the $g\to 0$ limit,
we can rephrase this in the language of the thimble decomposition
that the perturbative thimble
and the $|n|$ thimbles associated with the nonperturbative saddles 
contribute to the partition function in this regime. 

(iii) We have shown that 
the relation between each of the thimble integrals in the thimble decomposition 
and each of building blocks of the transseries 
do not necessarily have one-to-one correspondence for finite $g$.
Each building block of the transseries 
can be expressed as the multiple thimble integrals in general.
For example, we have shown that 
a sum of the thimble integrals associated with the ``perturbative saddle"
and one of the ``nonperturbative saddles"
gives the perturbative Borel resummation along $\mathbb{R}_+$ for $\tilde{m}<m<(2n+1)\pi$, 
where $\tilde{m}$ is a certain value smaller than $(2n+1)\pi$.

(iv) We have proposed path integral interpretations of the nonperturbative contributions 
appearing in the transseries.
We interpret the nonperturbative effects as
the complexified SUSY solutions constructed in \cite{Honda:2017qdb},
up to the three subtleties discussed in Sec.~\ref{Int}.
The contributions from the complex SUSY solutions 
should be shown to yield the nonperturbative exponential contributions 
in the full transseries of the partition function 
by calculating their one-loop or quasi-zero-mode integrals (thimble integral). 
We leave this task for future works,
which will test our interpretation.

(v) Based on our results, 
one may expect that, 
even if a perturbative series of a physical quantity 
is Borel-summable along $\varphi ={\rm arg}(g)$ (e.g. $\varphi=0$ for real positive $g$)
and its resurgent structure is trivial, 
one could obtain its exact result 
by including the residues of ``some" of perturbative Borel singularities 
on the right-half Borel plane which correspond to the nonperturbative contributions. 
In other words, one may be able to obtain the exact result
by deforming a contour in the Borel resummation. 
However, in general, we cannot know which Borel singularities contribute to the exact result only from the perturbative series 
and we may need to perform the thimble decomposition of the path integral. 
In the examples of this paper, we have easily found that 
the Borel singularities in the fourth quadrant are relevant while those in the first quadrant are not
since we have rewritten the integral representation for the exact result 
directly in terms of the Borel resummation.
Note that, even if we do not know this representation of the exact result,
the Lefschetz-thimble analysis enables us to derive the correct contour
as we have explicitly demonstrated.

We conclude this paper with discussing possible future studies.
It is known that, in the Coulomb branch localization formula, 
picking up poles of the one-loop determinant 
gives rise to Higgs branch representation of the partition function
which includes a product of vortex and anti-vortex partition functions 
for some theories \cite{Pasquetti:2011fj,Fujitsuka:2013fga}.  
Since we know that the poles correspond to the bosonic CSS,
it is natural to expect that the CSS are closely related to the Higgs branch representation.
It would be illuminating to make this expectation more precise.

It is interesting to see
whether the resurgent structures become simplified for higher SUSY theories
such as 3d $\mathcal{N}=4$ CS matter theories.
For example, it is known that
sphere partition function of 
the 3d $\mathcal{N}=6$ $U(2)\times U(2)$ ABJM theory (without mass)
has a Borel summable series along $\mathbb{R}_+$ \cite{Russo:2012kj}.
This implies simplifications of the resurgent structure
for the theories with higher SUSY.

We also make a comment on the paper \cite{Gukov:2017kmk}, which discusses the resurgent structure for expansions by the geometric parameter $q=e^\hbar$ in 3d $\mathcal{N}=2$ theories on $D^2 \times_q S^1$.
In that analysis, critical points in the localization formula for their partition functions 
are determined by the twisted effective potentials of 2d $\mathcal{N} = (2,2)$ theories with infinite KK towers
or equivalently so-called Bethe vacua.
Although the authors of \cite{Gukov:2017kmk} consider the expansion and the space that differ from ours,
we expect that some aspects in their problem are also of importance in our problem
since $D^2 \times_q S^1$ is the building block of 3d manifolds including spheres 
\cite{Pasquetti:2011fj,Fujitsuka:2013fga}.
It would be nice to see connections between the analysis in \cite{Gukov:2017kmk} and ours.
Perhaps the detailed analysis of the case for the squashed sphere with $b\rightarrow 0$ may shed some lights on 
this question.

Finally, although this paper has focused on weak coupling expansions,
it is also very interesting to study $1/N$-expansion in the context of AdS/CFT correspondence
which should correspond to perturbative expansion in quantum gravity.
Technically one can study resurgence structures of $1/N$-expansion of 3d $\mathcal{N}=2$ supersymmetric theories
by using the localization formula
but some important simplifications used in this paper are not available.
Most crucially we cannot naively use the technique in \cite{Honda:2016vmv} and sec.~\ref{sec:SQED_trans},
which enables us to rewrite the exact results in terms of Borel resummations
without explicitly computing perturbative coefficients.
Of course Lefschetz thimble decomposition of the (Coulomb branch) localization formula is still applicable
but it would be much harder than the examples in this paper.
Therefore it currently seems that
we need to perform heavy numerical computations of thimble decompositions
or invent some new techniques
unless we work in examples such that we can get explicit closed expressions for exact results.
There are some expectations on resurgence structures of $1/N$-expansions
in theories with string/M-theory duals.
Probably the most understood example is the ABJM theory \cite{Aharony:2008ug},
which is the 3d $\mathcal{N}=6$ superconformal theory dual to type IIA superstring on $AdS_4 \times \mathbb{CP}^3$.
In this case, the gravity dual has D2-brane instantons \cite{Drukker:2011zy}
which are non-perturbative effects of string coupling expansion.
It has been checked that
the D2-brane instantons appear 
in $1/N$-expansions of the ABJ(M) theory 
under the parametrization of AdS/CFT dictionary \cite{Drukker:2011zy,Marino:2011eh}.
Therefore it is natural to expect that
$1/N$-expansions of SUSY Chern-Simons matter theories with string duals
have Borel ambiguities which are related to brane instantons and canceled by ambiguities 
in non-perturbative sectors (in the sense of string theory)
if exact results are resurgent.
Indeed $1/N$-expansion of sphere partition function of the ABJM theory
has Borel ambiguities which are naturally interpreted as D2-brane instantons \cite{Grassi:2014cla} .
There is also a class of Chern-Simons matter theories 
which is expected to be dual to Vasiliev higher spin theories on $AdS_4$.
In this type of correspondence,
$1/N$ corresponds to Newton constant in Vasiliev theory as in standard AdS/CFT
but we currently do not have expectations on resurgence structures 
because there are no works so far on expected non-perturbative effects in Vasiliev theory as far as we know.
Therefore studying $1/N$-expansions in theories with Vasiliev duals would give some insights 
on possible non-perturbative effects in Vasiliev theory.

\begin{acknowledgments}
We thank Muneto Nitta for his early collaboration and discussions.
We are grateful to Sergei Gukov for useful comments to the draft in the first version of arXiv. 
Part of this work has been completed during the workshop 
``Resurgent Asymptotics in Physics and Mathematics"
at Kavli Institute for Theoretical Physics from October 2017.
The authors are also grateful to the organizers and participants 
of ``RIMS-iTHEMS International Workshop on Resurgence Theory" at RIKEN, Kobe. 
This work is supported by 
MEXT-Supported Program for the Strategic Research Foundation
at Private Universities ``Topological Science" (Grant No. S1511006).
This work is also supported in part 
by the Japan Society for the 
Promotion of Science (JSPS) 
Grant-in-Aid for Scientific Research
(KAKENHI) Grant Numbers 18K03627 (T.\ F.), 
16K17677 (T.\ M.) and 18H01217 (N.\ S.)
This work is supported in part by the US Department of Energy Grant No. DE-FG02-03ER41260 (S. \ K.).
\end{acknowledgments}

\appendix
\section{Supersymmetric actions in 3D $\mathcal{N}=2$ theory on $S^3$}
In this appendix
we write down supersymmetric actions in 3D $\mathcal{N}=2$ theory on $S^3$
known in literature.

\subsection{$\mathcal{N}=2$ vector multiplet}
The 3D $\mathcal{N}=2$ vector multiplet 
is dimensional reduction of 4D $\mathcal{N}=1$ vector multiplet and
consists of gauge field $A_\mu$, adjoint scalar $\sigma$, auxiliary field $D$
and gaugino $(\bar{\lambda},\lambda )$.
The 3D $\mathcal{N}=2$ SYM has the following action
\begin{\eqa}
S_{\rm YM}
&=&  \frac{1}{g_{\text{YM}}^2}\int_{S^3} d^3 x \sqrt{g}\,\Tr \Big{[}
     \frac{1}{4} F_{\mu\nu}F^{\mu\nu}+\frac12 D_\mu \sigma D^\mu \sigma 
     +\frac{1}{2} \Big{(} D+\frac{\sigma}{R_{S^3}} \Big{)}^2 \NN\\
&& +\frac{i}{2}\blam \gamma^\mu D_\mu \lambda 
 +\frac{i}{2}\blam [\sigma, \lambda] -\frac{1}{4R_{S^3}}\blam \lambda \Big{]} ,
\end{\eqa}
while the SUSY CS term is given by 
\begin{\eq}
S_{\rm CS} 
= \frac{ik}{4\pi}\int_{S^3} d^3 x \sqrt{g} {\rm Tr} 
                  \Biggl[ \epsilon^{\mu\nu\rho} \left( A_\mu \partial_\nu A_\rho +\frac{2i}{3}A_\mu A_\nu A_\rho \right) 
                            -\blam\lambda +2D\sigma  \Biggr] ,
\end{\eq}
If gauge group includes $U(1)$,
we can add the FI term
\begin{\eq}
S_{\rm FI} 
= -\frac{i\zeta}{2\pi R_{S^3}} \int_{S^3} d^3 x \sqrt{g}\; 
{\rm Tr} \left( D -\frac{\sigma}{R_{S^3}}  \right).
\end{\eq}

\subsection{$\mathcal{N}=2$ chiral multiplet}
The 3D $\mathcal{N}=2$ chiral multiplet 
is dimensional reduction of 4D $\mathcal{N}=1$ chiral multiplet and
consists of scalars $(\phi ,\bar{\phi})$, auxiliary field $(F,\bar{F})$
and fermions $(\psi, \bar{\psi})$.
The SUSY action of the chiral multiplet without superpotential is given by
\begin{\eqa}
S_{\rm chiral}
&=&  \int_{S^3} d^3x \sqrt{g}\,\Big{(}
          D_\mu \bar{\phi}D^\mu \phi +\bar{\phi}\sigma^2 \phi +\frac{i(2\Delta-1)}{R_{S^3}}\bar{\phi}\sigma\phi 
  +\frac{\Delta(2-\Delta)}{R_{S^3}^2}\bar{\phi}\phi +i \bar{\phi}D \phi +\bar{F}F  \NN\\
&& -i \bpsi \gamma^\mu D_\mu \psi 
  +i \bpsi \sigma\psi -\frac{2\Delta-1}{2R_{S^3} }\bpsi \psi 
          +i\bpsi \lambda \phi -i \bar{\phi}\blam \psi  \Big{)} , 
\end{\eqa}
where $\Delta$ is the $U(1)_R$ charge.

\section{Details on computation of perturbative series}
In this appendix 
we compute the perturbative coefficients in the standard way
while we have derived the same results in main text 
by Taylor expanding the Borel transformations.

\subsection{$\mathcal{N}=3$ CS SQED}
\label{QEDpert}
In terms of Euler number and the binomial theorem,
we rewrite the hyper multiplet contribution as
\begin{\eq}
\frac{1}{2\cosh{\frac{\sigma -m}{2}}}
= \frac{1}{2} \sum_{q=0}^\infty \sum_{a =0}^{2q} 
  \frac{E_{2q}}{2^{2q} \Gamma (2q-a +1)\Gamma (a +1)} 
    \sigma^{2q-a} (-m)^{a} .
\label{eq:cosh}
\end{\eq}
Then the perturbative part of the partition function for $N_f =1$ is given by
\begin{\eq}
Z_{\rm pt}
= \frac{1}{2} \sum_{q=0}^\infty \sum_{a =0}^{q} 
  \frac{E_{2q}}{2^{2q} \Gamma (2q-2a +1) \Gamma (2a +1)}  m^{2a} 
  \int_{-\infty}^\infty d\sigma \ \sigma^{2q-2a} e^{\frac{i}{g}\sigma^2} .
\end{\eq}
Using
\begin{\eq}
 \int_{-\infty}^\infty dx \ x^{2n} e^{\frac{i}{g}x^2} 
=\Gamma \left( n+\frac{1}{2} \right) (ig)^{\frac{2n+1}{2}} , 
\end{\eq}
we find
\begin{align}
Z_{\rm pt}
&= \frac{\sqrt{ig}}{2} \sum_{q=0}^\infty \sum_{a =0}^{q} 
  \frac{E_{2q}\Gamma \left( q-a +\frac{1}{2}\right)}
  {2^{2q} \Gamma (2q-2a +1) \Gamma (2a +1)}
  m^{2l} (ig)^{q-a} 
\nonumber\\
&= \frac{\sqrt{ig}}{2} \sum_{q=0}^\infty \sum_{a =0}^\infty 
\frac{E_{2q+2a}\Gamma \left( q +\frac{1}{2}\right)}
  {2^{2q+2a} \Gamma (2q +1) \Gamma (2a +1)} 
  m^{2a} (ig)^q .
\end{align}

For general $N_f$,
applying \eqref{eq:cosh} 
to the contribution from each hyper multiplet 
($\frac{1}{2\cosh{\frac{\sigma -m_a}{2}}}$),
we obtain 
\be
Z_{\rm pt} 
&=& \frac{\sqrt{i g}}{2} 
\sum_{\{q_a \} =0}^{\infty} \sum_{\{l_a \} =0}^{\infty} 
 \frac{\Gamma( \bar{q} + 1/2)}{2^{2(\bar{q}+\bar{l})}} 
\Biggl[ \prod_{a=1}^{N_f} \frac{ E_{2(q_a +l_a)} }
{\Gamma(2 q_a + 1) \Gamma(2l_a + 1)} m_a^{2l_a} \Biggr]  (i g)^{\bar{q}} ,
\ee
with $\bar{q} = \sum_{a=1}^{N_f} q_a$ and $\bar{l} = \sum_{a=1}^{N_f} l_a$.

\subsection{$\mathcal{N}=3$ $SU(2)$ CS SQCD}
\label{QCDpert}
The only difference from SQED is the presence of $(2\sinh{\sigma})^2$.
We first consider $N_f=2$ case.
Using Taylor expansion of this factor and 
applying \eqref{eq:cosh} to each $\frac{1}{2\cosh{\frac{\sigma \pm m_a}{2}}}$,
we find
\be
Z_{\rm pt} &=& \frac{\sqrt{i}}{2^2}
\sum_{s_b=0}^{\infty}\sum_{q_b=0}^{\infty}  \sum_{p_b=0}^{\infty} \sum_{l_b=0}^{2q_b} \sum_{k_b=0}^{2p_b} 
\int^{\infty}_{0} dx \, e^{- x^2 /g } \, \left\{ \prod_{b=1}^{2}
\begin{pmatrix}
  2q_b \\
  l_b
\end{pmatrix}
\begin{pmatrix}
  2p_b \\
  k_b
\end{pmatrix} \right. \nl
&& \times \left.
\frac{2^{-2(q_{b} + p_{b})}  E_{2q_{b}} E_{2p_{b}}}{\Gamma(2s_b+2) \Gamma(2q_{b}+1) \Gamma(2p_{b}+1)} (\sqrt{i}x)^{2(q_b+p_b+s_b) - (l_b+k_b)+1} m_b^{l_b+k_b} \right\} \nl
&=& \frac{\sqrt{i g}}{8}\sum_{s_b=0}^{\infty}\sum_{q_b=0}^{\infty}  \sum_{p_b=0}^{\infty} \sum_{l_b=0}^{2q_b} \sum_{k_b=0}^{2p_b}  2^{-2\bar{Q}} \Gamma(\bar{Q} - \bar{L}/2 + \bar{s} + 3/2 ) (i g)^{\bar{Q} - \bar{L}/2 + \bar{s}+1} \, \delta_{\bar{L} \, {\rm mod}2, 0} \nl
&& \times  \prod_{b=1}^{2}
\begin{pmatrix}
  2q_b \\
  l_b
\end{pmatrix}
\begin{pmatrix}
  2p_b \\
  k_b
\end{pmatrix} 
\frac{E_{2q_{b}} E_{2p_{b}}}{\Gamma(2s_b+2) \Gamma(2q_{b}+1) \Gamma(2p_{b}+1)} m_b^{l_b+k_b} \nl
&=& \frac{\sqrt{i g}}{8} \sum_{s_b=0}^{\infty} \sum_{q_b=0}^{\infty} \sum_{p_b=0}^{\infty} \sum_{l_b=0}^{2q_b} \sum_{k_b=0}^{2p_b}2^{-2\bar{Q}} \Gamma(\bar{Q} - \bar{L}/2 + \bar{s} + 3/2 ) (i g)^{\bar{Q} - \bar{L}/2 + \bar{s}+1} \, \delta_{\bar{L} \, {\rm mod}2, 0} \nl 
&& \times  \prod_{b=1}^2 \frac{E_{2q_b} E_{2p_b}}{\Gamma(2s_b+2) \Gamma(2q_b-l_b+1) \Gamma(2p_b-k_b+1) \Gamma(l_b+1) \Gamma(k_b+1)} m_b^{l_b+k_b} , 
\ee
with $\bar{Q} = q_1 + q_2 + p_1 + p_2$, $\bar{L} =  l_1 + l_2 + k_1 + k_2$, and $\bar{s} = s_1+s_2$.

For general $N_f$ it is expressed as
\be
Z_{\rm pt} &=& \frac{\sqrt{i g}}{2^{2N_f-1}} \sum_{s_b=0}^{\infty} \sum_{q_b=0}^{\infty} \sum_{p_b=0}^{\infty} \sum_{l_b=0}^{2q_b} \sum_{k_b=0}^{2p_b}2^{-2\bar{Q}} \Gamma(\bar{Q} - \bar{L}/2 + \bar{s} + 3/2 ) (i g)^{\bar{Q} - \bar{L}/2 + \bar{s}+1} \, \delta_{\bar{L} \, {\rm mod}2, 0} \nl 
&& \times  \left( \prod_{b=1}^{2} \frac{1}{\Gamma(2s_b+2)} \right) \prod_{b=1}^{N_f} \frac{E_{2q_b} E_{2p_b}}{\Gamma(2q_b-l_b+1) \Gamma(2p_b-k_b+1) \Gamma(l_b+1) \Gamma(k_b+1)} m_b^{l_b+k_b} , 
\ee
with $\bar{Q} = \sum_{b=1}^{N_f} (q_b+p_b)$, $\bar{L} = \sum_{b=1}^{N_f} (l_b+k_b)$, and $\bar{s} = \sum_{b=1}^{2}s_b$.

\section{Brief review of the thimble analysis}
\label{LT}
In the thimble analysis, 
we firstly extend the real variable $x \in {\mathbb R}$ to the complex one $z \in {\mathbb C}$, and then, we obtain the steepest descent given by
\begin{align}
  \frac{ dz(s)}{d s} \,=\, \overline{F(z(s))}, \quad
   F(z) = \frac{\partial S[z]}{\partial z},
\end{align}
where $s$ is a real flow parameter and 
$\overline{F}$ 
denotes the complex conjugation of $F$. 
The critical points $z^{\rm c}_{\sigma}$ are obtained by solving 
$F(z^c)=0$.
The thimble ${\cal J}_{\sigma}$ associated 
with the critical point $\sigma$ is determined as a particular 
flow with the initial condition given by
\begin{align}
  \lim_{s \rightarrow - \infty} z_{\sigma}(s) = z^{\rm c}_{\sigma},
\end{align}
whereas the dual thimble ${\cal K}_{\sigma}$ is defined by a 
flow with the condition 
$\lim_{s \rightarrow + \infty} z_{\sigma}(s) = z^{\rm c}_{\sigma}$. 
One can easily find that
\begin{align}
  \frac{d {\rm Re} S[z(s)]}{ds} \, \ge \, 0  \quad{\rm and}\quad
  \frac{d {\rm Im} S[z(s)]}{ds} \, = \, 0.
\end{align}
The original integration contour ${\cal C}_{\mathbb R}$ 
can be reproduced by a linear combination of the thimbles 
as
\begin{align}
{\cal C}_{\mathbb R} \,=\, \sum_{ \sigma \in \Sigma} n_{\sigma} 
{\cal J}_{\sigma}, \label{eq:CeqnJ}
\end{align}
where $\Sigma$ is a set of the critical points and $n_{\sigma}$ 
is an integer, called the intersection number.
The intersection number is determined by each of dual-thimbles 
${\cal K_{\sigma}}$ so as to have the same homology class as the real contour:
$n_{\sigma} = \pm 1$ if the dual-thimble has an intersection with 
the real contour, $n_{\sigma}=0$ otherwise.

By choosing particular values of parameters, one might encounter the Stokes phenomenon, which is defined as
\begin{align}
 \lim_{s \rightarrow +\infty}  z_{\sigma_1}(s) \, = \, 
z^{\rm c}_{\sigma_2}, \quad (\sigma_1 \ne \sigma_2),
\end{align}
and (\ref{eq:CeqnJ}) becomes ill-defined.
Even if the Stokes phenomenon occurs, one can avoid the phenomenon
by introducing a sufficiently small complex phase to a parameter.
This fact implies that there is an ambiguity regarding choices of the 
modified contours to avoid the Stokes phenomenon.

The complexified configuration space generally has not only critical points but also other objects such as singularities(sources) and zero-points(sinks) defined as
\be
&& {\rm Re} S[z] =
\begin{cases}
-\infty  & \quad \mbox{for singularities} \\
+\infty  & \quad \mbox{for zero-points}
\end{cases}.
\ee
These points have the role of end-points of the thimbles.


\end{document}